\documentclass[12pt,preprint]{aastex}
\usepackage{graphics}

\shorttitle{Bright Rimmed Clouds, I}
\shortauthors{De Vries, Narayanan, \& Snell}

\renewcommand{\deg}{\mbox{$^{\circ}$}}
\renewcommand{\min}{\mbox{$^{\prime}$}}
\renewcommand{\sec}{\mbox{$^{\prime\prime}$}}
\newcommand{\jequals}[2]{\mbox{$J = {#1}\rightarrow{#2}$}}
\newcommand{\nthp}{N$_{2}$H$^{+}$}
\newcommand{\hcop}{HCO$^{+}$}
\newcommand{\hthcop}{H$^{13}$CO$^{+}$}

\newcommand{\ceio}{C$^{18}$O}

\begin{document}

\title{Star Formation in Bright Rimmed Clouds. I. Millimeter and
  Submillimeter Molecular Line Surveys}
\author{Christopher H. De Vries, Gopal Narayanan, Ronald L. Snell}
\affil{Five College Radio Astronomy Observatory, 619 LGRC, University
  of Massachusetts, Amherst, MA 01003}

\begin{abstract}
We present the results of the first detailed millimeter and submillimeter
molecular line survey of bright rimmed clouds, observed at FCRAO in the
CO~(\jequals{1}{0}), \ceio~(\jequals{1}{0}), \hcop~(\jequals{1}{0}),
\hthcop~(\jequals{1}{0}), and \nthp~(\jequals{1}{0}) transitions, and at the
HHT in the CO~(\jequals{2}{1}), \hcop~(\jequals{3}{2}), \hcop~(\jequals{4}{3}),
\hthcop~(\jequals{3}{2}), and \hthcop~(\jequals{4}{3}) molecular line
transitions. The source list is composed of a selection of bright rimmed clouds
from the catalog of such objects compiled by \citet{sfo}. We also present
observations of
three Bok globules done for comparison with the bright rimmed clouds. We find
that the appearance of the millimeter CO and \hcop\ emission is dominated by
the morphology of the shock front in the bright rimmed clouds. The
\hcop~(\jequals{1}{0}) emission tends to trace the swept up gas ridge and
overdense regions which may be triggered to collapse as a result of sequential
star formation. Five of the seven bright rimmed clouds we observe seem to have
an outflow, however only one shows the spectral line blue-asymmetric signature
that is indicative of infall, in
the optically thick \hcop\ emission. We also present evidence that in bright
rimmed clouds the nearby shock front may heat the core from outside-in thereby
washing out the normally observed line infall signatures seen in isolated star
forming regions. 
We find that the derived
core masses of these bright rimmed clouds are similar to other low and
intermediate mass star forming regions.
\end{abstract}


\section{Introduction}
Triggered Star Formation is the process by which star formation is initiated or
accelerated through compression of a clump in a molecular cloud by a shock
front \citep{elm1}. Although a basic understanding of the process of triggered
star formation
exists, only recently have numerical models becoming
sophisticated enough to yield detailed comparisons with
observations. Comparison of hydrodynamic shock induced star formation models to
observations of potential triggered star formation regions is complicated by
the fact that such regions can be very chaotic. Shocks, outflows,
and core rotation all have kinematic signatures which may overwhelm the
signatures associated with the triggering mechanism \citep{elm2}. In order to
alleviate this
confusion somewhat, we choose to observe regions of potential small scale
triggering. These regions typically involve one core in a cometary
cloud, and a well defined shock front geometry. 

Several numerical studies of shock triggered collapse have been undertaken
recently \citep[hereafter VC]{bos,fb1,fb2,vc}. These studies have found that
shocks with
velocities less than 45 km s$^{-1}$ can cause cores to collapse, but fast
shocks ($> 100$ km s$^{-1}$) tend to destroy the principal cooling molecules,
leading ultimately to the destruction of the cloud \citetext{VC}. These models
also yield an appreciable number of binary and multiple star systems, which are
common in our galaxy, yet difficult to form by spontaneous, isolated star
formation. The VC models tend to reproduce characteristics of observed
cometary
clouds thought to be small scale triggered star formation regions
\citep{elm1,elm2}. This comparison between the models and the observations will
be explored further in a forthcoming paper. 

\begin{figure}
\epsscale{1.0}
\plotone{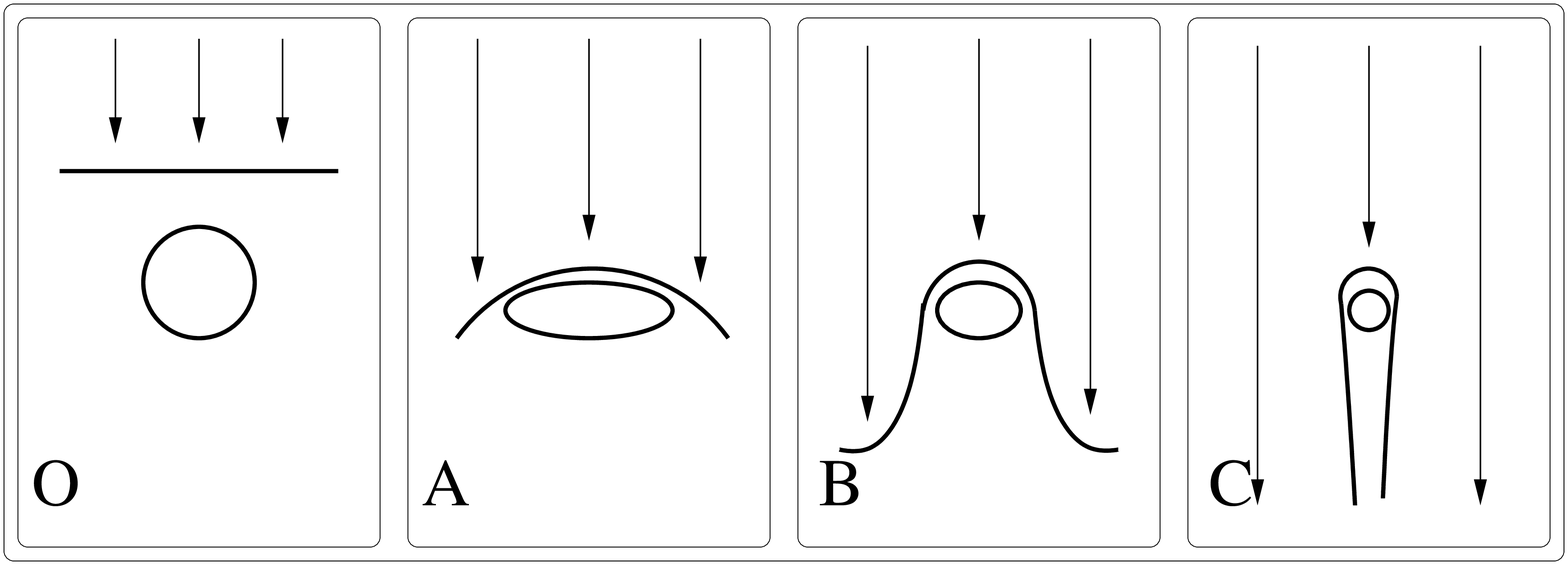}
\caption{\label{schematic} This schematic shows a planar shock from approaching
  a dense clump (0). As the shock front hits the clump it slows down, and the
  clump flattens out, this corresponds to the type~A morphology. The clump
  contracts under the influence of the shock front, and shields the gas behind
  it from ionization as the shock front passes, this corresponds to the type~B
  morphology. The clump continues to contract and shield the gas behind it,
  forming a cometary morphology, this correspondes to the type~C morphology.}
\end{figure}

In order to perform a detailed study of sites of triggered star formation,
which can be
compared to models, we conducted a survey of bright rimmed clouds identified by
\citet[hereafter, SFO]{sfo}. These are all molecular clouds which border
expanding HII regions. At the boundary between the ionized gas within the
Stromgren sphere, and the molecular cloud is an ionization front, which shows
up as
the bright rim identified by SFO. Embedded near the rim, and just within
the molecular cloud in each of these sources is an IRAS source. The fact that
the ionization front is collecting gas into a ridge or core in many of these
sources
makes them excellent candidates for the globule squeezing or collect and
collapse mode of triggered star formation \citep{elm1,elm2}. These clouds are
also divided into three morphological
types, shown schematically in Figure~\ref{schematic}. The first is type~A,
which has a moderately curved rim, and looks much like a shield in three
dimensions. The second is type~B, which has a more tightly curved rim near the
head of the cloud, but which tends to broaden near the tail. Type~B is also
known as an elephant trunk morphology. The third is
type~C which has a very tightly curved rim and a well defined tail. Type~C is
often called a cometary cloud. The shock induced collapse models of VC as well
as observations of the expanding ionization front of the Orion OB 1 association
\citep{os2} suggest that these morphological types of bright rimmed clouds may
actually be a time evolution sequence with clouds evolving from type~A through
type~B to type~C. 

\section{Observational Approach}
\subsection{Source Selection}
Potential triggered star formation regions have been selected from the catalog
of bright-rimmed clouds with associated IRAS point sources by SFO. Each
of these sources is associated with an HII region and is suspected to be
forming stars via triggering. In all of these sources the geometry of the
incoming shock region is obvious, as the shock emanates from the interface of
the cloud and HII region. Since there are three different morphologies, or
bright rim cloud types, we decided to chose at least two of each type. We also
chose nearby sources, both to get the best possible resolution, and because
most Bok globules that have been studied have been nearby objects. The bright
rimmed clouds we observed have distances between 190 and 1900 pc, and include
two type~A bright rimmed clouds (SFO~16 and SFO~18), three type~B bright rimmed
clouds (SFO~4, SFO~13, and SFO~25), and two
type~C bright rimmed clouds (SFO~20 and SFO~37). We cannot hope to resolve the
cores of the more
distant
sources with single dish millimeter or submillimeter observations, however we
do resolve the structure of the swept up molecular gas cloud surrounding the
core, which is greatly affected by the impacting ionization front. The embedded
IRAS sources in the selected bright rimmed clouds have infrared luminosities
which range from 5 to 1300 $L_{\odot}$, though all but one of the sources have
a luminosity less that 110 $L_{\odot}$. These luminosities tend to be typical
of low or intermediate mass star formation. Finally all but one of the bright
rimmed clouds were chosen to have the IRAS colors of their embedded IRAS point
source be consistent with Young Stellar Objects (YSOs) according to the color
criteria presented in \citet{bei}. There are two
exceptions to this rule. The first, SFO~4, was chosen
because it is exceptionally nearby (190 pc). 
The second, SFO~13, was chosen because it is thought to be a region of cluster
formation \citep{chs}. Although SFO~4 and SFO~13 do not meet the IRAS color
criteria of \citet{bei} for inclusion in the population of YSOs, they do meet
the criteria of \citet{chs} for embedded star forming regions.
Digitized Sky Survey observations
of the bright rimmed clouds are shown in Figure~\ref{sfo_dss}. In most of the
optical images the bright rim is clearly visible, enclosed within the rim is
the molecular cloud. Several of the sources show high extinction on one side of
the rim indicating the presence of the molecular cloud. We have drawn in an
arrow indicating the direction of propagation of the ionization front from the
ionizing source creating the HII region.

In order to compare potential triggered star formation regions with quiescent
star formation regions, we also observed Bok globules with embedded IRAS
sources in their cores from the \citet{cb} catalog
of small,
optically selected molecular clouds. We chose Bok globules with properties that
are similar to our bright rimmed clouds. They lie at distances between 250 and
2500 pc from us. They are believed to contain one star forming core with an
embedded IRAS
source, and their luminosities between 3 and 930 $L_{\odot}$. Observations of
these sources allow for detailed
comparisons between the observed properties of triggered and spontaneous star
formation regions, and will allow us to isolate those observed properties
unique to triggered star formation. The three Bok globules we have observed are
B335, CB~3, and CB~224. The source list is summarized in table~\ref{sources}. 

\subsection{Choice of Molecular Transitions}
One cannot obtain a complete picture of the physical conditions in a star
forming cloud by observing one molecular line. It is only by combining
observations of
several molecular line transitions, sensitive to various physical conditions,
that one can approach an accurate physical representation of the star forming
cloud. As a result we chose several molecular transitions, in both the
millimeter and submillimeter wave bands, in order to probe the conditions we
felt would provide us with a basis for comparing bright rimmed clouds with Bok
globules. The transitions we observed are summarized in
table~\ref{transitions}. 

CO has traditionally been the main tracer of molecular clouds. It has a fairly
high abundance and a low critical density, making it a very useful tracer of
molecular gas. Our observations of CO were driven by the fact that both the
CO~(\jequals{1}{0}) and the CO~(\jequals{2}{1}) emission are good tracers of
molecular outflows. \hcop is a good tracer of moderate density ($\geq 10^{4}$
\mbox{cm$^{-3}$}). We observe the millimeter \hcop~(\jequals{1}{0}) line
to trace the morphology of the molecular gas swept up by the ionization
front. The submillimeter \hcop~(\jequals{3}{2}) and \hcop~(\jequals{4}{3})
trace the denser clumps within the swept up gas. \hcop emission in all three
tracers tends to be optically thick in the cores, making it a good tracer of
infall as discussed in \S\ref{infall}. \nthp\ has been shown to be a good
tracer
of the dense core material as it can only form in well-shielded regions
\citep{tur}. \nthp\ also remains relatively undepleated in cores \citep{bl},
making it especially useful in probing the densest regions of star forming
clouds. The \nthp~(\jequals{1}{0}) transition has seven well measured hyperfine
components resulting from the interaction of the molecular electric field
gradient and the electric quadrupole moments of the nitrogen nuclei
\citep{cmt}. These components can be used to determine the optical depth of
\nthp\ from which we can derive the column density of \nthp. From this
measurement we can infer the density of star forming cores.

\subsection{Presentation of the Data}

In \S\ref{results} we will discuss our observations of each source in a
systematic way. This section describes the order in which various observations
will be described and also points out various features which we look for in
these sources. For each source we present figures illustrating the line
profiles at the IRAS source position, an integrated intensity map of each
tracer which we mapped, a map of the centroid velocity of the
\hcop~(\jequals{1}{0}) transition, and a map of the molecular outflow if one is
detected. Our rationale for presenting these figures is discussed below.

We begin by 
presenting  the central line profiles of all the observed molecular
transitions. These profiles are taken at the position of the IRAS source. There
are
several features in the line profiles which are of interest to us. The CO lines
often have broad low intensity features called wings. These wings, which are
especially prominent in the CO~(\jequals{2}{1}) transition, are usually the
result of an outflow from the forming protostellar object. These outflows are
often bipolar and are thought to arise from megnetohydrodynamic winds driven by
infall and interaction with the forming star's surrounding disk (see
\citet{bac} and references therein). The shape of the line profile can also be
of interest. The \hcop\ transitions often have profiles which deviate
significantly from gaussian shapes at line center. This is due to the high
optical depth which results in self-absorption near the line center. Often this
results in a profile with two distinct peaks. When this is indeed the result of
self-absorption, the lower abundance \hthcop\ lines are often still optically
thin and tends to show their greatest intensity at a velocity between the twin
peaks of their self-absorbed isotopomer \hcop. The relative size of each peak
often indicates the kinematics of the gas within the beam. When the gas is
collapsing to a point within the beam, the red-shifted gas is preferentially
self-absorbed leaving a more intense blue-shifted peak and an overall
blue-asymmetry with respect to the \hthcop\ line. The reverse is true for gas
expanding from a point within a beam \citep{sl,zhou2,zhou}. 

The integrated intensity maps provide a look at the two-dimensional morphology
of the molecular cloud and cores as seen by the different tracers we mapped
over the region surrounding the IRAS sources. The CO integrated intensity map
gives a global look at the molecular gas in the region. One thing to note in
the bright rimmed clouds we observe is that the boundary between the molecular
cloud and the ionized gas is very sharp and well traced by the CO. The \hcop\
tends to trace denser regions, but is also excited by shocks \citep{bpg}. As a
result
\hcop~(\jequals{1}{0}) tends to trace both the star forming cores as well as
the rim of the ionization front. The higher level \hcop\ transitions tend to
trace the dense gas. We overlay the \nthp\ half-power contour on the integrated
intensity map of a higher level \hcop\ transition in order to show the
correspondence between the dense gas traced by \nthp\ and that traced by
\hcop. There is often good agreement between the two tracers. Following the
integrated intensity maps we present \hcop~(\jequals{1}{0}) centroid velocity
maps for most of the sources. Centroid velocity maps have been shown to be very
effective in disentangling the underlying velocity fields in regions of complex
kinematics \citep{nw}. Often there is a
velocity gradient which may correspond to rotation of the cloud or be a result
of acceleration of the molecular gas due to the rocket effect which arises at
the molecular cloud boundary. 

The kinematic signature suggested by the \hcop\ is often interesting when
compared to the integrated intensity maps of the CO outflows, which we present
next for many of the sources we observe. Often the bipolar outflow we detect in
CO is perpendicular to the velocity gradient we detect in \hcop. This is
evidence that we may be detecting rotation along the direction predicted in
current models of star formation. We do not have the angular resolution,
however, to probe this rotation down to the small scales of the star forming
core.

\section{Observations}

\subsection{FCRAO 14m}
We performed observations at the Five College Radio Astronomy
Observatory\footnote{FCRAO is supported in part by the National Science
  Foundation under grant AST 01-00793.} (FCRAO) 14~m telescope over the
1999--2000 and 2000--2001 observing seasons using the Second Quabbin
Observatory Imaging Array
(SEQUOIA) 16-beam array receiver, and the Focal Plane Array Autocorrelation
Spectrometers (FAAS) which consists of 16 autocorrelating spectrometers. A
summary of the observed molecular line transitions is shown in
table~\ref{transitions}. The \ceio, \hcop, \hthcop, and \nthp\ transitions were
observed using the frequency-switched mode, while CO was observed by position
switching to a region that was reasonably clear of CO emission, though some
CO contamination was found and considered acceptable in the offset positions as
long as it was well separated in velocity from the line of interest. The \hcop
and \hthcop\ were frequency folded and 3$^{\rm rd}$ order baselines were
removed, while
for the \nthp\ data a 4$^{\rm th}$ order baseline was used. A 2$^{\rm nd}$
order
baseline was subtracted from the position-switched CO data. \hcop\ and \nthp\
were mapped with half-beam spacing, while CO and \hthcop\ were mapped with
full-beam spacing. The extent of the region mapped varies depending on the
spatial extent of the source's emission.

\subsection{HHT 10m}
We also performed submillimeter observations in April and December 2000 and in
April 2001 with the 10~m
Heinrich Hertz Telescope\footnote{The HHT is operated by the Submillimeter
  Telescope Observatory (SMTO), and is a joint facility for the University of
  Arizona's Steward Observatory and the Max-Planck-Institut f\"{u}r
  Radioastronomie (Bonn).} (HHT). CO \jequals{2}{1}, \hcop, and \hthcop
\jequals{3}{2} observations were conducted using the
facility 230 GHz SIS receiver, 
while \hcop\ and \hthcop\ observations were conducted using the facility dual
polarization
345 GHz SIS receiver. Several backends were used simultaneously, including a 1
GHz wide ($\sim 1$ MHz resolution) acousto-optical spectrometer (AOS), and
three
filterbanks with 1 MHz, 250 kHz, and 62.5 kHz resolutions. The results
presented in this paper were processed using the 250 kHz filterbank. All
observations were position-switched, and second order baselines were
removed. \hthcop\ observations were made of the star forming core at the IRAS
source position, while \hcop\ and CO observations were made using the 
on-the-fly (OTF) mapping technique. Observations of $120\sec \times 60\sec$
regions in OTF mode were combined into maps of various spatial extents
depending on the source. 

\subsection{CSO 10.4m}
We observed B335 in May of 1996 at the Caltech Submillimeter
Observatory\footnote{The CSO is operated by the California Institute of
Technology under funding from the National Science Foundation, grant number AST
93-13929.} (CSO). \hcop~(\jequals{4}{3}) mapping was performed using the OTF
mapping technique. The OTF data was gridded into 49 positions with a spacing of
10\sec. \hthcop~(\jequals{4}{3}) was observed at the IRAS source position. All
observations were carried out with low-noise SIS waveguide receivers and a 1000
channel, 50 MHz wide acousto-optical spectrometer. The main beam efficiency at
345 GHz is $\sim 0.65$. Calibration of observations was done by the chopper
wheel technique.

\section{Results}
\label{results}
\subsection{Bright Rimmed Clouds}
\subsubsection{SFO~4}
SFO~4 is a bright rimmed cloud with a type~B
morphology on the edge of the HII region S185 ($\gamma$ Cas) (SFO). IRAS source
00560+6037 is embedded within SFO~4. This IRAS source has been classified by
SFO as having IRAS colors consistent with hot cirrus sources, and not a forming
or
newly formed star (SFO). However according to the criteria of \citet{chs} SFO~4
fits the IRAS color profile of an embedded star forming region. SFO~4 is nearby
(190 pc), and has no known associated outflows or HH objects. 

\begin{figure}
\epsscale{1.0}
\plotone{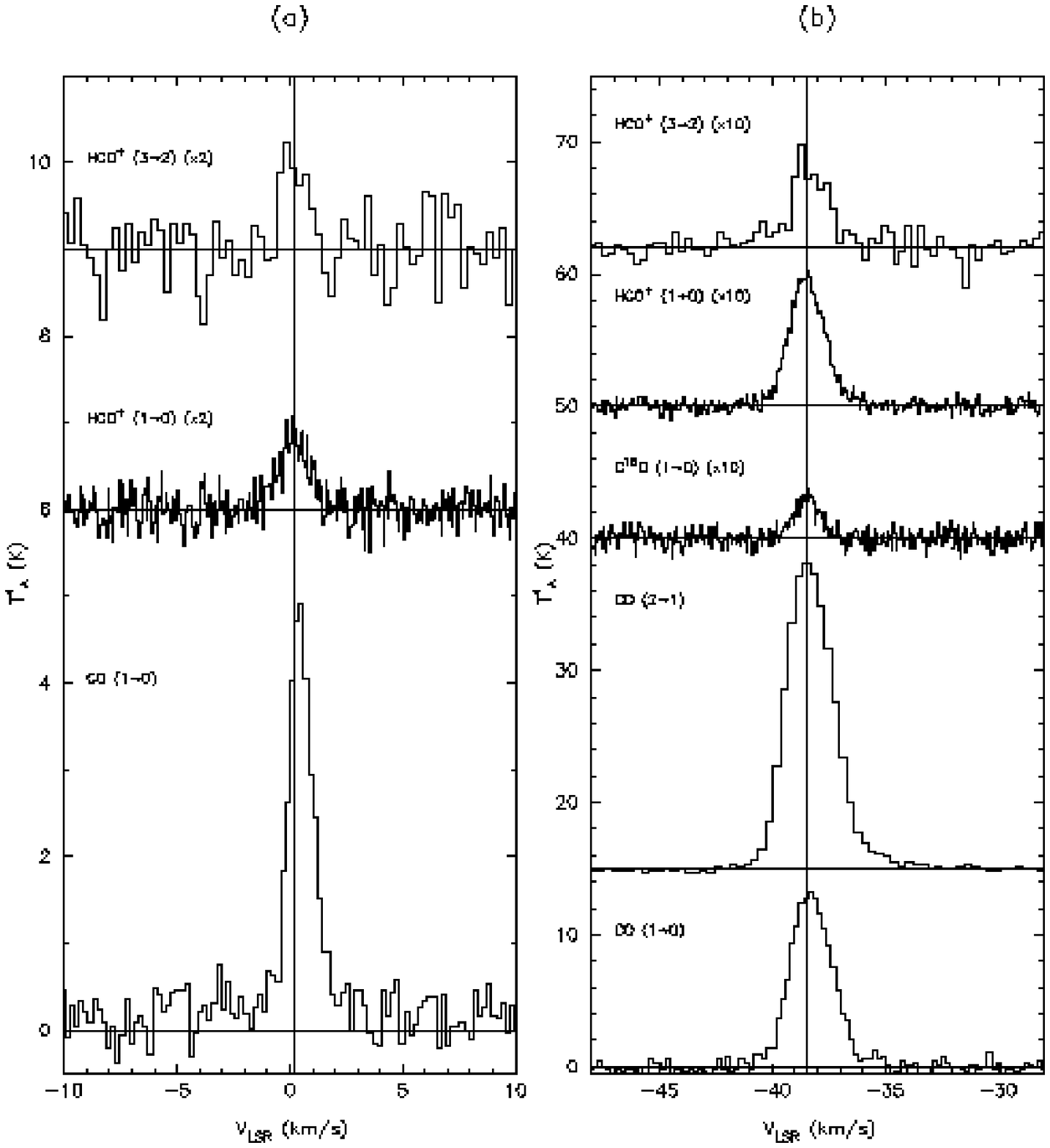}
\caption{\label{sfoiv_xiii_profiles} Line profiles of molecular transitions in
  the direction of the central IRAS source of SFO~4 (a) and SFO~13 (b).}
\end{figure}

Figure~\ref{sfoiv_xiii_profiles}a shows the spectra of our observations towards
the IRAS source in
SFO~4. Although \hthcop\ (\jequals{1}{0}) was not detected, our observations
limit the integrated intensity of that transition to less that
0.037~K~km~s$^{-1}$. Based on this limit, and the integrated intensity
measurement of \hcop\ (\jequals{1}{0}), and an assumed C/$^{13}$C abundance
ratio of 64 we derive an upper limit to the column density of \hcop\ to be
$2\times
10^{13}$ cm$^{-2}$. The line profiles appear fairly gaussian, in all the
observed transitions, with nearly identical centroid velocities. None of the
isotopic lines of either CO or \hcop\ were detected.

\begin{figure}
\epsscale{0.9}
\plotone{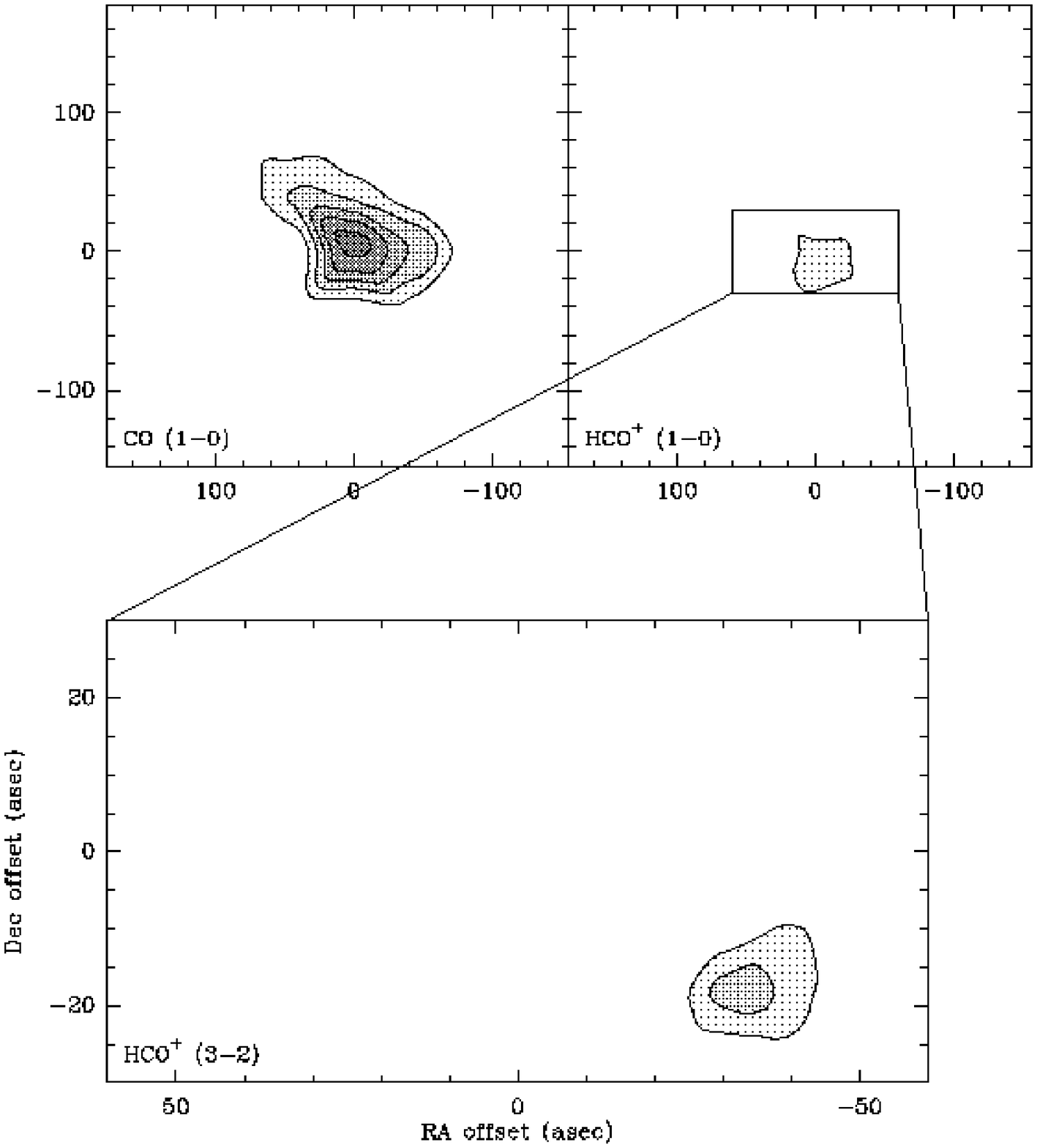}
\caption{\label{sfoiv_itty} Integrated intensity maps of SFO~4 in various
  transitions and isotopomers of \hcop and CO. The IRAS source 00560+6037 is
  located at the center of each map. The CO~(\jequals{1}{0}) map has a lowest
  contour of 1.9 K~\mbox{km~s$^{-1}$} ($3\sigma$) and increments of 1.3
  K~\mbox{km~s$^{-1}$} ($2\sigma$). The \hcop~(\jequals{1}{0}) map has a
  lowest contour of 0.6 K~\mbox{km~s$^{-1}$} ($3\sigma$) and increments of 0.2
  K~\mbox{km~s$^{-1}$} ($1\sigma$). The \hcop~(\jequals{3}{2}) map has a lowest
  contour of 0.7 K~\mbox{km~s$^{-1}$} ($2\sigma$) and increments of 0.3
  K~\mbox{km~s$^{-1}$} ($1\sigma$).}
\end{figure}

Figure~\ref{sfoiv_itty} shows the \hcop\ and CO integrated intensity
maps. Neither the CO (\jequals{1}{0}) map, nor the \hcop\ maps show much
structure. In the CO (\jequals{1}{0}) map and the \hcop\ (\jequals{1}{0}) map
the cloud core surrounding the embedded IRAS source is visible, but no gas
traces the ionization front or the elephant trunk morphology seen in the DSS
image
in Figure~\ref{sfo_dss}. The \hcop\ (\jequals{3}{2}) map shows only one point
whose integrated intensity is above 1.0 K \mbox{km~s$^{-1}$}, the three
sigma threshold for the map. This source is unusually weak compared to the
other sources in the sample, which is unfortunate as it is also the most nearby
source in the sample.

\subsubsection{SFO~13}
\begin{figure}
\epsscale{1.0}
\plotone{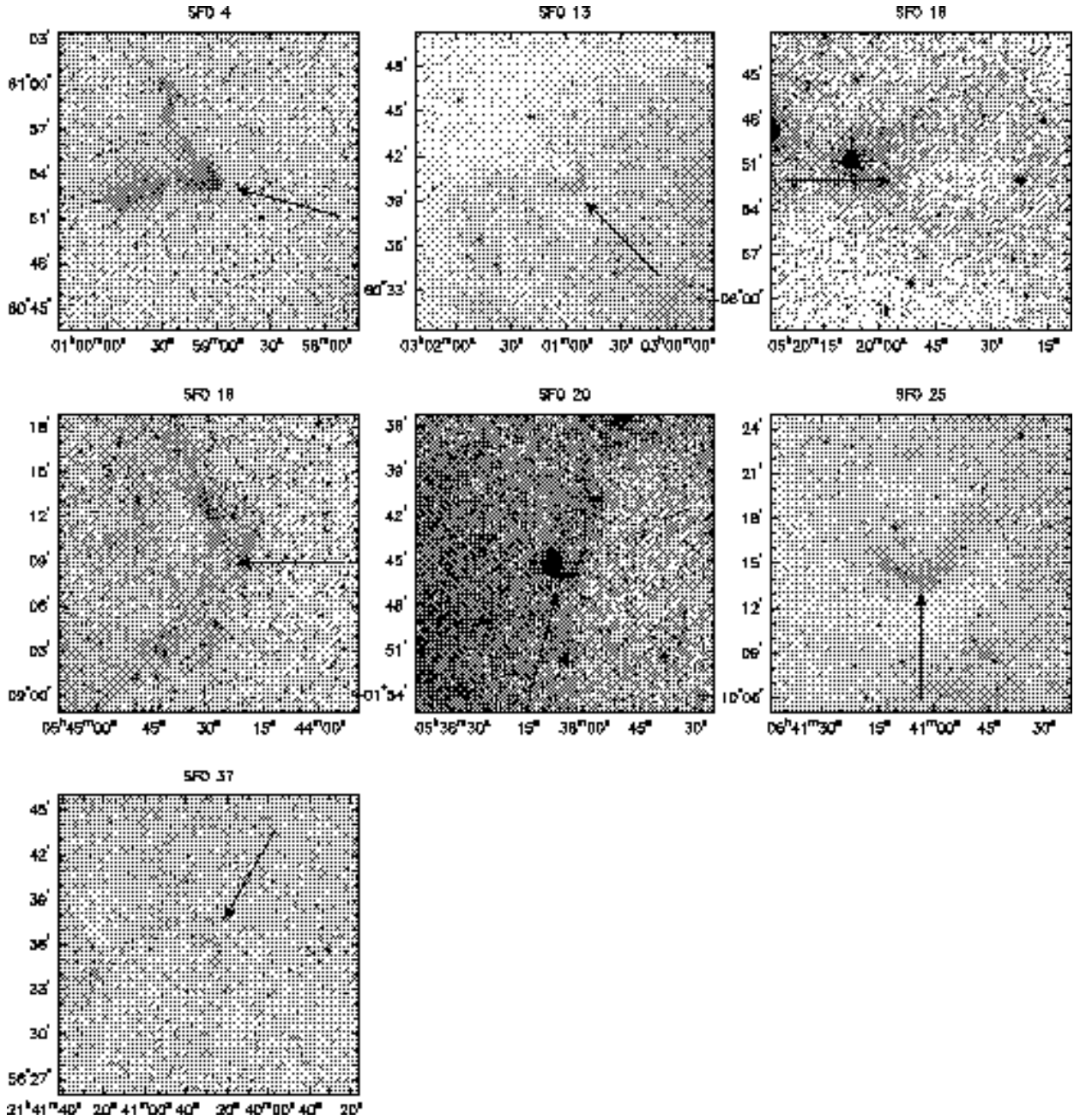}
\caption{\label{sfo_dss} Digitized Sky Survey Image of SFO objects. These
  images show 20\min\ by 20\min\ sections of the sky centered on the embedded
  IRAS object. SFO used these images to classify these objects into their
  morphological scheme. The arrows indicates the direction of motion of the
  ionization front.}
\end{figure}

SFO~13 is classified as a type~B bright rimmed cloud by
SFO. Figure~\ref{sfo_dss} shows the elephant trunk morphology characteristic of
type~B
clouds. The initially cometary head of the cloud broadens into a wider
structure as one follows the cloud away from the ionizing source. SFO~13
borders the HII region IC1848, and lies at a distance of 1.9 kpc from
us, making it the most distant SFO source in the sample. A recent near-infrared
survey reveals $\sim20$ point sources whose spectral energy distributions are
consistent with pre-main-sequence stars \citep{sto2,sto}. $K^{\prime}$ band
observations by \citet{chs} also confirm the presence an embedded stellar
cluster in this molecular cloud.
These sources are
distributed preferentially towards the HII region with respect to the IRAS
source, and are probably older than the IRAS source indicating that this cloud
may be an example of small scale sequential star formation. 

The millimeter and submillimeter line profiles for SFO~13, shown in
Figure~\ref{sfoiv_xiii_profiles}b, are all fairly gaussian and also peak at
approximately the same velocity. This is not entirely unexpected as the average
infall region for individual cores in a star formation region is smaller than
the FCRAO beamwidth, which is  0.5~pc at this distance, and the HHT beamwidth,
of 0.3~pc, making it difficult to detect infall. 
The CO~(\jequals{2}{1}) line strength
is almost twice as much as that of CO~(\jequals{1}{0}). 
This intensity difference may be due to excitation effects, but may also be the
result of the small angular size of the core. Since the 
CO~(\jequals{2}{1}) transition is 
observed with a much smaller beam than the CO~(\jequals{1}{0}) transition, the
beam dilution is much greater for the lower transition and can have significant
effects on the intensities observed with such distant sources.
The \hcop\ and \ceio\
line profiles are all fairly gaussian. \nthp~(\jequals{1}{0}) was not detected
in this source, down to an rms level of 0.017 K~km~s$^{-1}$. 

\begin{figure}
\epsscale{1.0}
\plotone{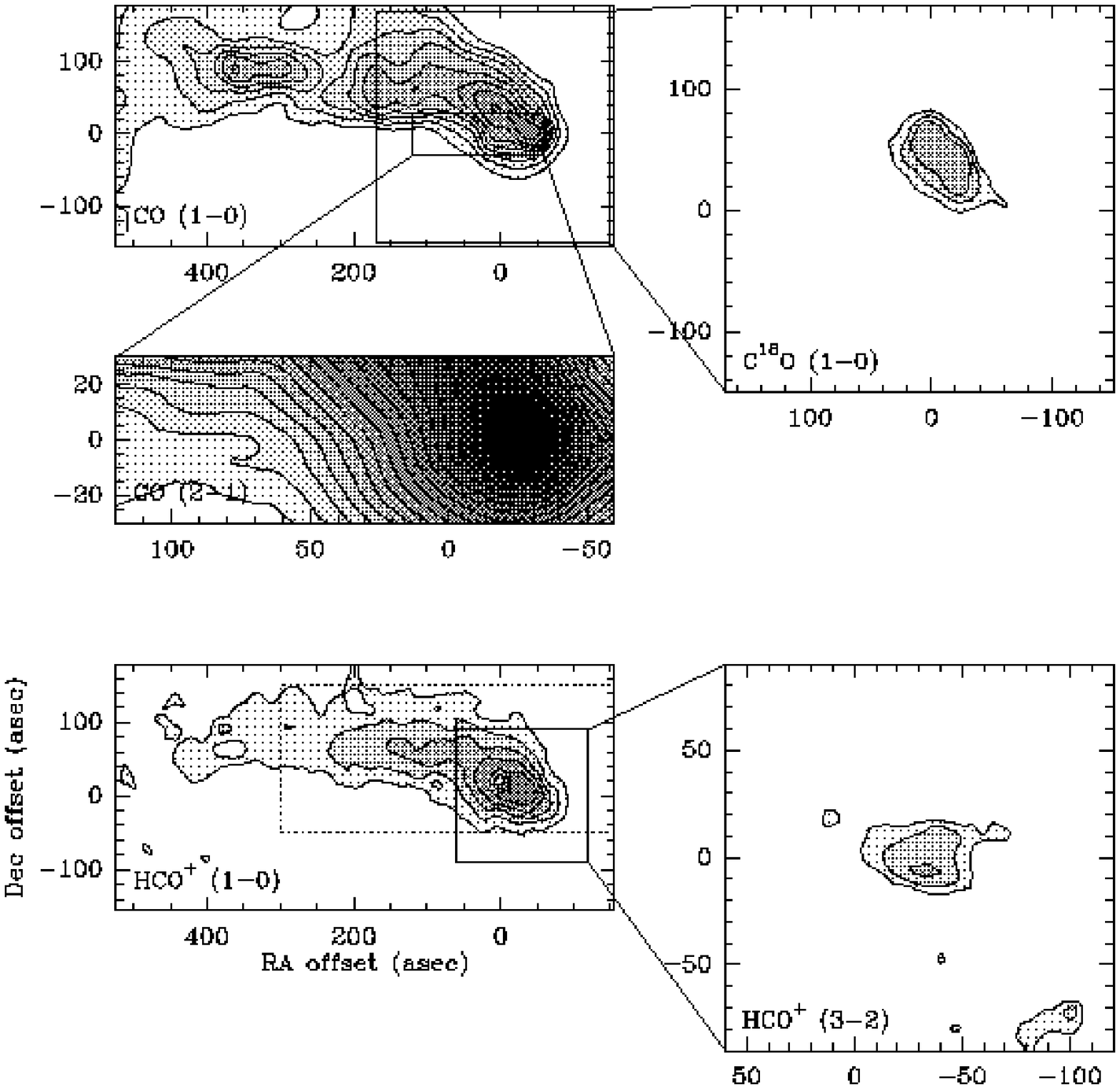}
\caption{\label{sfoxiii_itty} Integrated intensity maps of SFO~13 in various
  transitions and isotopomers of \hcop and CO. The IRAS source 02570+6028 is
  located at the center of each map. The CO~(\jequals{1}{0}) map has a lowest
  contour of 4.6 K~\mbox{km~s$^{-1}$} ($3\sigma$) and increments of 4.6
  K~\mbox{km~s$^{-1}$} ($3\sigma$). The CO~(\jequals{2}{1}) map has a lowest
  contour of 2.1 K~\mbox{km~s$^{-1}$} ($3\sigma$) and increments of 5.0
  K~\mbox{km~s$^{-1}$} ($7\sigma$). The \ceio~(\jequals{1}{0}) map has a lowest
  contour of 0.6 K~\mbox{km~s$^{-1}$} ($3\sigma$) and increments of 0.2
  K~\mbox{km~s$^{-1}$} ($1\sigma$). The \hcop~(\jequals{1}{0}) map has a lowest
  contour of 0.4 K~\mbox{km~s$^{-1}$} ($3\sigma$) and increments of 0.4
  K~\mbox{km~s$^{-1}$} ($3\sigma$). The \hcop~(\jequals{3}{2}) map has a lowest
  contour of 1.7 K~\mbox{km~s$^{-1}$} ($3\sigma$) and increments of 0.6
  K~\mbox{km~s$^{-1}$} ($1\sigma$). The dotted rectangle in the
  \hcop~(\jequals{1}{0}) map indicates the region over which the \hcop\
  centroid is shown in Figure~\ref{sfoxiii_centroid}.}
\end{figure}

Although the spatial resolution of our observations are not good enough to
trace individual infall regions, the
overall gas morphology is resolved by the beam. Figure~\ref{sfoxiii_itty} shows
integrated intensity maps of various transitions and isotopes of
CO and \hcop. The CO~(\jequals{1}{0}) shows a distinct head, located at the
IRAS source position upon which the map is centered. Lower intensity
CO~(\jequals{1}{0}) emission continues from the core and spreads out to the
north-east of the image, similar to the elephant trunk morphology seen in the
optical (Figure~\ref{sfo_dss}). This is also consistent with the fact that the
ionizing source is located to the south-west of the core. The region mapped in
CO (\jequals{2}{1}) shows a similar morphology to the CO (\jequals{1}{0})
emission, however the core position is much more intense than the surrounding
gas. 
The \ceio\ (\jequals{1}{0}) map shows
emission mostly north of the \hcop\ core, with
very little north-east extension. This indicates that the
large column density molecular gas may not correspond exactly with the star
forming core in this source.

\begin{figure}
\epsscale{1.0}
\plotone{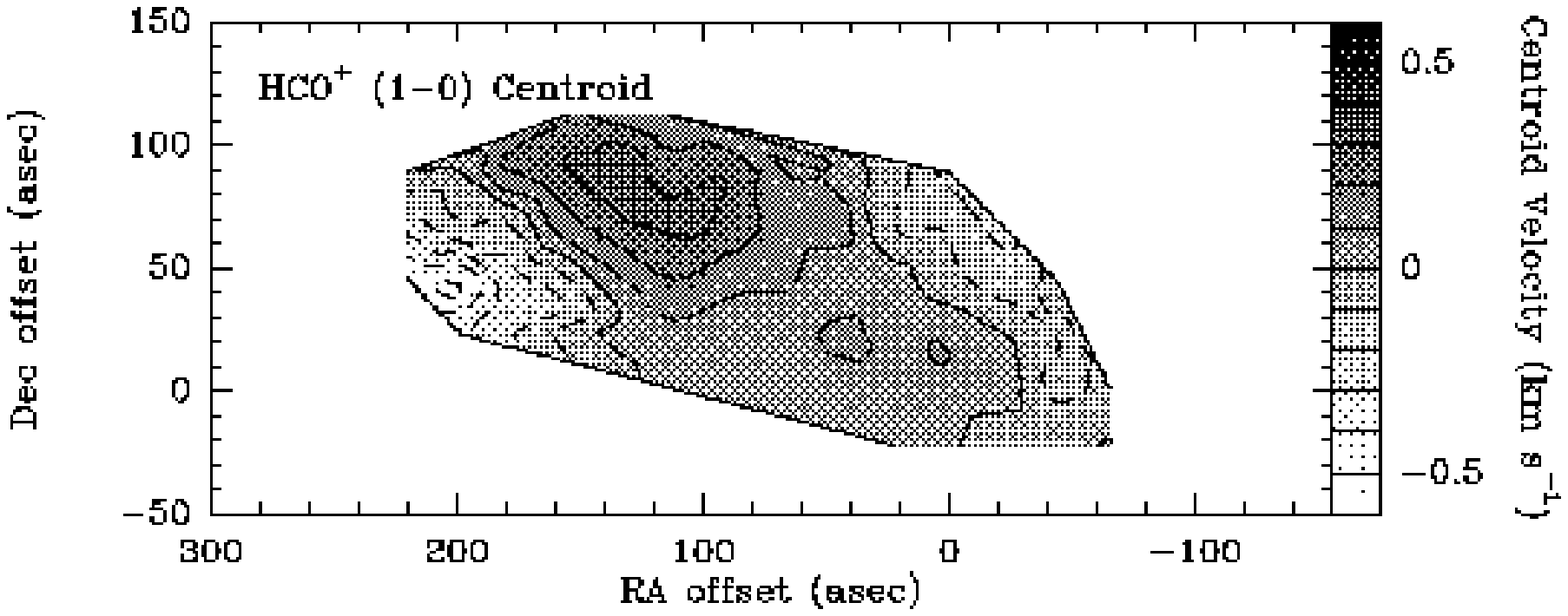}
\caption{\label{sfoxiii_centroid} The SFO~13 centroid velocity integrated over
  the line core of \hcop~(\jequals{1}{0}). The line of sight velocity has been
  subtracted out and the contours and greyscale are indicated on the wedge to
  the right of the figure.}
\end{figure}

\begin{figure}
\epsscale{1.0}
\plotone{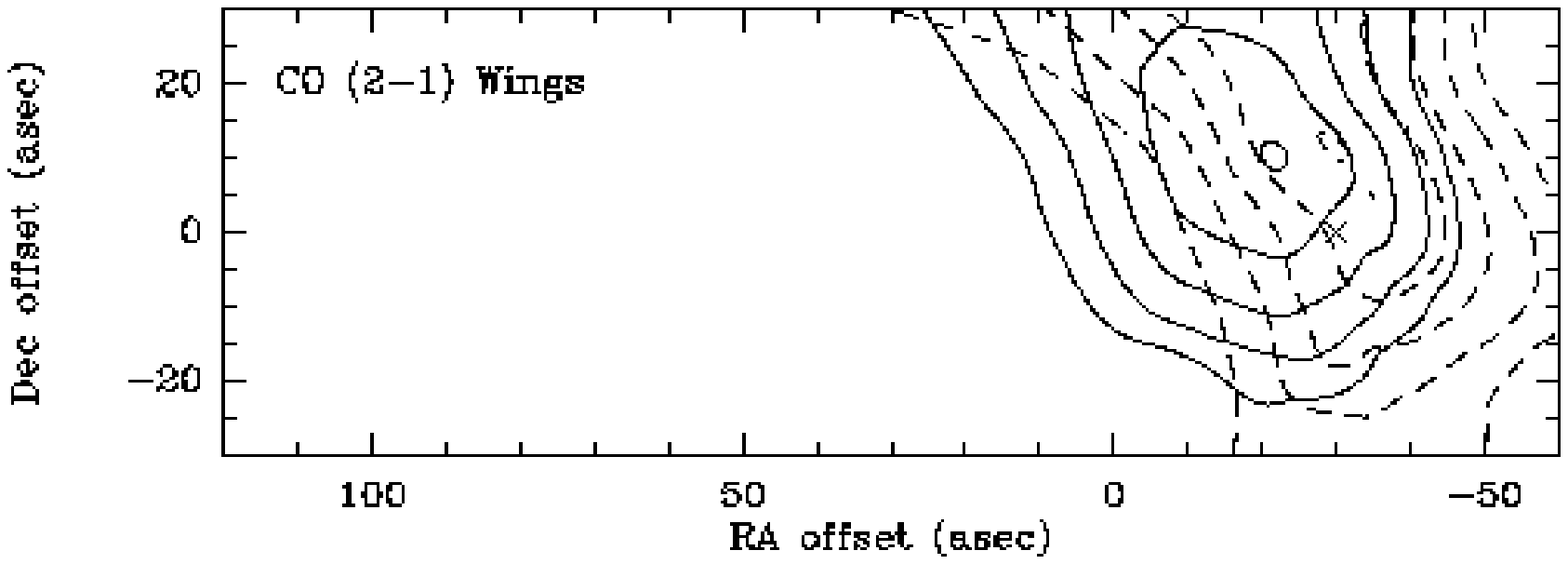}
\caption{\label{sfoxiii_outflow} The SFO~13 CO~(\jequals{2}{1}) line wing
  emission. The blue lobe, indicated by dotted lines, is the integrated
  intensity in the range of --48 to --39.5 \mbox{km~s$^{-1}$}. The red lobe,
  indicated by solid lines, is the integrated intensity in the range from
  --37.5 to --28 \mbox{km~s$^{-1}$}. The lowest contour in each case is the
  half power contour. The x indicates the \hcop~(\jequals{3}{2}) peak
  integrated intensity position.}
\end{figure}

The \hcop~(\jequals{1}{0}) integrated intensity map (Figure~\ref{sfoxiii_itty})
shows a more cometary morphology than the digitized sky survey image in
Figure~\ref{sfo_dss}. The dense gas appears to be preferentially located along
a ridge near the southern edge of the cloud. This could be a result of the
preexisting density structure of the molecular cloud which was swept up by the
ionization front or an effect of the geometry of the expanding HII
region. However,
these observations can not determine how the gas was swept into that ridge. The
\hcop~(\jequals{3}{2}) integrated intensity map only shows significant emission
at the position of the star forming core. The \hcop~(\jequals{1}{0}) centroid
map (Figure~\ref{sfoxiii_centroid}) does not show a very coherent structure,
however there is a significant velocity gradient across the tail in the
\hcop~(\jequals{1}{0}) map along the northwest-southeast direction.
The CO~(\jequals{2}{1}) line profiles show distinct wings, which seem to
form two lobes of an outflow (Figure~\ref{sfoxiii_outflow}). This outflow has
not previously been observed, although SFO found line wings indicative of an
outflow in their CO~(\jequals{2}{1}) observation of this source.

\subsubsection{SFO~16}
SFO~16, also known as L1634 and RNO40, is classified as a type~A bright rimmed
cloud by SFO, as it is a broad ionization front, with no tail. SFO~16 borders
Barnard's Loop, and is at a distance of 400 pc. The embedded IRAS source
05173--0555 has a spectral energy distribution in the IRAS bands consistent
with an embedded YSO (SFO). \citet{coh} describes this object as a purely
nebulous elongated very Red source typical of Herbig-Haro objects. Emission
lines which \citet{coh} detects in this source include H$\alpha$, H$\beta$,
H$\gamma$, [NII], [SII], [OI], [NI], [OIII], and HeI. 

\begin{figure}
\epsscale{1.0}
\plotone{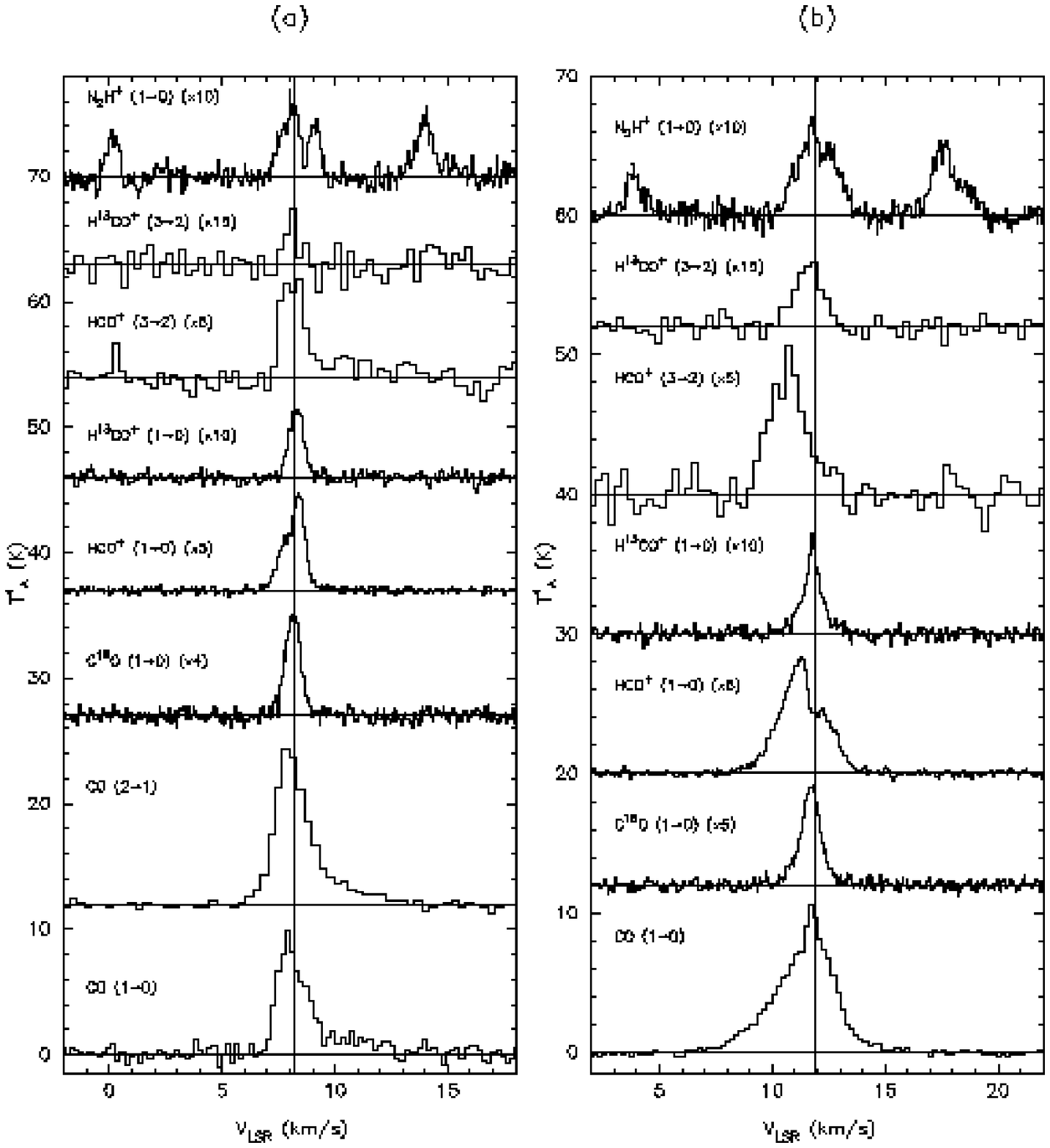}
\caption{\label{sfoxvi_xviii_profiles} Line profiles of molecular transitions
  in the direction of the central IRAS source of SFO~16 (a) and SFO~18 (b).} 
\end{figure}

Figure~\ref{sfoxvi_xviii_profiles}a shows the central line profiles towards
the IRAS source 05173--0555. The vertical line indicates the best fit velocity
of the \nthp~(\jequals{1}{0}) hyperfine ensemble. Both the CO~(\jequals{1}{0})
and
the CO~(\jequals{2}{1}) lines show wings. These wings often indicate the
presence of outflows. The
\ceio~(\jequals{1}{0}) line, however appears to have a fairly gaussian
shape. 
The \hcop\ line profiles, shown in
Figure~\ref{sfoxvi_xviii_profiles}a, are a
bit more enigmatic. The \hcop~(\jequals{3}{2}) and \hthcop~(\jequals{3}{2})
lines seem fairly well centered on the $V_{\rm LSR}$, but the
\hcop~(\jequals{1}{0}) shows a pronounced asymmetry. This asymmetric profile is
not shared by the \hthcop~(\jequals{1}{0}) line, indicating that the asymmetry
observed in the \hcop\ line is probably the result of a radiative transfer
effect, and not an effect of the superposition of clouds along the line of
sight. Line asymmetries can be produced by infall \citep{wnb}, however this
produces a line which has greater emission blue-shifted from line center, and
our example is red-shifted from line center. It is possible that the
\hcop~(\jequals{1}{0}) profile is caused by an outflow. Another
possibility is that this
red-asymmetric profile is due to infall in a region where the shock heating has
inverted the excitation temperature profile of the infalling gas, so that is
has a higher excitation temperature far from the core and a lower excitation
temperature near the core. This hypothesis will be fully explored in
\S\ref{infall}.

\begin{figure}
\epsscale{0.8}
\plotone{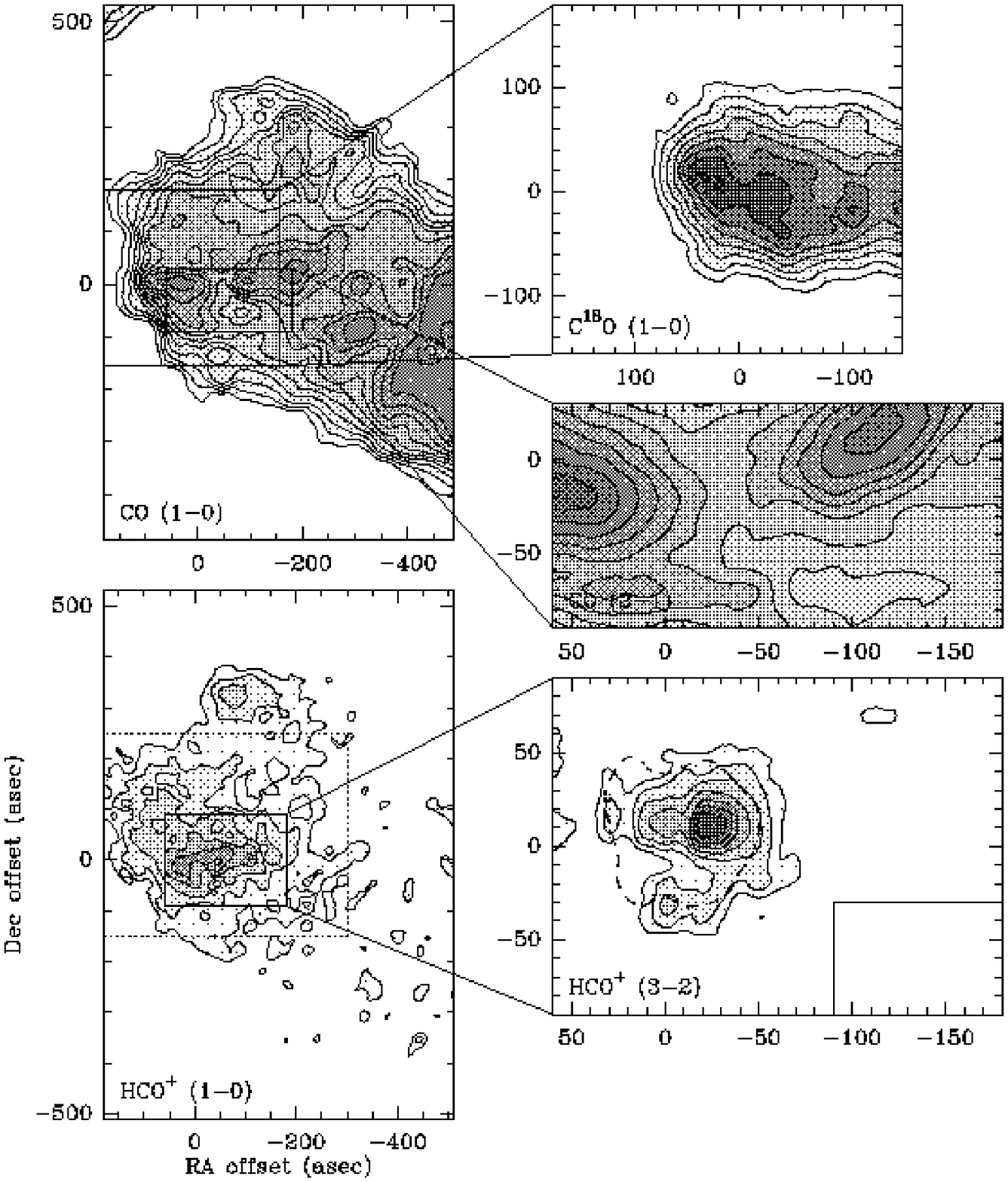}
\caption{\label{sfoxvi_itty} Integrated intensity maps of SFO~16 in various
  transitions and isotopomers of \hcop and CO. The IRAS source 05173--0555 is
  located at the center of each map. The CO~(\jequals{1}{0}) map has a lowest
  contour of 4.3 K~\mbox{km~s$^{-1}$} ($3\sigma$) and increments of 1.4
  K~\mbox{km~s$^{-1}$} ($1\sigma$). The CO~(\jequals{2}{1}) map has a lowest
  contour of 6.0 K~\mbox{km~s$^{-1}$} ($12\sigma$) and increments of 2.5
  K~\mbox{km~s$^{-1}$} ($5\sigma$). The \ceio~(\jequals{1}{0}) map has a lowest
  contour of 0.6 K~\mbox{km~s$^{-1}$} ($3\sigma$) and increments of 0.2
  K~\mbox{km~s$^{-1}$} ($1\sigma$). The \hcop~(\jequals{1}{0}) map has a lowest
  contour of 0.4 K~\mbox{km~s$^{-1}$} ($3\sigma$) and increments of 0.3
  K~\mbox{km~s$^{-1}$} ($2\sigma$). The \hcop~(\jequals{3}{2}) map has a lowest
  contour of 0.9 K~\mbox{km~s$^{-1}$} ($3\sigma$) and increments of 0.3
  K~\mbox{km~s$^{-1}$} ($1\sigma$). The dotted rectangle in the
  \hcop~(\jequals{1}{0}) map indicates the region over which the \hcop\
  centroid is shown in figure~\ref{sfoxvi_centroid}. The dashed contour in the
  \hcop~(\jequals{3}{2}) indicates the half power contour of the
  \nthp~(\jequals{1}{0}) emission.}
\end{figure}

\begin{figure}
\epsscale{1.0}
\plotone{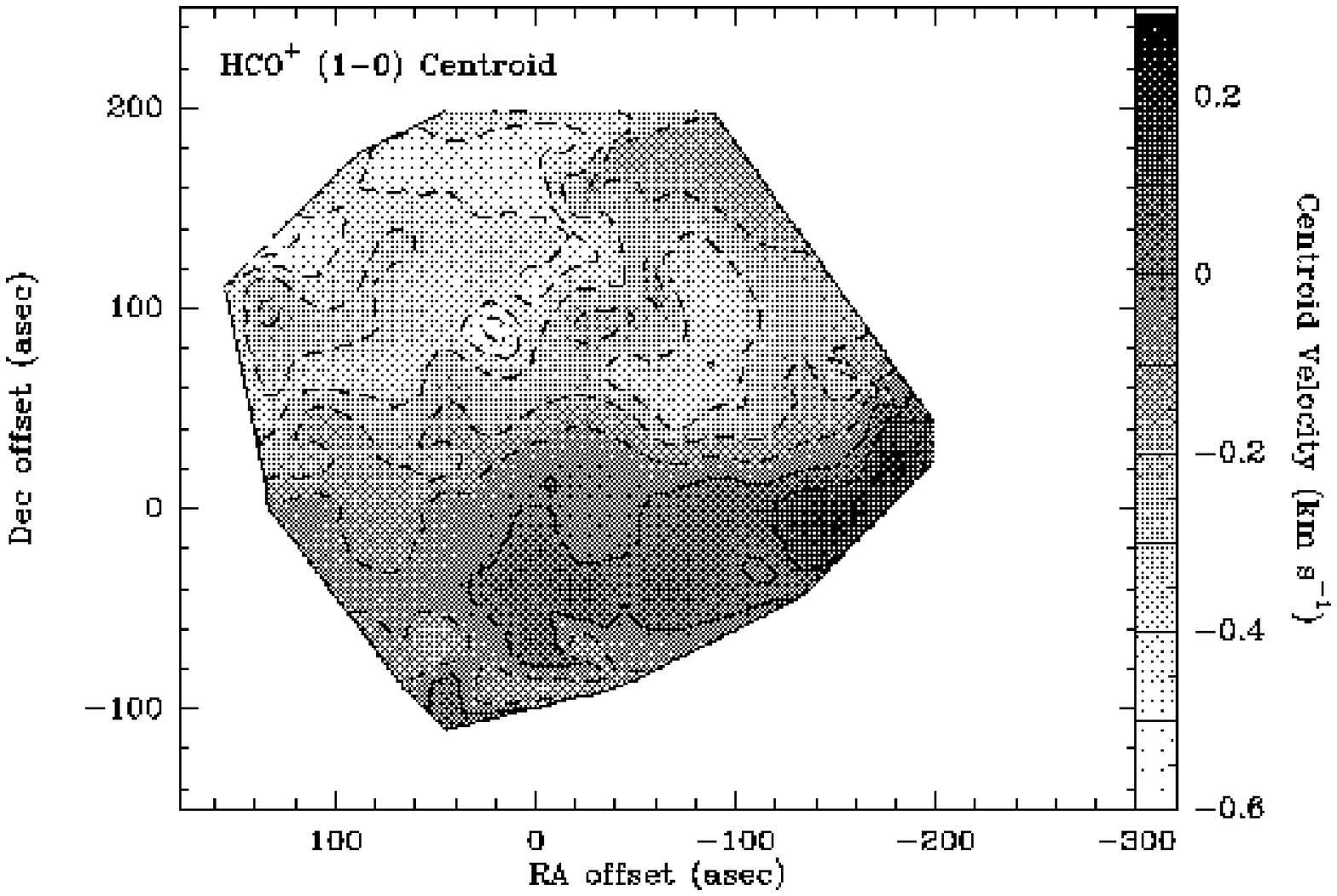}
\caption{\label{sfoxvi_centroid} The SFO~16 centroid velocity integrated over
  the line core of \hcop~(\jequals{1}{0}). The line of sight velocity has been
  subtracted out and the contours and greyscale are indicated on the wedge to
  the right of the figure.}
\end{figure}

The CO and \hcop\ integrated intensity maps are shown in
Figure~\ref{sfoxvi_itty}.
Although classified as a type~A cloud by SFO, the
CO~(\jequals{1}{0}) seems to show a well defined tail moving off to the
south-west, however the ionization source is located directly to the east of
the core. The
\ceio~(\jequals{1}{0}) integrated intensity map (Figure~\ref{sfoxvi_itty})
shows the molecular core surrounding the IRAS source,
and no evidence of the outflow lobes. It does appear that there is a tail
traced by the \ceio\ elongated to the west of the core.
The \hcop~(\jequals{1}{0}) integrated intensity image in
Figure~\ref{sfoxvi_itty} is more indicative of a type~A morphology that the CO
integrated intensity maps. 
The majority of the gas traced by the \hcop\ emission is in
a fairly broad regions which is densest around the IRAS core, but shows another
possible core 300\sec\ north of the IRAS core. The \hcop~(\jequals{3}{2})
integrated
intensity map highlights the location of the core. Figure~\ref{sfoxvi_centroid}
shows the centroid velocity of the \hcop~(\jequals{1}{0}) emission in the dense
regions of the source. This centroid was calculated over the line core
velocities, and do not include contributions from line wings if they are
present. 
On average there appears to be a north-south velocity gradient which could
indicate rotation of the cloud. 

\begin{figure}
\epsscale{1.0}
\plotone{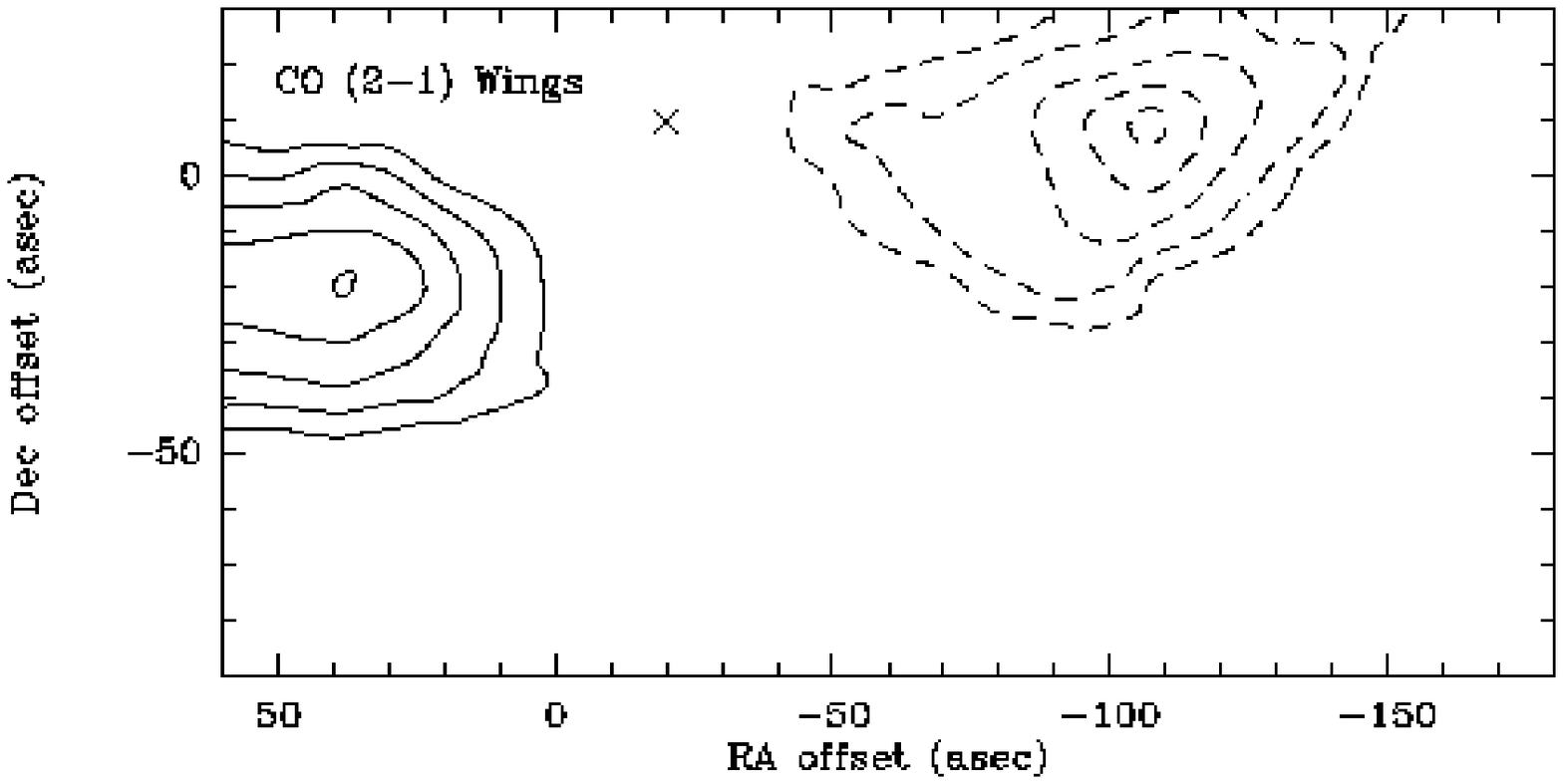}
\caption{\label{sfoxvi_outflow} The SFO~16 CO~(\jequals{2}{1}) line wing
  emission. The blue lobe, indicated by dotted lines, is the integrated
  intensity in the range of --2 to 7.2 \mbox{km~s$^{-1}$}. The red lobe,
  indicated by solid lines, is the integrated intensity in the range from
  9 to 18 \mbox{km~s$^{-1}$}. The lowest contour in each case is the
  half power contour. The x indicates the \hcop~(\jequals{3}{2}) peak
  integrated intensity position.}
\end{figure}

In both the
CO~(\jequals{1}{0}) and CO~(\jequals{2}{1}) integrated intensity maps, lobes
indicative of a bipolar outflow are evident. 
Figure~\ref{sfoxvi_outflow} shows an integrated intensity map of the
CO~(\jequals{2}{1}) line wings. This figure shows that there is an outflow,
with the blue shifted lobe on the west side of the core and the red shifted
lobe to the east of the core. Previous studies by \citet{hl} and \citet{fuk}
indicate the presence of this outflow and a second outflow, which we do not
detect, aligned in the northwest-southeast direction. 
The axis of rotation, derived from the \hcop\ centroid map, is
aligned east-west, matching the alignment of outflow we detect. This alignment
is predicted by
current star formation theories that posit that outflows are located along the
polar direction of a rotating protostar, in order to shed some angular momentum
before material can fall into the protostar.

\subsubsection{SFO~18}
SFO~18, also known as B35A, is also a type~A bright rimmed cloud
\citetext{SFO}. It borders the HII region $\lambda$ Ori at a distance of about
400 pc. The embedded IRAS source 05417+0907 has a spectral energy
distribution consistent with an embedded YSO \citetext{SFO}. \citet{mhsg} have
detected an outflow in CO~(\jequals{1}{0}) which extends in the
northwest-southeast direction around the embedded IRAS source. 

Our line profiles shown in Figure~\ref{sfoxvi_xviii_profiles}b show bright,
extensive wings on the CO~(\jequals{1}{0}) line, when compared to its \ceio\ 
counterpart. The
\hcop~(\jequals{1}{0}) and \hcop~(\jequals{3}{2}) line profiles
both show a significant blue asymmetry when compared to their \hthcop\ isotopic
counterparts. This may indicate that the gas is infalling around the core or
cores \citep{wnb}. 

\begin{figure}
\epsscale{0.8}
\plotone{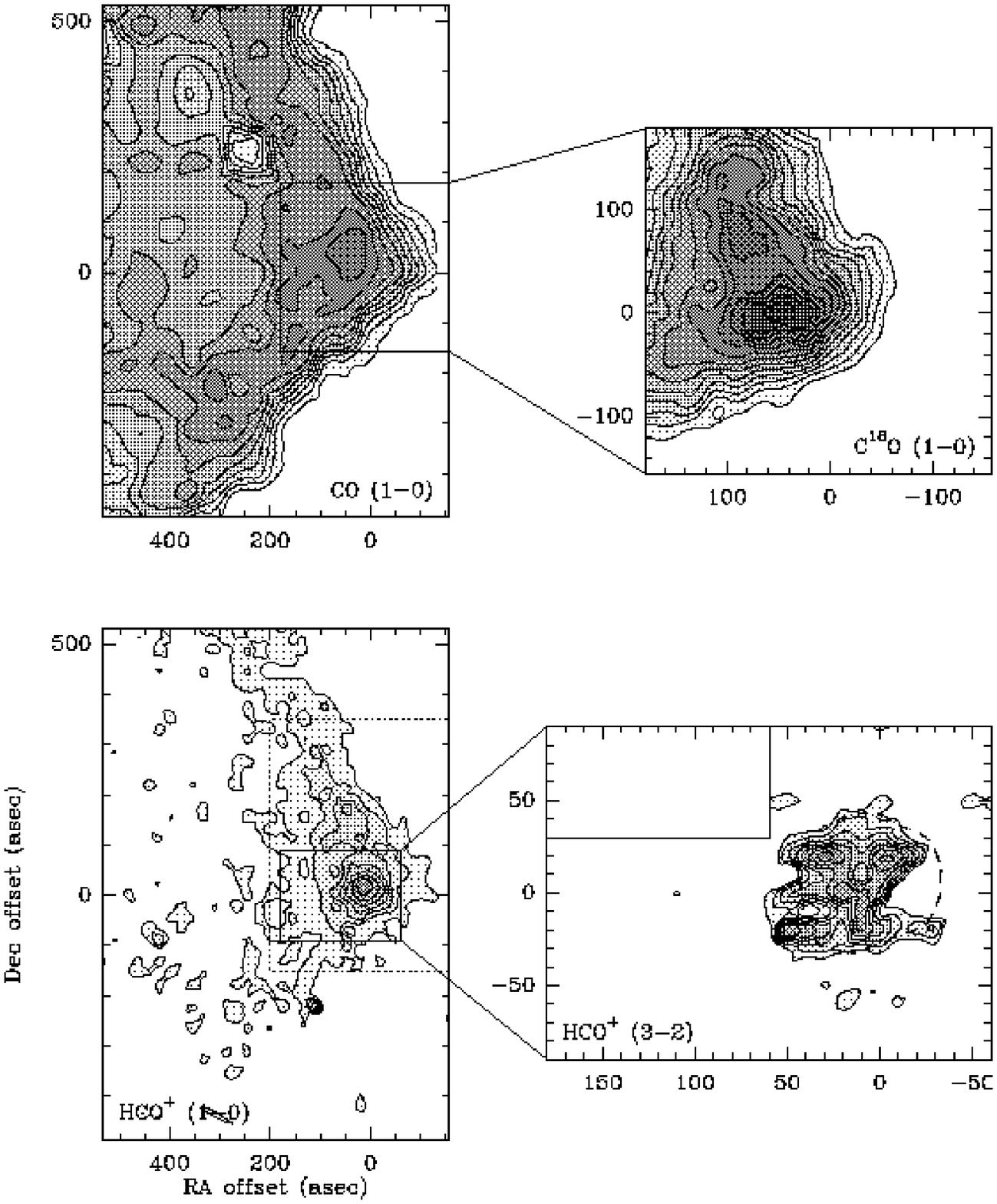}
\caption{\label{sfoxviii_itty} Integrated intensity maps of SFO~18 in various
  transitions and isotopomers of \hcop and CO. The IRAS source 05417+0907 is
  located at the center of each map. The CO~(\jequals{1}{0}) map has a lowest
  contour of 1.8 K~\mbox{km~s$^{-1}$} ($3\sigma$) and increments of 1.8
  K~\mbox{km~s$^{-1}$} ($3\sigma$). The \ceio~(\jequals{1}{0}) map has a lowest
  contour of 0.4 K~\mbox{km~s$^{-1}$} ($3\sigma$) and increments of 0.1
  K~\mbox{km~s$^{-1}$} ($1\sigma$). The \hcop~(\jequals{1}{0}) map has a lowest
  contour of 0.7 K~\mbox{km~s$^{-1}$} ($3\sigma$) and increments of 0.5
  K~\mbox{km~s$^{-1}$} ($2\sigma$). The \hcop~(\jequals{3}{2}) map has a lowest
  contour of 1.6 K~\mbox{km~s$^{-1}$} ($3\sigma$) and increments of 0.5
  K~\mbox{km~s$^{-1}$} ($1\sigma$). The dotted rectangle in the
  \hcop~(\jequals{1}{0}) map indicates the region over which the \hcop\
  centroid is shown in figure~\ref{sfoxviii_centroid}. The dashed contour in
  the \hcop~(\jequals{3}{2}) indicates the half power contour of the
  \nthp~(\jequals{1}{0}) emission.}
\end{figure}

The
CO~\jequals{1}{0} integrated intensity map (Figure~\ref{sfoxviii_itty}) shows
the broad ionization front with a core at its apex, typical of type~A
sources. The
CO emission cuts off fairly abruptly on the western edge of the cloud, where
the ionization front is located. The ionizing source is located directly to the
west of the core. Besides the increased emission at the core, one can see a
general increase in CO~(\jequals{1}{0}) emission adjacent to the ionization
front which is probably due to an increase in column density as gas gets swept
up by the ionization front, a factor which ultimately enhances the density and
leads
to self-gravitating clumps \citep{elm1}. The \ceio~(\jequals{1}{0}) better
traces the high column density gas around the core and shows what may be
another core to the northeast of the first.
The \hcop~(\jequals{1}{0}) emission better traces the dense ridge, as seen in
the integrated intensity map in Figure~\ref{sfoxviii_itty}, and the core in
which the IRAS source is embedded. This ridge of dense gas is consistent with
the type~A morphological classification by \citetext{SFO}. Higher resolution
\hcop~(\jequals{3}{2}) observations of the core show several emission peaks,
indicating that there may be several low mass stars forming simultaneously in
that region. This is consistent with models of shock driven collapse
\citep{vc} in which multiple stars can be formed. 

\begin{figure}
\epsscale{1.0}
\plotone{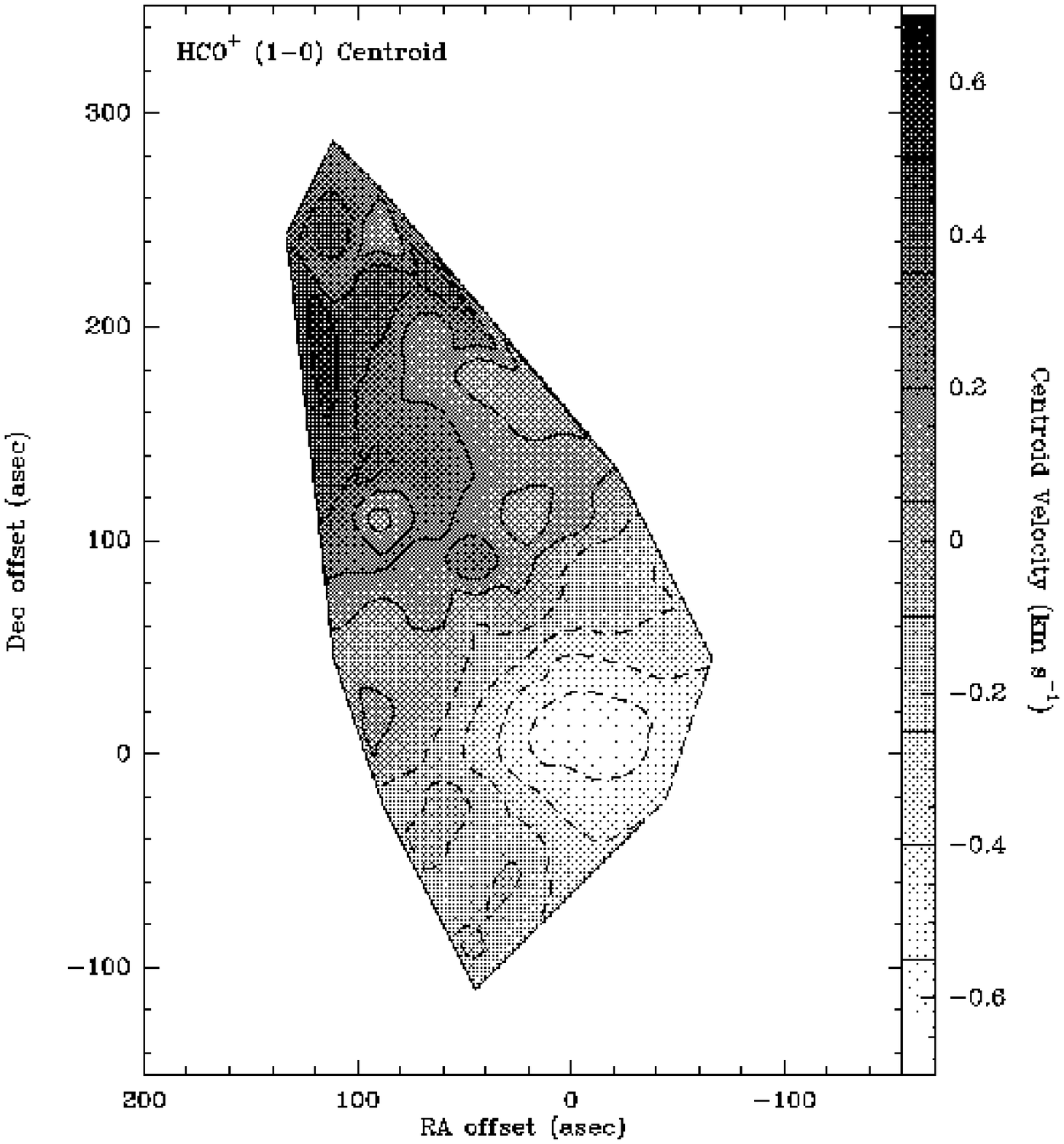}
\caption{\label{sfoxviii_centroid} The SFO~18 centroid velocity integrated over
  the line core of \hcop~(\jequals{1}{0}). The line of sight velocity has been
  subtracted out and the contours and greyscale are indicated on the wedge to
  the right of the figure.}
\end{figure}

\begin{figure}
\plotone{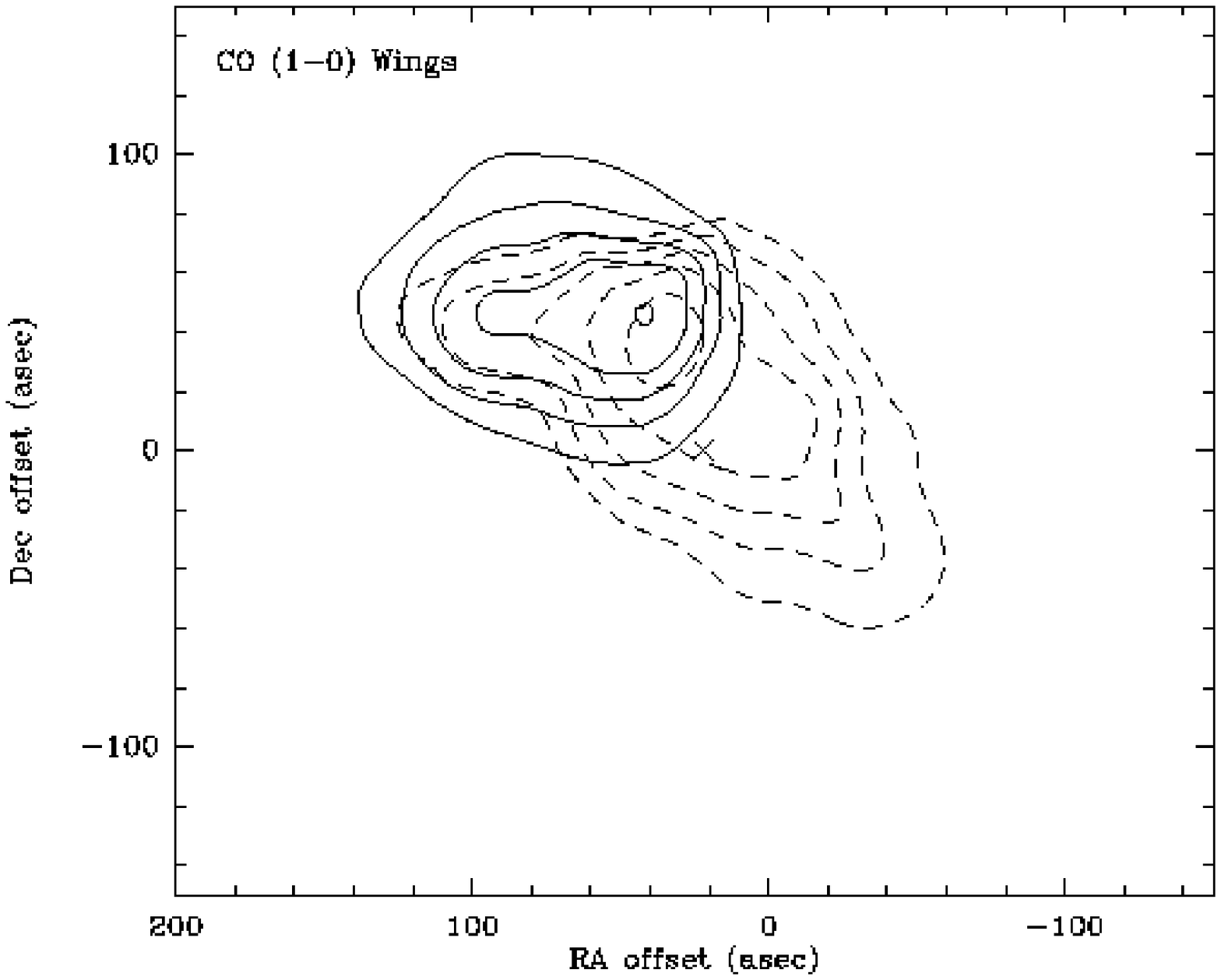}
\caption{\label{sfoxviii_outflow} The SFO~18 CO~(\jequals{1}{0}) line wing
  emission. The blue lobe, indicated by dotted lines, is the integrated
  intensity in the range of 2 to 11 \mbox{km~s$^{-1}$}. The red lobe,
  indicated by solid lines, is the integrated intensity in the range from
  13 to 22 \mbox{km~s$^{-1}$}. The lowest contour in each case is the
  half power contour. The x indicates the \ceio~(\jequals{1}{0}) peak
  integrated intensity position.}
\end{figure}

Further evidence supporting a region
of infall is seen in
Figure~\ref{sfoxviii_centroid} where we see that the entire core region has
blue shifted \hcop~(\jequals{1}{0}) centroid
velocities relative to the rest of
the cloud where emission is strong enough to accurately measure a
centroid. This match between the core emission and the region of possible
infall is further evidence of a collapsing, star forming, core \citep{nmwb}.
Figure~\ref{sfoxviii_outflow} shows the integrated intensity map of
the CO~(\jequals{1}{0}) line wings, and  we see the outflow in the same
direction as detected by \citet{mhsg}. 

\subsubsection{SFO~20}
SFO~20 is a type~C bright rimmed cloud. The long tail is visible in the optical
image (Figure~\ref{sfo_dss}) pointing northwards, away from the HII region
IC434, which is approximately 400 parsecs from us \citetext{SFO}. The
infrared spectral energy distribution of this source is consistent with an
embedded YSO
\citetext{SFO}. \citet{sfmo} discovered and mapped an outflow around the
embedded IRAS source in SFO~20 using the NRO 45m telescope in the
CO~(\jequals{1}{0}) transition.

\begin{figure}
\epsscale{1.0}
\plotone{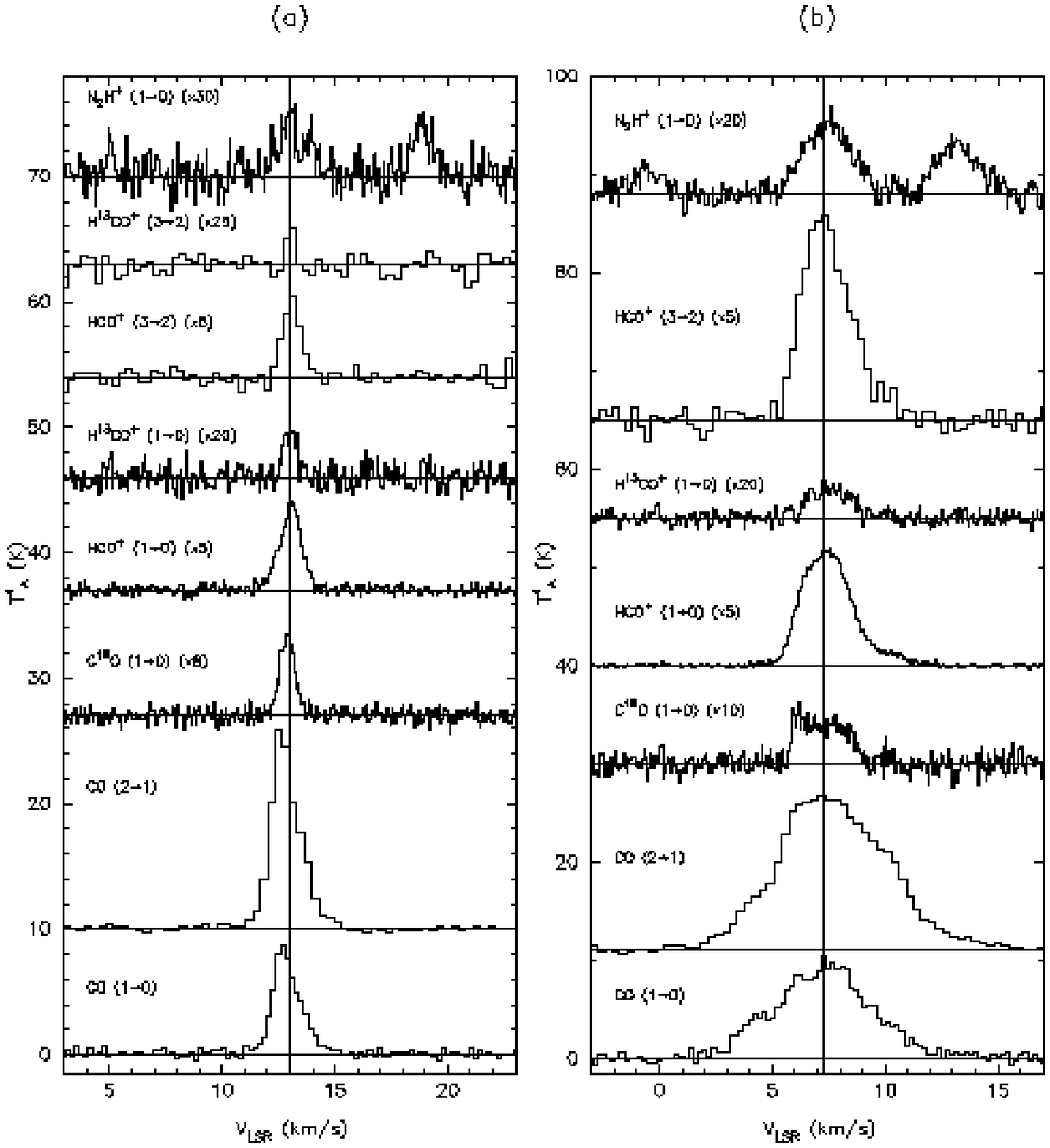}
\caption{\label{sfoxx_xxv_profiles} Line profiles of molecular transitions
  in the direction of the central IRAS source of SFO~20 (a) and SFO~25 (b).}
\end{figure}

The CO~(\jequals{1}{0}) and CO~(\jequals{2}{1}) line profiles shown in
Figure~\ref{sfoxx_xxv_profiles}a show a blue asymmetry and wings on
the \jequals{2}{1} transition. 
The \hcop\ line profiles are all fairly gaussian, and fairly well centered on
the \nthp~(\jequals{1}{0}) $V_{\rm LSR}$. There is no clear evidence of infall
from the \hcop\ transitions. 

\begin{figure}
\epsscale{0.8}
\plotone{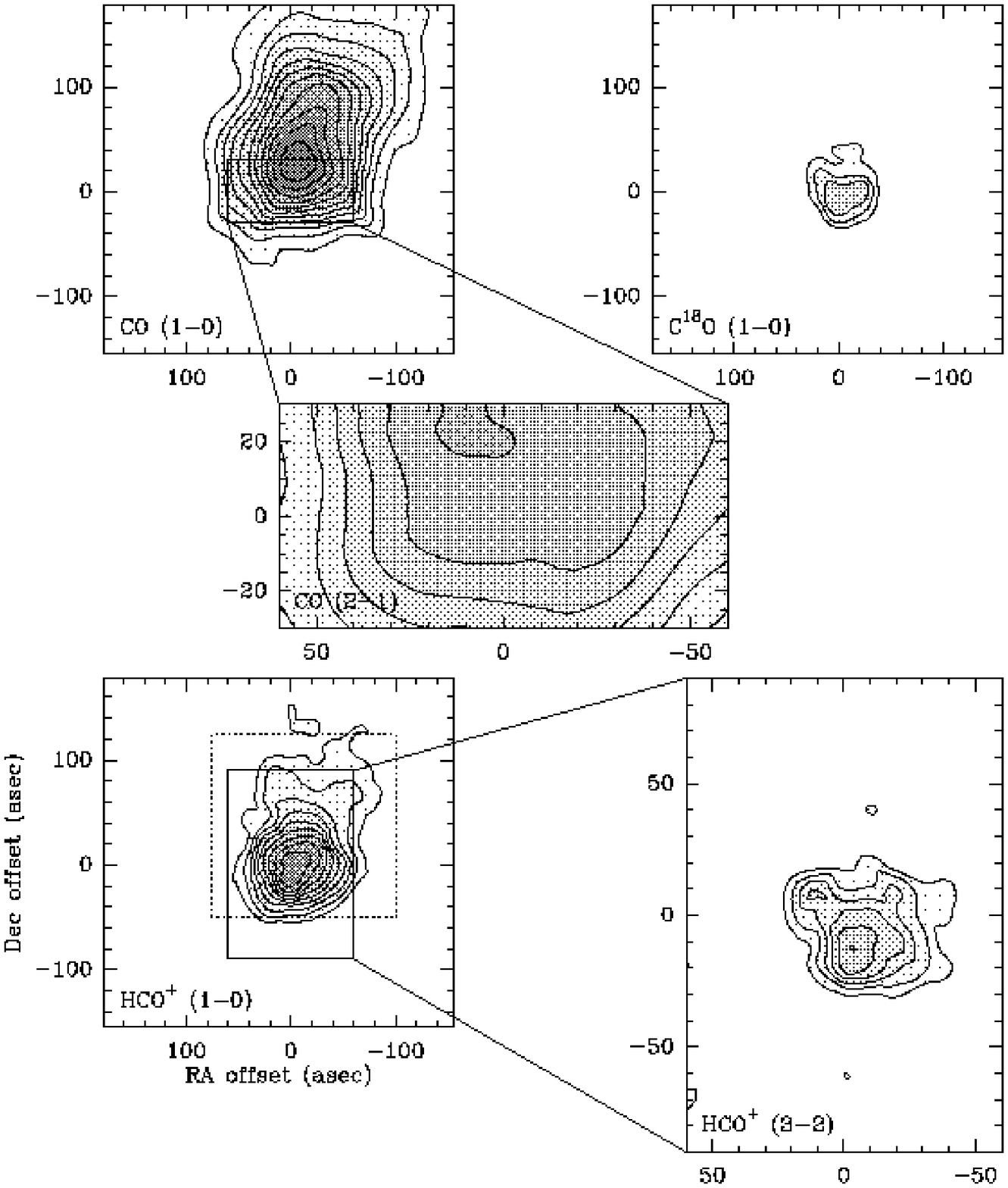}
\caption{\label{sfoxx_itty} Integrated intensity maps of SFO~20 in various
  transitions and isotopomers of \hcop and CO. The IRAS source 05173--0555 is
  located at the center of each map. The CO~(\jequals{1}{0}) map has a lowest
  contour of 1.1 K~\mbox{km~s$^{-1}$} ($3\sigma$) and increments of 1.1
  K~\mbox{km~s$^{-1}$} ($3\sigma$). The CO~(\jequals{2}{1}) map has a lowest
  contour of 1.0 K~\mbox{km~s$^{-1}$} ($3\sigma$) and increments of 3.4
  K~\mbox{km~s$^{-1}$} ($10\sigma$). The \ceio~(\jequals{1}{0}) map has a lowest
  contour of 0.4 K~\mbox{km~s$^{-1}$} ($3\sigma$) and increments of 0.1
  K~\mbox{km~s$^{-1}$} ($1\sigma$). The \hcop~(\jequals{1}{0}) map has a lowest
  contour of 0.4 K~\mbox{km~s$^{-1}$} ($3\sigma$) and increments of 0.1
  K~\mbox{km~s$^{-1}$} ($1\sigma$). The \hcop~(\jequals{3}{2}) map has a lowest
  contour of 0.6 K~\mbox{km~s$^{-1}$} ($3\sigma$) and increments of 0.2
  K~\mbox{km~s$^{-1}$} ($1\sigma$). The dotted rectangle in the
  \hcop~(\jequals{1}{0}) map indicates the region over which the \hcop\
  centroid is shown in figure~\ref{sfoxx_centroid}.}
\end{figure}

The CO~(\jequals{1}{0}) integrated
intensity map (Figure~\ref{sfoxx_itty}) shows the type~C morphology, with a
strong core and evidence of a
tail trailing off to the north. The CO~(\jequals{2}{1}) emission also traces
the core, though the emission peak does not match the \hcop~(\jequals{3}{2})
core position exactly. The \ceio~(\jequals{1}{0}) emission is also peaked on
the core and relatively absent in the tail.
The
\hcop~(\jequals{1}{0}) emission is mainly centered on the core with some
emission tracing the tail. While many of the previous sources had substantial
emission outside of the core region, this one does not. Under the assumption
that the type~C morphology is the most temporally evolved morphology, 
the ionization front may have long since passed this region and it is
surrounded by
ionized gas. As the core gas collapses the tail can be eaten away as it becomes
less and less shielded. The \hcop~(\jequals{3}{2}) traces only the core
material. 

\begin{figure}
\epsscale{1.0}
\plotone{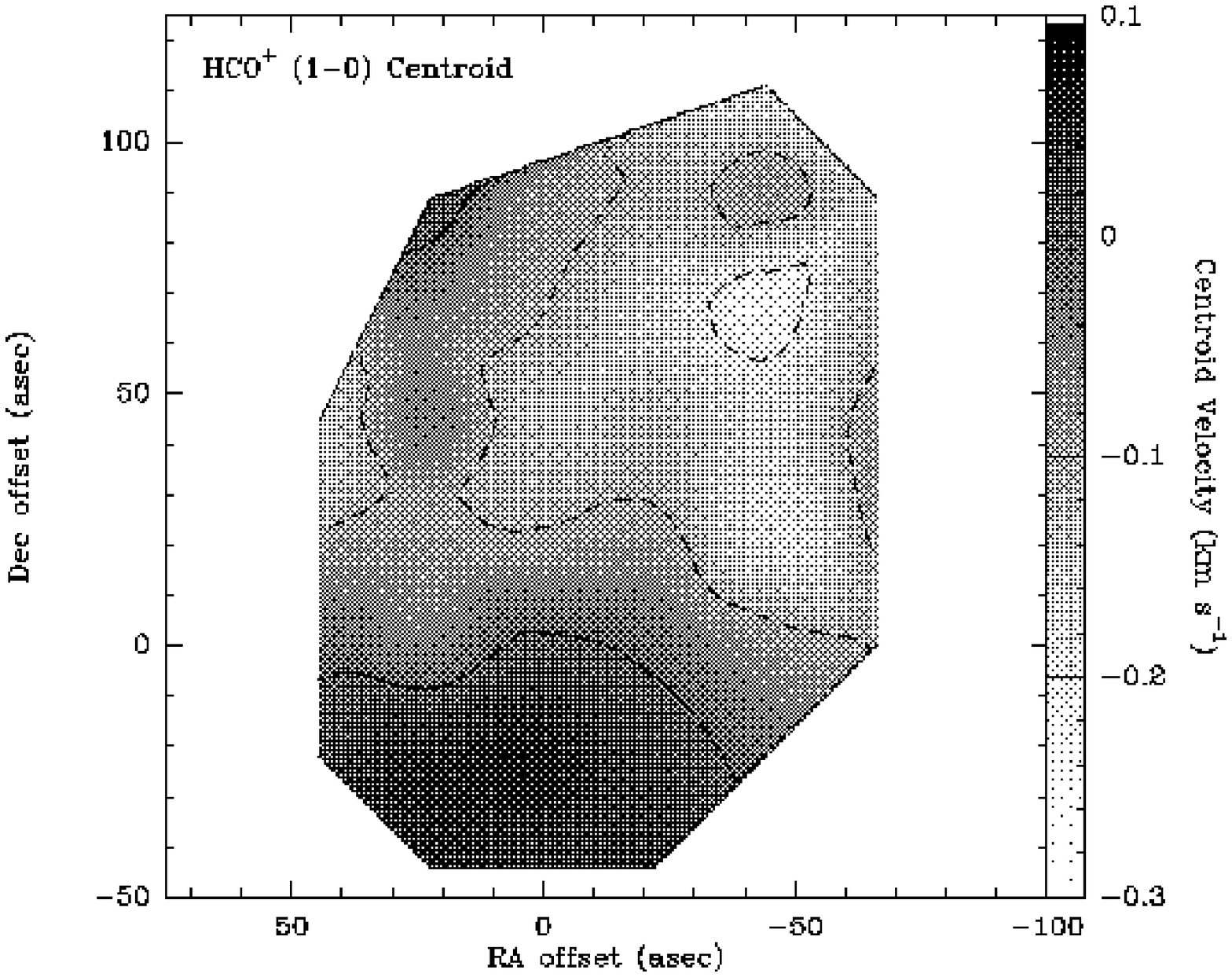}
\caption{\label{sfoxx_centroid} The SFO~20 centroid velocity integrated over
  the line core of \hcop~(\jequals{1}{0}). The line of sight velocity has been
  subtracted out and the contours and greyscale are indicated on the wedge to
  the right of the figure.}
\end{figure}

\begin{figure}
\epsscale{1.0}
\plotone{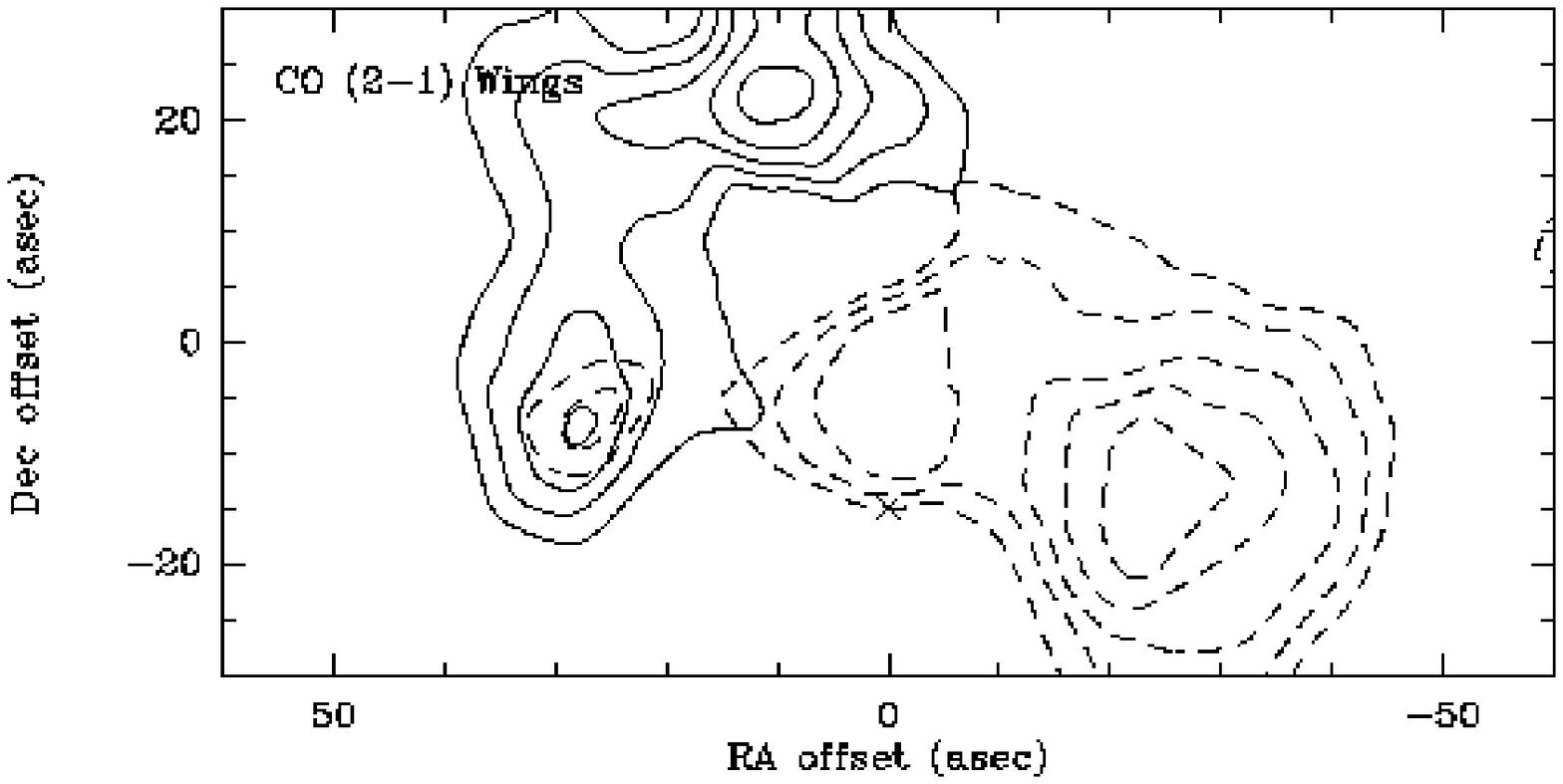}
\caption{\label{sfoxx_outflow} The SFO~20 CO~(\jequals{2}{1}) line wing
  emission. The blue lobe, indicated by dotted lines, is the integrated
  intensity in the range of 3 to 12 \mbox{km~s$^{-1}$}. The red lobe,
  indicated by solid lines, is the integrated intensity in the range from
  14 to 23 \mbox{km~s$^{-1}$}. The lowest contour in each case is the
  half power contour. The x indicates the \hcop~(\jequals{3}{2}) peak
  integrated intensity position.}
\end{figure}

Analysis of the \hcop~(\jequals{1}{0}) centroid velocity, shown in
Figure~\ref{sfoxx_centroid}  indicates a small, but statistically significant
gradient across the core. This gradient may indicate the cloud's rotation.
As an outflow has already been detected in this source by \citet{sfmo} we chose
to look at the CO~(\jequals{2}{1}) wing integrated intensity in order to try to
verify the existing observation. Figure~\ref{sfoxx_outflow} shows
that the line wings do show two lobes of emission, with the red lobe being
predominantly northeast of the core and the blue lobe predominantly southwest
of the core, similar to previous results. The cloud rotation shown in
Figure~\ref{sfoxx_centroid} is perpendicular to the outflow, which is a
prediction of current theories of star formation.

\subsubsection{SFO~25}
SFO~25, also known as HH124 \citep{wor}, is a type B cloud located at a
distance of 780~pc. Embedded within
is the IRAS source 06382+1017, whose spectral energy distribution is consistent
with an embedded YSO \citetext{SFO}. SFO~25 is associated with the HII region
NGC~2264. Figure~\ref{sfo_dss} shows a moderately curved rim arching northwards
away from the ionizing source. However, although there is no tail visible in
the optical, there is also no point at which the structure begins to broaden
away from the ionizing source as there is for the other type~B sources, SFO~4
and SFO~13. As we shall see below, this source bears many similarities with
type~C cometary clouds as well.

The CO~(\jequals{1}{0}) and CO~(\jequals{2}{1}) profiles show a wide
(5~\mbox{km~s$^{-1}$}), slightly asymmetric line
(Figure~\ref{sfoxx_xxv_profiles}b). The CO~(\jequals{2}{1}) profile
shows line wings, which may indicate the presence of an outflow.
The \hcop\ line profiles show some
asymmetry as well, mostly in the form of a redshifted tail. 

\begin{figure}
\epsscale{0.75}
\plotone{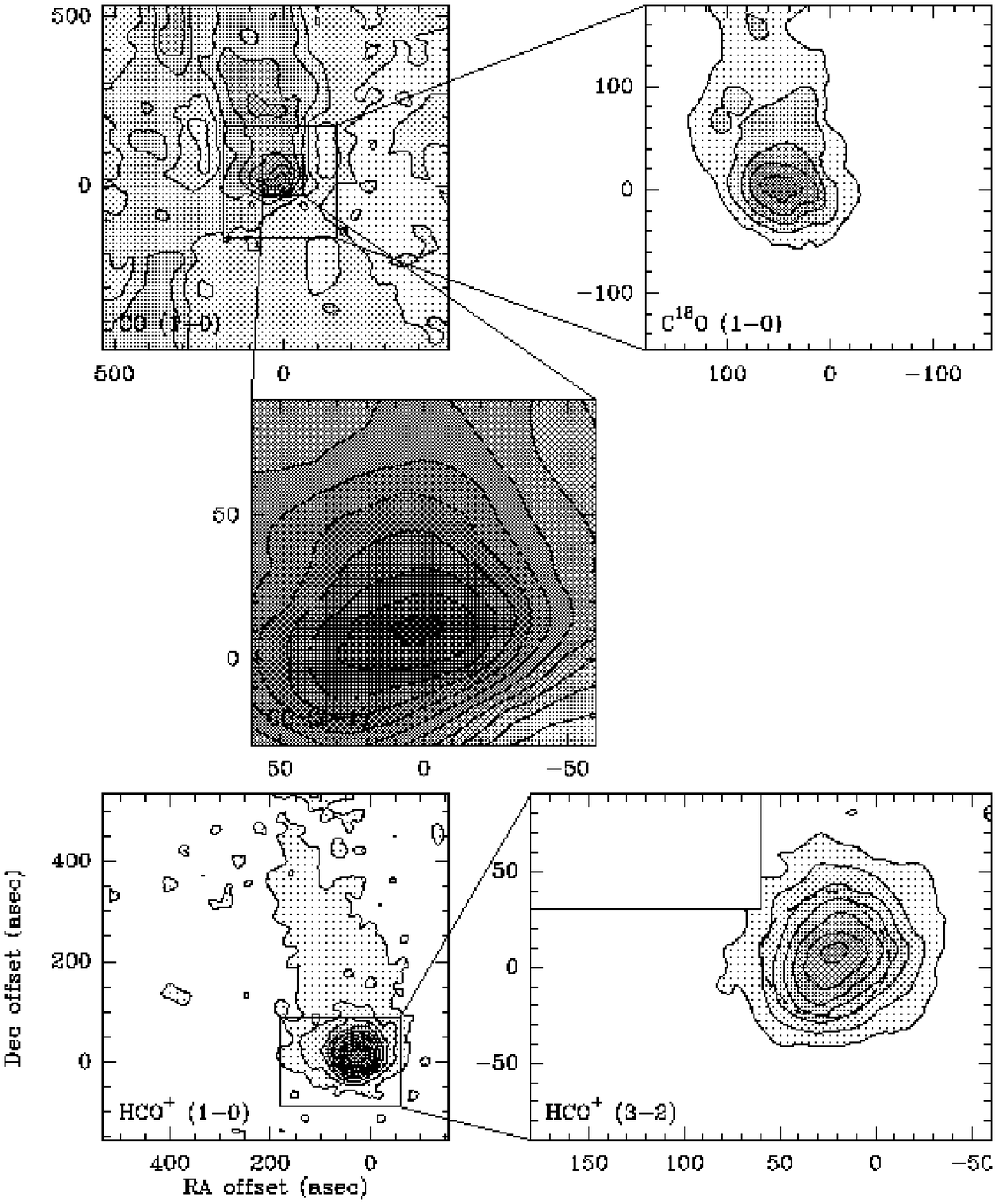}
\caption{\label{sfoxxv_itty} Integrated intensity maps of SFO~25 in various
  transitions and isotopomers of \hcop and CO. The IRAS source 06382+1017 is
  located at the center of each map. The CO~(\jequals{1}{0}) map has a lowest
  contour of 3.6 K~\mbox{km~s$^{-1}$} ($3\sigma$) and increments of 6.1
  K~\mbox{km~s$^{-1}$} ($5\sigma$). The CO~(\jequals{2}{1}) map has a lowest
  contour of 1.7 K~\mbox{km~s$^{-1}$} ($3\sigma$) and increments of 5.6
  K~\mbox{km~s$^{-1}$} ($10\sigma$). The \ceio~(\jequals{1}{0}) map has a lowest
  contour of 0.5 K~\mbox{km~s$^{-1}$} ($3\sigma$) and increments of 0.4
  K~\mbox{km~s$^{-1}$} ($2\sigma$). The \hcop~(\jequals{1}{0}) map has a lowest
  contour of 0.5 K~\mbox{km~s$^{-1}$} ($3\sigma$) and increments of 0.8
  K~\mbox{km~s$^{-1}$} ($5\sigma$). The \hcop~(\jequals{3}{2}) map has a lowest
  contour of 1.2 K~\mbox{km~s$^{-1}$} ($3\sigma$) and increments of 2.0
  K~\mbox{km~s$^{-1}$} ($5\sigma$). The dashed contour in the
  \hcop~(\jequals{3}{2}) indicates the half power contour of the
  \nthp~(\jequals{1}{0}) emission.}
\end{figure}

The CO~(\jequals{1}{0}) integrated
intensity map (Figure~\ref{sfoxxv_itty}) shows a low level of CO emission
throughout the region, which may be the result of contamination from nearby
diffuse
clouds which lie at nearly the same velocity of SFO~25. The
\ceio~(\jequals{1}{0}) traces the higher column density gas which forms the
SFO~25 core and
shows a molecular gas tail to the north. This morphology is more typical of a
type~C cometary cloud than a type~B source. The CO~(\jequals{2}{1}) emission
peaks strongly at the IRAS source position. 
The
\hcop~(\jequals{1}{0}) integrated intensity map (Figure~\ref{sfoxxv_itty})
shows strongly peaked emission around the core with much weaker emission along
the tail to the north of the source, which looks more typical of a type~C
source than a type~B source. The \hcop~(\jequals{3}{2}) emission is
also strongly peaked at the IRAS source position. 

\begin{figure}
\epsscale{1.0}
\plotone{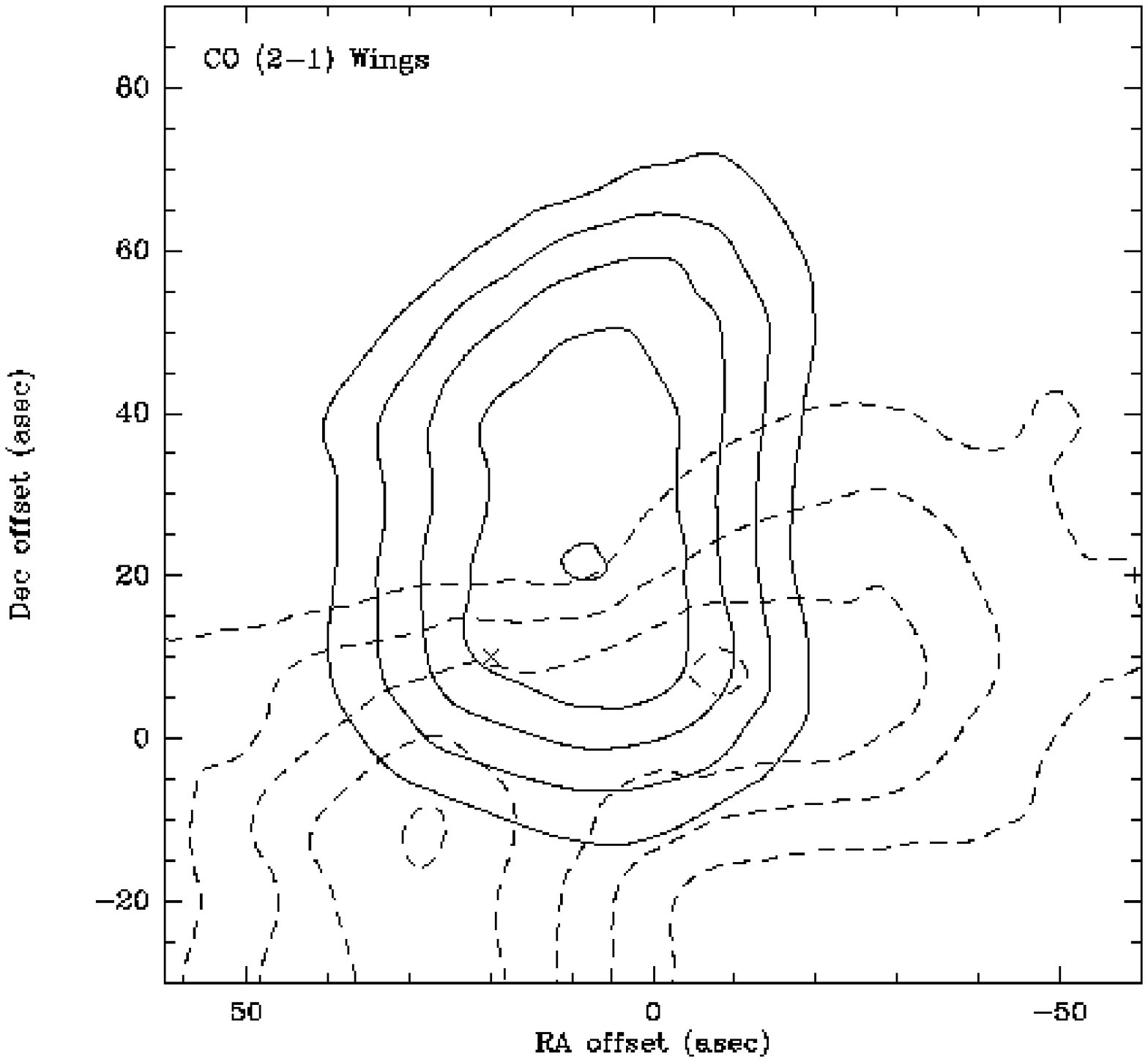}
\caption{\label{sfoxxv_outflow} The SFO~25 CO~(\jequals{2}{1}) line wing
  emission. The blue lobe, indicated by dotted lines, is the integrated
  intensity in the range of --2 to --5.5 \mbox{km~s$^{-1}$}. The red lobe,
  indicated by solid lines, is the integrated intensity in the range from
  9 to 18 \mbox{km~s$^{-1}$}. The lowest contour in each case is the
  half power contour. The x indicates the \hcop~(\jequals{3}{2}) peak
  integrated intensity position.}
\end{figure}

We detected no coherent velocity structure in the \hcop~(\jequals{1}{0})
velocity centroid observations.
However the CO~(\jequals{2}{1})
line wings show evidence of an outflow oriented in the
north-south direction (Figure~\ref{sfoxxv_outflow}), though unrelated emission
near the same velocity makes it hard to separate the outflow emission from
emission due to other sources. 

\subsubsection{SFO~37}
SFO~37, also known as GN 21.38.9,  is a type~C cloud (SFO). It's
cometary structure is quite striking in the optical
(Figure~\ref{sfo_dss}). SFO~37 is associated with the HII region IC1396 located
at a distance of 750~pc. Embedded within is the IRAS source 21388+5622
whose spectral energy distribution is consistent with an embedded YSO. 
An investigation of this
region by \citet{dcbg} revealed H$\alpha$ and [SII] emission on the northeast
side of the molecular emission core. This emission is consistent with an
ionization front driven by the strong UV field of a neighboring O6 star. 

\begin{figure}
\epsscale{1.0}
\plotone{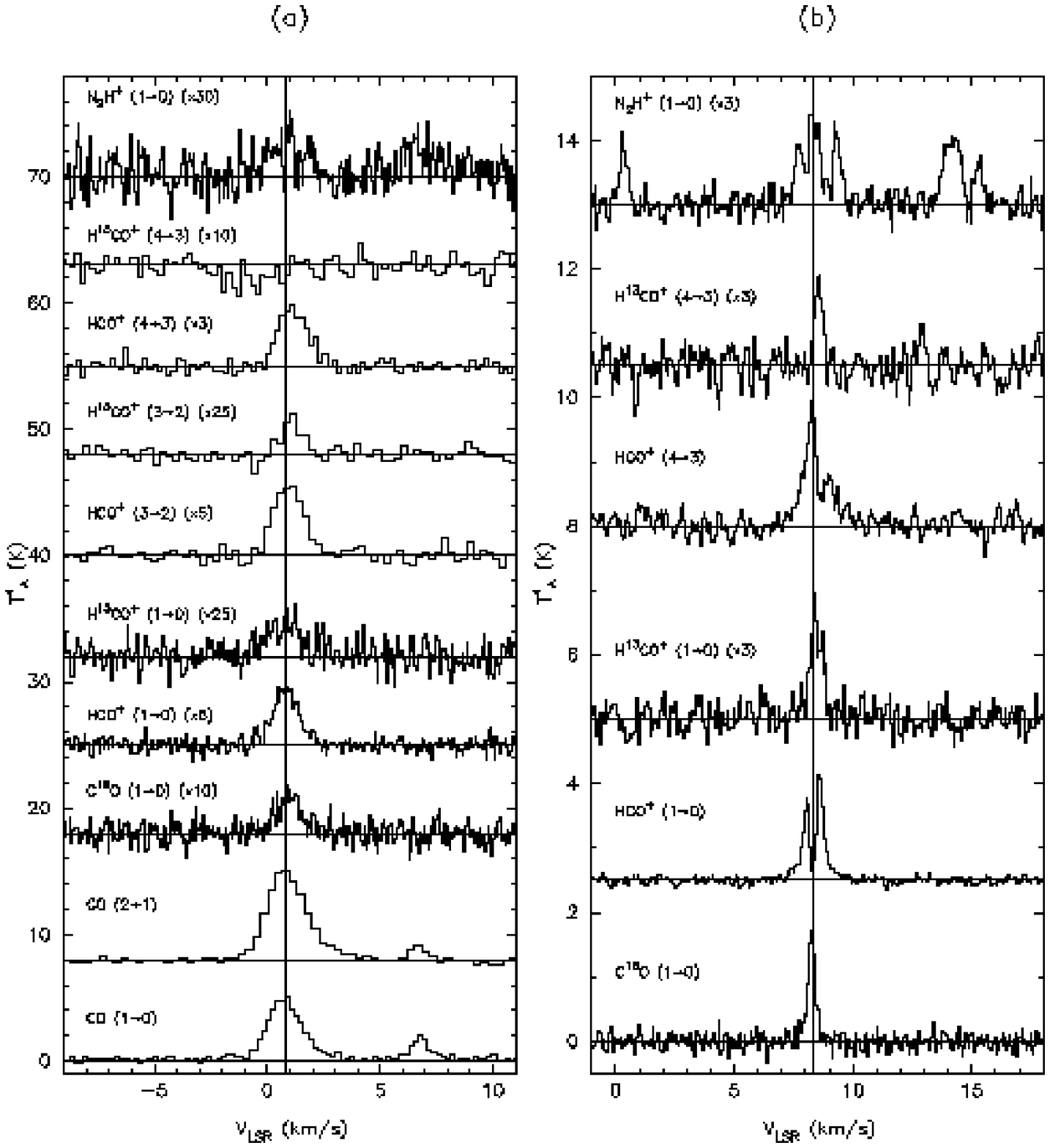}
\caption{\label{sfoxxxvii_bcccxxxv_profiles} Line profiles of molecular
  transitions in the direction of the central IRAS source of SFO~37 (a) and
  B~335 (b).}
\end{figure}

The CO~(\jequals{1}{0}) and CO~(\jequals{2}{1}) line profiles
(Figure~\ref{sfoxxxvii_bcccxxxv_profiles}a) shows a fairly gaussian shape
centered at a line of sight velocity of approximately 1
\mbox{km~s$^{-1}$}. A second component is visible at a line of sight velocity
of approximately 7 \mbox{km~s$^{-1}$}. \citet{dcbg} speculate that this second
component may be a remnant of the low density gas shell in which the core
formed, however we find this to be unlikely; the second component's spatial
distribution appears unrelated to that of the main component, and the low
density gas surrounding the SFO~37 core has most likely been ionized. 
Both the \hcop\ and \hthcop\ profiles in
Figure~\ref{sfoxxxvii_bcccxxxv_profiles}a appear
fairly gaussian and share approximately the same centroid velocity. As is the
case for many of these bright rimmed clouds, there is little or no indication
of infall, despite the presence of a previously detected outflow \citep{dcbg}. 

\begin{figure}
\epsscale{0.7}
\plotone{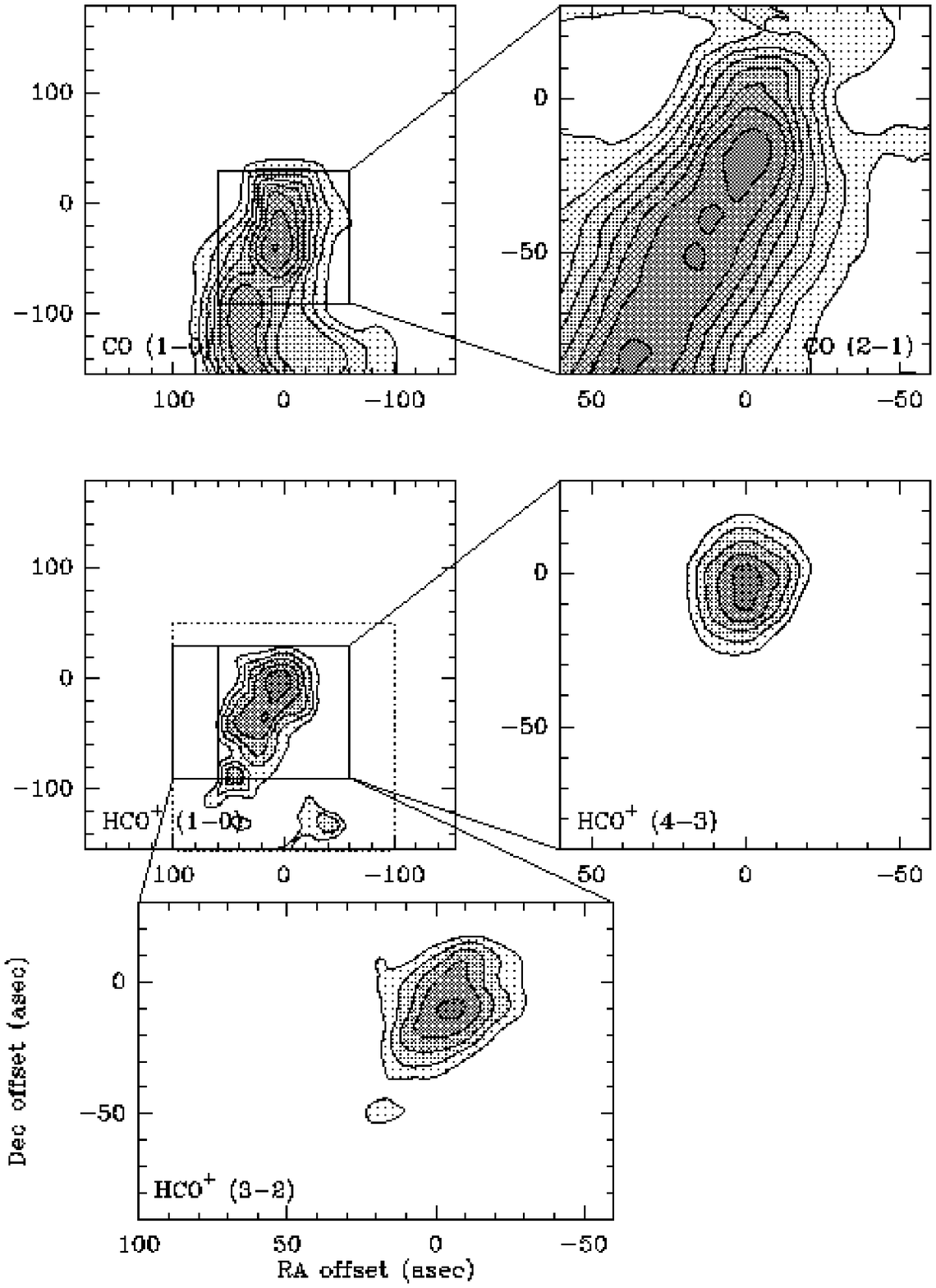}
\caption{\label{sfoxxxvii_itty} Integrated intensity maps of SFO~37 in various
  transitions and isotopomers of \hcop and CO. The IRAS source 21388+5622 is
  located at the center of each map. The CO~(\jequals{1}{0}) map has a lowest
  contour of 1.4 K~\mbox{km~s$^{-1}$} ($3\sigma$) and increments of 1.4
  K~\mbox{km~s$^{-1}$} ($3\sigma$). The CO~(\jequals{2}{1}) map has a lowest
  contour of 1.6 K~\mbox{km~s$^{-1}$} ($3\sigma$) and increments of 2.6
  K~\mbox{km~s$^{-1}$} ($5\sigma$). The \hcop~(\jequals{1}{0}) map has a lowest
  contour of 0.4 K~\mbox{km~s$^{-1}$} ($3\sigma$) and increments of 0.2
  K~\mbox{km~s$^{-1}$} ($1\sigma$). The \hcop~(\jequals{3}{2}) map has a lowest
  contour of 0.8 K~\mbox{km~s$^{-1}$} ($3\sigma$) and increments of 0.3
  K~\mbox{km~s$^{-1}$} ($1\sigma$). The \hcop~(\jequals{4}{3}) map has a lowest
  contour of 0.9 K~\mbox{km~s$^{-1}$} ($3\sigma$) and increments of 0.3
  K~\mbox{km~s$^{-1}$} ($1\sigma$). The dotted rectangle in the
  \hcop~(\jequals{1}{0}) map indicates the region over which the \hcop\
  centroid is shown in figure~\ref{sfoxxxvii_centroid}.}
\end{figure}

The integrated intensity map of the
CO~(\jequals{1}{0}) emission shows the cometary cloud structure stretching from
the center of the plot down to the southwest. There also appears to be a spur
of gas 120\sec\ south of the core extending west out of the tail of the
cometary cloud (Figure~\ref{sfoxxxvii_itty}).
The CO~(\jequals{2}{1})
integrated intensity map also shows a cometary structure running
northwest-southeast. Although we detected no outflow, the higher resolution
survey of \citet{dcbg} detected a bipolar outflow with a 20\sec\ separation
between the red and blue lobes and a north-south orientation.

\begin{figure}
\epsscale{1.0}
\plotone{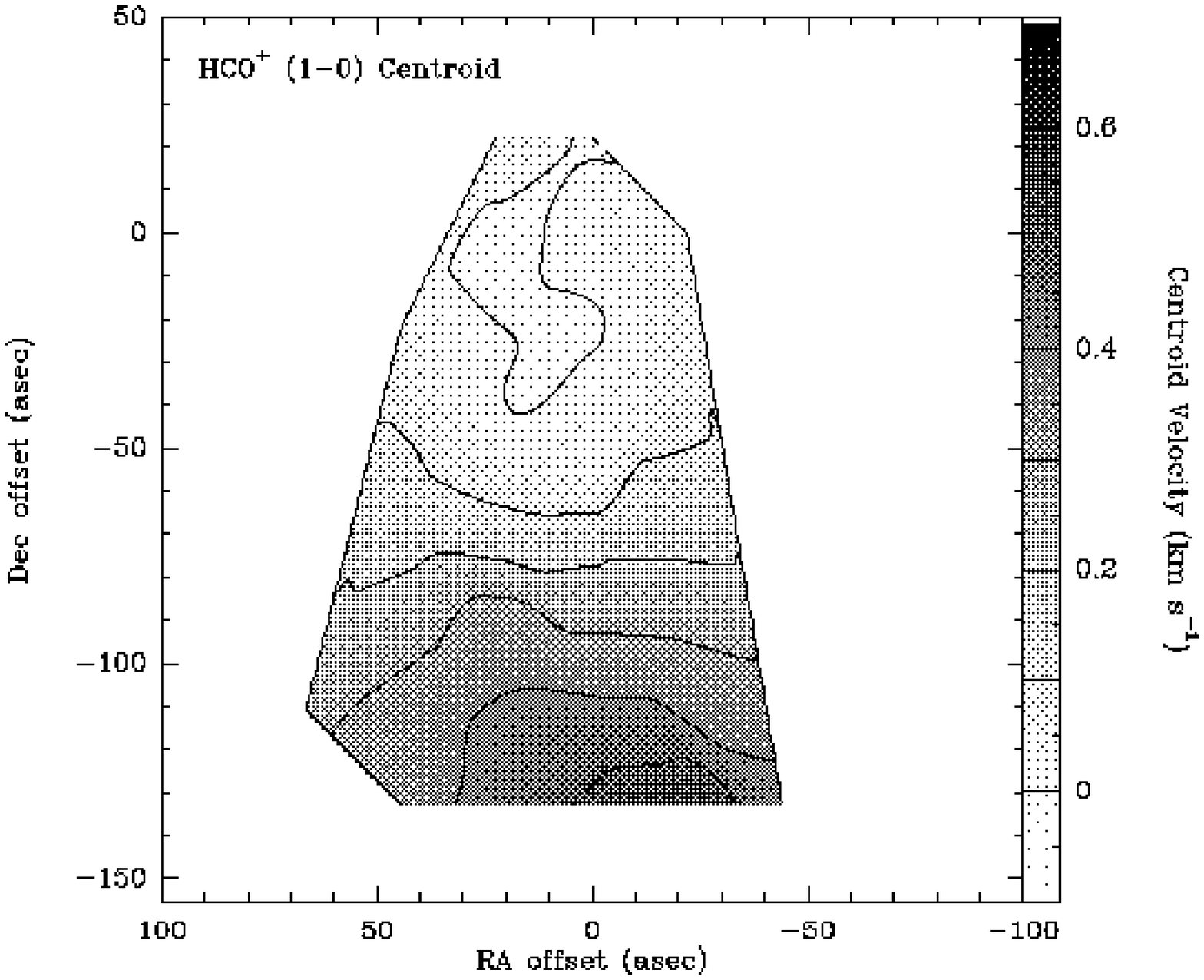}
\caption{\label{sfoxxxvii_centroid} The SFO~37 centroid velocity integrated
  over
  the line core of \hcop~(\jequals{1}{0}). The line of sight velocity has been
  subtracted out and the contours and greyscale are indicated on the wedge to
  the right of the figure.}
\end{figure}

The \hcop~(\jequals{1}{0})
integrated intensity map (Figure~\ref{sfoxxxvii_itty}) shows emission from the
IRAS core, as well as an
extension to the southeast. 
There is a centroid velocity gradient of the
\hcop~(\jequals{1}{0}) emission along the length of the cometary cloud
(Figure~\ref{sfoxxxvii_centroid}). This velocity gradient may be due to the
interaction of the ionization front and the cometary cloud. 
Recent near-infrared
observations of this source have also shown several point sources which may be
pre-main-sequence stars which lie on a line extending from the embedded IRAS
source in the opposite direction as the tail \citep{sto2,sto}. These young
stars, and their extremely regular arrangement along the line which bisects the
ionization front, which would be the line along which the dense core would
travel in
a radiation driven cometary globule \citep{ll}, make this source an excellent
candidate for small scale sequential star formation \citep{sto}. The
\hcop~(\jequals{1}{0}) map shows evidence of some fragmentation of the tail
of the cometary cloud into what could potentially seed the cores of new
sequentially driven star formation. The \hcop~(\jequals{3}{2}) and
\hcop~(\jequals{4}{3}) integrated intensity maps pick up the dense gas
surrounding the central IRAS source. The second core, located approximately
100\sec\ behind the IRAS source, has a lower excitation temperature than the
core surrounding the IRAS source, which may indicate that it is much less
dense. Once the ionization front moves past the IRAS source, it could continue
on to
this second core, inducing star formation once again. The IRAS source would
become another young star in the line of young stars which originated in this
sequential star forming region. 

\subsection{Bok Globules}
\subsubsection{B335}
The dark cloud B335 has been well studied. The embedded IRAS source 19345+0727
appears to be a protostellar core which is the source of a bipolar outflow
which is oriented in the east-west direction on the plane of the
sky. The central source shows an elongation in the far-infrared and
submillimeter continuum in the north-south direction, orthogonal to the outflow
\citep{cgshdgh}. The high visual extinction and lack of near-infrared emission
\citep{cs,arcet} indicates that it may be a class~0 source. \citet{zekw}
provide evidence for protostellar collapse in B335 using H$_{2}$CO and CS
observations. Subsequently, \citet{zhou} improved upon the infall models for
B335 by including rotation and confirmed the earlier suggestion of
infall. B335 is one of the best studied infall candidates, which is why we
chose to observe it and compare it with the bright rimmed clouds.

The B335 line profiles (Figure~\ref{sfoxxxvii_bcccxxxv_profiles}b) are very
narrow compared to the other sources we've observed. The optically thin lines
show a velocity width of only about 1 \mbox{km~s$^{-1}$}. This comparably
narrow line width is more typical of nearby isolated star forming regions than
the large line widths we see in the bright rimmed clouds. This is probably due
to the fact that bright rimmed clouds are on average more distant and larger
than many of the Bok globules observed in millimeter and submillimeter
wavelengths. As a result of the relationship between cloud size and velocity
dispersion \citep{larson,mg} we expect the bright rimmed clouds to have
generally larger line widths than smaller clouds such as B335 or CB~224.

The
\hcop~(\jequals{1}{0}) profile
shows a large self-absorption trough, however the centroid is not blue shifted,
as one would expect for infall, but red-shifted. This is a result of the core
rotation \citep{zhou} and the large beam of FCRAO. The \hcop~(\jequals{4}{3})
line profile also shows self absorption, which results in a double-peaked
spectrum. This spectrum, however, shows a considerable blue-shifted centroid
relative to its isotopic counterpart. The \hthcop~(\jequals{1}{0}) and
\hthcop~(\jequals{4}{3}) show single
peaks right at the point where the \hcop\ lines are most self-absorbed,
indicating that there is a high optical depth of \hcop\ at those velocities,
and not that there are two different molecular cloud structures with slightly
differing velocities in that direction. 

\begin{figure}
\epsscale{1.0}
\plotone{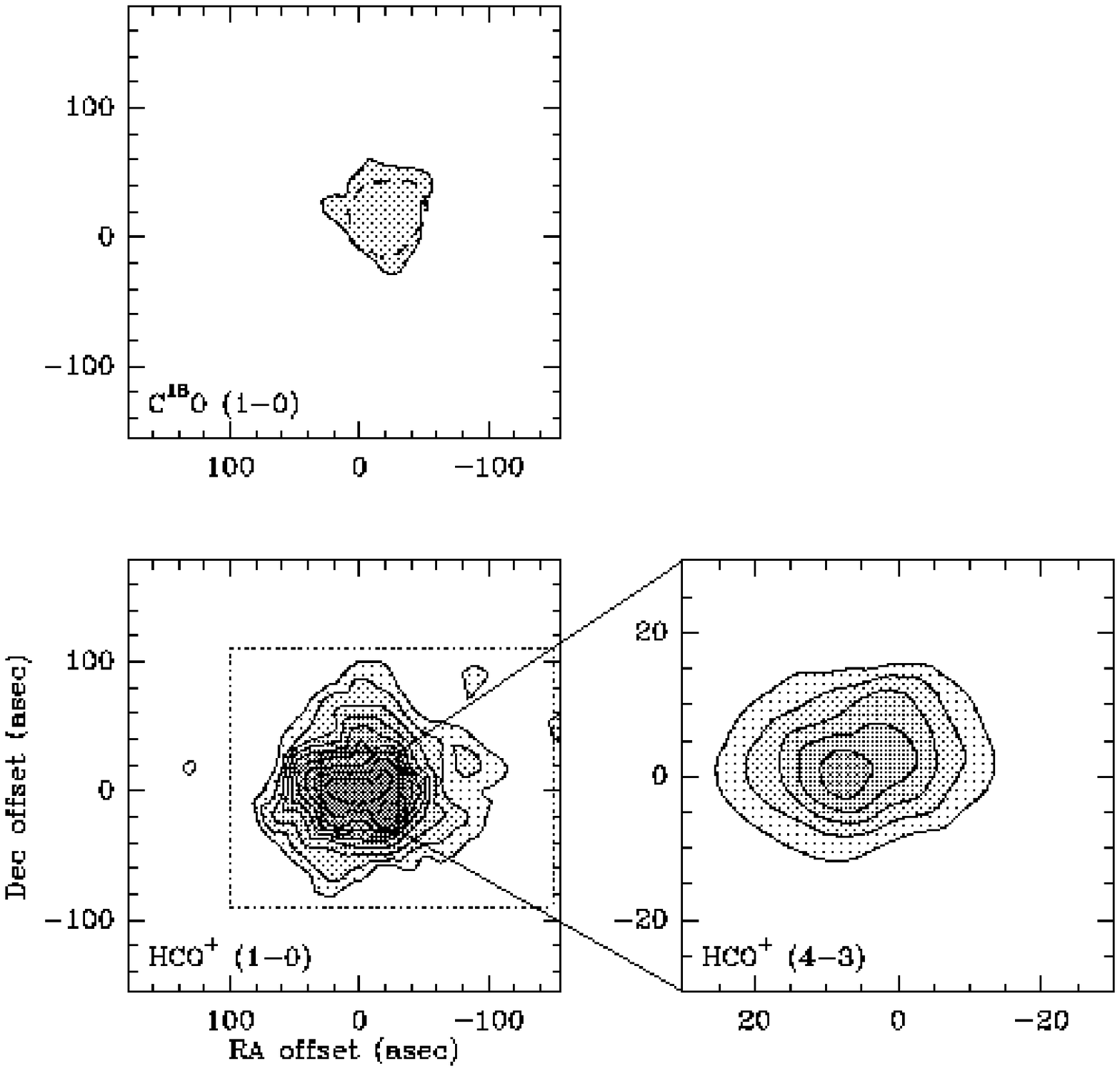}
\caption{\label{b_itty} Integrated intensity maps of B335 in various
  transitions and isotopomers of \hcop and CO. The IRAS source 19345+0727 is
  located at the center of each map. The \ceio~(\jequals{1}{0}) map has a lowest
  contour of 0.6 K~\mbox{km~s$^{-1}$} ($3\sigma$) and increments of 0.2
  K~\mbox{km~s$^{-1}$} ($1\sigma$). The \hcop~(\jequals{1}{0}) map has a lowest
  contour of 0.3 K~\mbox{km~s$^{-1}$} ($3\sigma$) and increments of 0.1
  K~\mbox{km~s$^{-1}$} ($1\sigma$). The \hcop~(\jequals{4}{3}) map has a lowest
  contour of 0.8 K~\mbox{km~s$^{-1}$} ($3\sigma$) and increments of 0.3
  K~\mbox{km~s$^{-1}$} ($1\sigma$). The dotted rectangle in the
  \hcop~(\jequals{1}{0}) map indicates the region over which the \hcop\
  centroid is shown in figure~\ref{b_centroid}. The dashed contour in the
  \ceio~(\jequals{1}{0}) indicates the half power contour of the
  \nthp~(\jequals{1}{0}) emission.}
\end{figure}

\begin{figure}
\epsscale{1.0}
\plotone{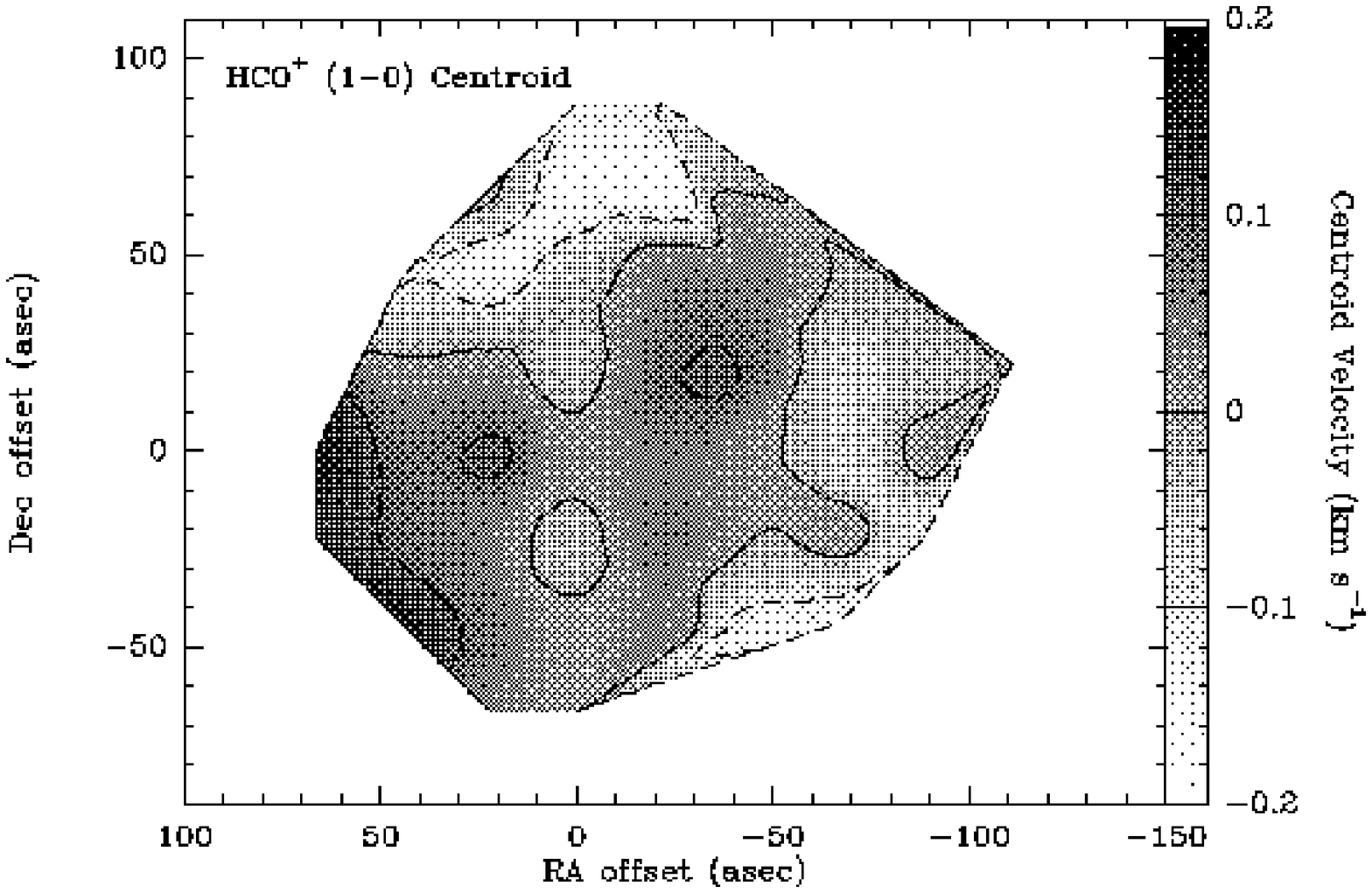}
\caption{\label{b_centroid} The B335 centroid velocity integrated over
  the line core of \hcop~(\jequals{1}{0}). The line of sight velocity has been
  subtracted out and the contours and greyscale are indicated on the wedge to
  the right of the figure.}
\end{figure}

The
\ceio~(\jequals{1}{0}) and \nthp~(\jequals{1}{0}) emission is confined to the
core, which does not seem to be resolved by our beam (Figure~\ref{b_itty}). The
\hcop~(\jequals{1}{0}) emission is
resolved by the FCRAO beam, and seems to peak at the position of the IRAS
core. The \hcop~(\jequals{1}{0}) centroid map (Figure~\ref{b_centroid}) does
not indicate any coherent structure across the cloud. The
\hcop~(\jequals{4}{3}) emission is elongated in the east-west direction, along
the outflow (Figure~\ref{b_itty}). The centroid velocity of
\hcop~(\jequals{4}{3}) has a coherent gradient running north-south indicating
that B335 may be rotating around the axis defined by its outflow \citep{nar}.

\subsubsection{CB~3}
CB~3 \citep{cb}, also known as L~594 \citep{lbn}, is a Bok globule located
approximately 2500 pc from us \citep{lh}. Embedded within it is the IRAS source
00259+5625, which has an extremely high luminosity (930 $L_{\sun}$) and appears
to be a source of isolated intermediate- or high- mass star formation, not
typically associated with Bok globules (see \citet{cb2} and references
therein). Because this source is so distant and has such a high luminosity IRAS
source it is not typically studied as a source of low-mass star formation, but
it may be a good counterpart for the more distant bright rimmed clouds, such as
SFO~13, SFO~25, or SFO~37.

\begin{figure}
\epsscale{1.0}
\plotone{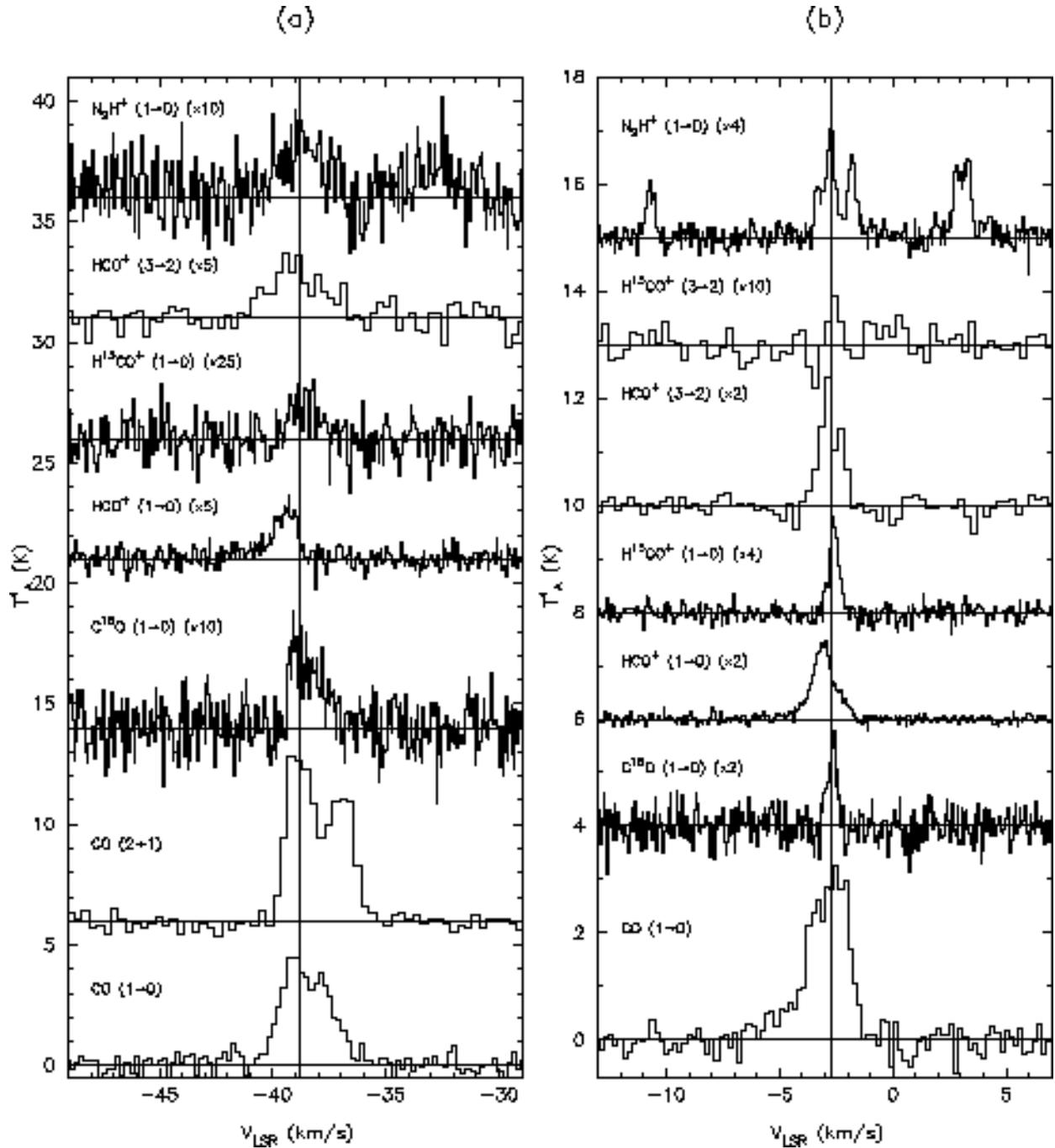}
\caption{\label{cb_lbn_profiles} Line profiles of molecular transitions in
  the direction of the central IRAS source of CB~3 (a) and CB~224 (b).}
\end{figure}

The CB~3 CO central line profiles (Figure~\ref{cb_lbn_profiles}a) are
more complex than many of the bright rimmed cloud's line profiles. Both the
CO~(\jequals{1}{0}) and CO~(\jequals{2}{1}) profiles show a component near the
\nthp\ line center velocity, and a second component red shifted by a couple of
\mbox{km~s$^{-1}$} from the first. The \ceio~(\jequals{1}{0}) line does not
show two components, but does have a significant red shifted tail. 
The \hcop~(\jequals{1}{0}) central line profile is blue-shifted relative to
both the \nthp~(\jequals{1}{0}) line as well as the \hthcop~(\jequals{1}{0})
line, which could indicate infall in the core.
The \hcop~(\jequals{3}{2}) line shows 
less self-absorption on the red-shifted side of the line, but still has a
centroid velocity which is slightly blue relative to the \nthp\ profile. 

\begin{figure}
\epsscale{0.75}
\plotone{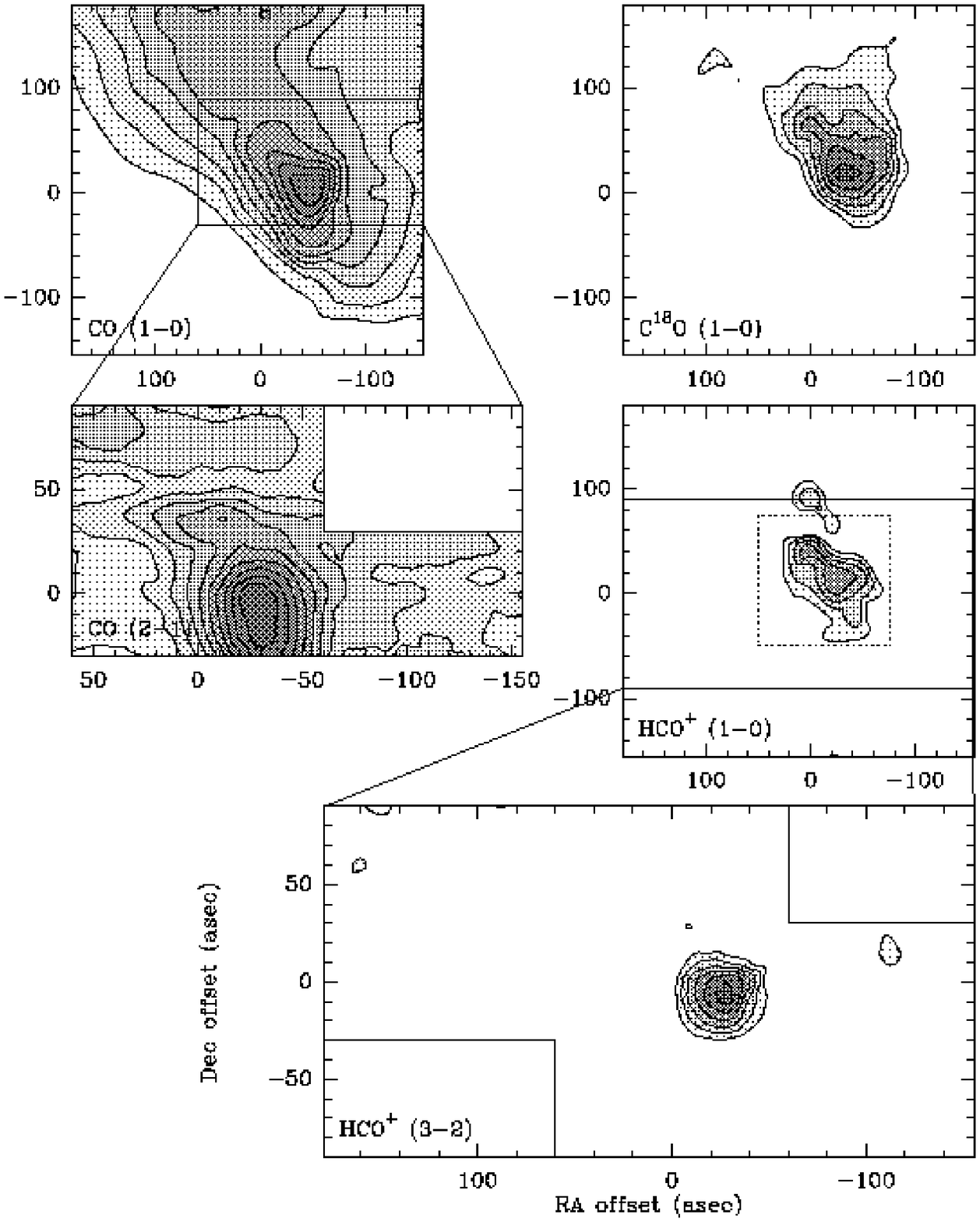}
\caption{\label{lbn_itty} Integrated intensity maps of CB~3 in various
  transitions and isotopomers of \hcop and CO. The IRAS source 00259+5625 is
  located at the center of each map. The CO~(\jequals{1}{0}) map has a lowest
  contour of 2.2 K~\mbox{km~s$^{-1}$} ($3\sigma$) and increments of 2.2
  K~\mbox{km~s$^{-1}$} ($3\sigma$). The CO~(\jequals{2}{1}) map has a lowest
  contour of 3.1 K~\mbox{km~s$^{-1}$} ($3\sigma$) and increments of 5.1
  K~\mbox{km~s$^{-1}$} ($5\sigma$). The \ceio~(\jequals{1}{0}) map has a lowest
  contour of 0.5 K~\mbox{km~s$^{-1}$} ($3\sigma$) and increments of 0.2
  K~\mbox{km~s$^{-1}$} ($1\sigma$). The \hcop~(\jequals{1}{0}) map has a lowest
  contour of 0.5 K~\mbox{km~s$^{-1}$} ($3\sigma$) and increments of 0.2
  K~\mbox{km~s$^{-1}$} ($1\sigma$). The \hcop~(\jequals{3}{2}) map has a lowest
  contour of 1.5 K~\mbox{km~s$^{-1}$} ($3\sigma$) and increments of 0.5
  K~\mbox{km~s$^{-1}$} ($3\sigma$). The dotted rectangle in the
  \hcop~(\jequals{1}{0}) map indicates the region over which the \hcop\
  centroid is shown in figure~\ref{lbn_centroid}.}
\end{figure}

The
CO~(\jequals{1}{0}) integrated intensity map (Figure~\ref{lbn_itty}) shows a
ridge of gas running northeast--southwest, but no peak at the core, however the
CO~(\jequals{2}{1}) map clearly 
shows the core, which is displaced from the (0,0) position, but remains
consistent with the error position of the IRAS source. The
\ceio~(\jequals{1}{0}) also highlights the position of the dense core along the
low-density molecular gas ridge. 
The
\hcop~(\jequals{1}{0}) integrated intensity map shows
the gas core, with some extension along the ridge, however the
\hcop~(\jequals{3}{2}) emission is centered only on the core position. 

\begin{figure}
\epsscale{1.0}
\plotone{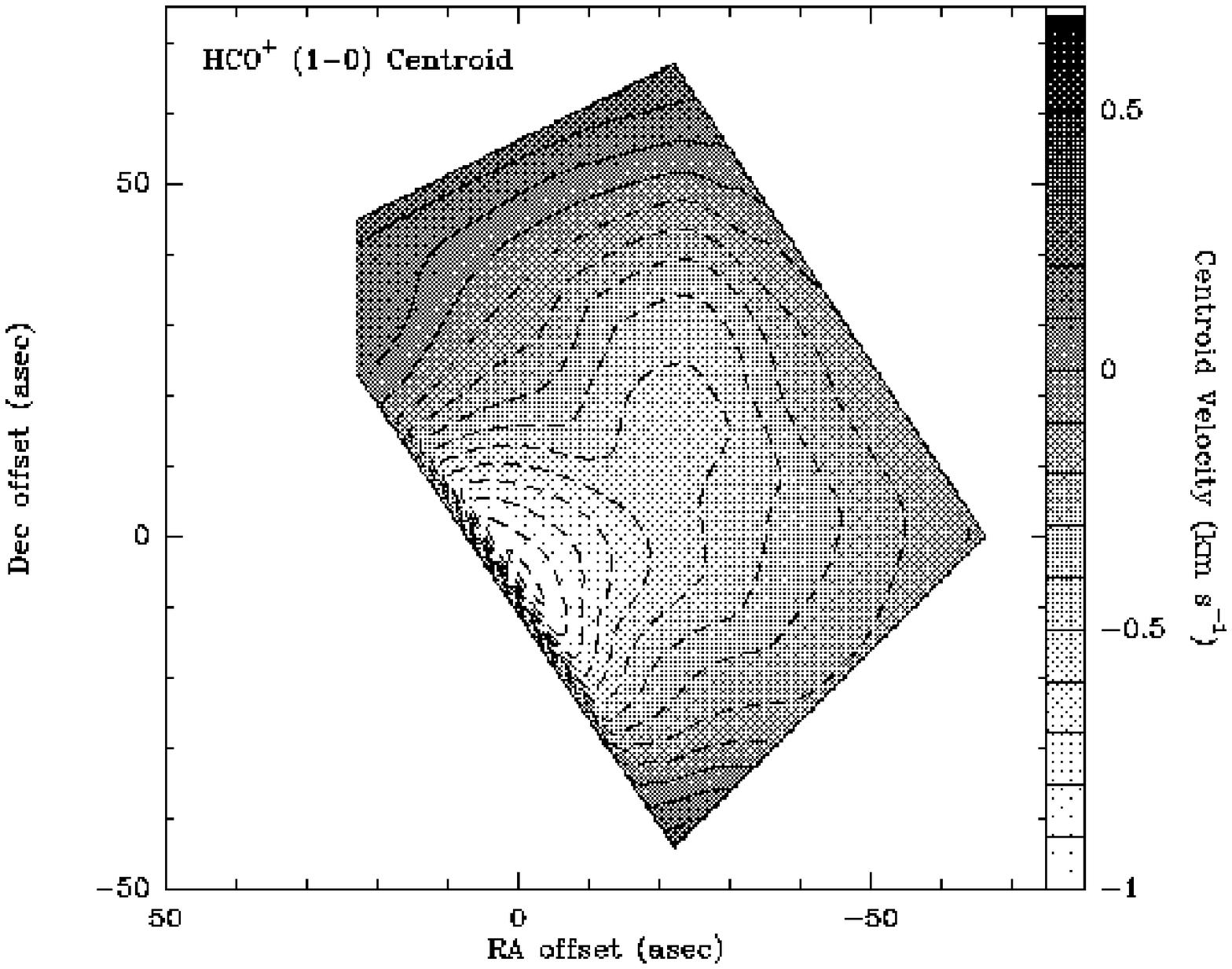}
\caption{\label{lbn_centroid} The CB~3 centroid velocity integrated over
  the line core of \hcop~(\jequals{1}{0}). The line of sight velocity has been
  subtracted out and the contours and greyscale are indicated on the wedge to
  the right of the figure.}
\end{figure}

\begin{figure}
\epsscale{1.0}
\plotone{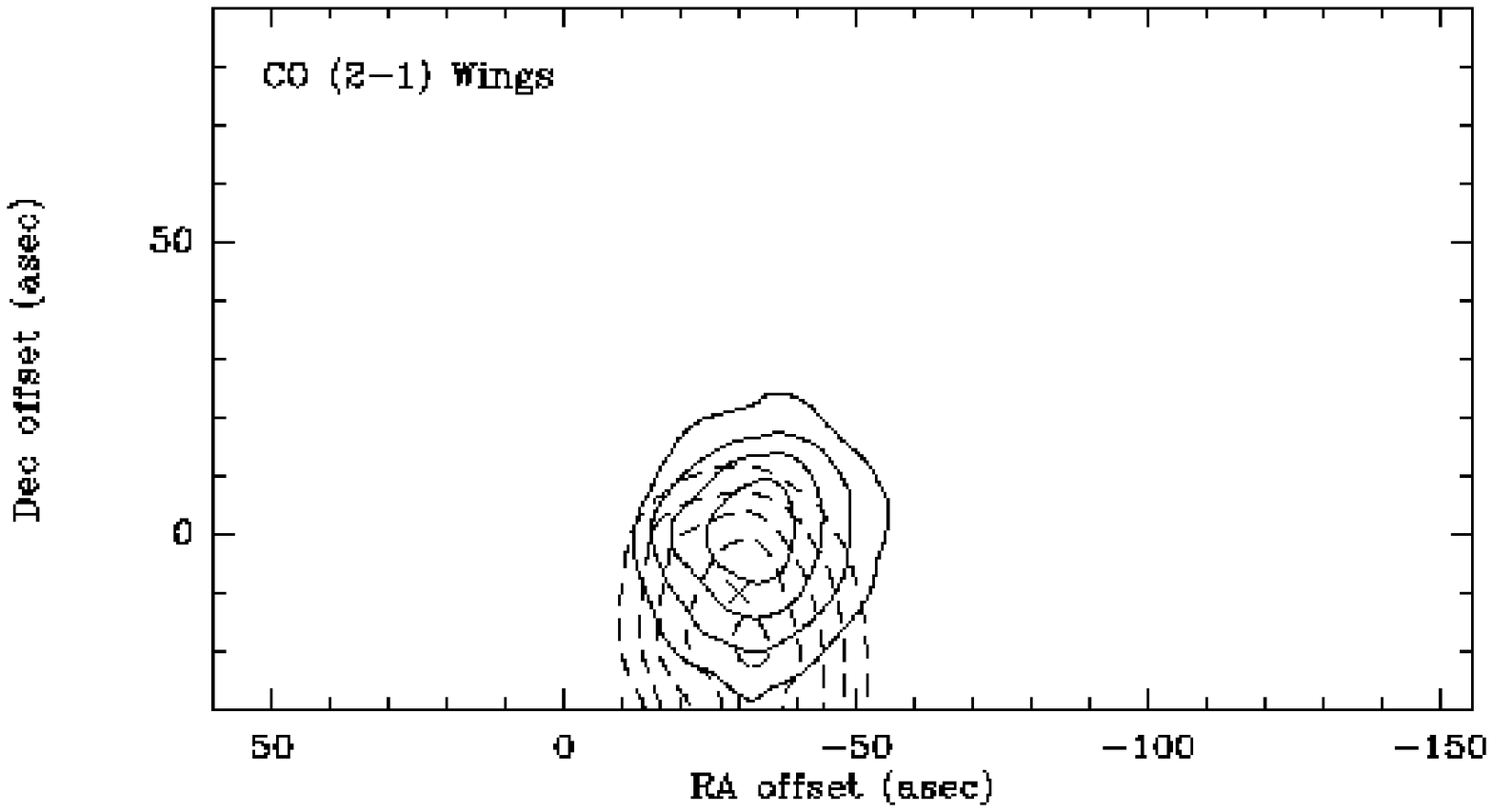}
\caption{\label{lbn_outflow} The CB~3 CO~(\jequals{2}{1}) line wing
  emission. The blue lobe, indicated by dotted lines, is the integrated
  intensity in the range of --49 to --39.5 \mbox{km~s$^{-1}$}. The red lobe,
  indicated by solid lines, is the integrated intensity in the range from
  --37 to --29 \mbox{km~s$^{-1}$}. The lowest contour in each case is the
  half power contour. The x indicates the \hcop~(\jequals{3}{2}) peak
  integrated intensity position.}
\end{figure}

The
\hcop~(\jequals{1}{0}) emission does not show a linear velocity gradient, but
does have some structure (Figure~\ref{lbn_centroid}).
The CO~(\jequals{2}{1}) line wing map
(Figure~\ref{lbn_outflow}) shows a bipolar structure with the blue-shifted line
wing to the south of the core and the red-shifted wing to the north. This
indicates the probable presence of a bipolar outflow in this core. Our
detection of this outflow is consistent with the outflow detected by
\citet{cb2}.

\subsubsection{CB~224}
CB~224 \citep{cb} is a Bok globule located at a distance of approximately 450
pc, and it's embedded IRAS source, 20355+6343, has a luminosity of 3.9
$L_{\sun}$ \citep{lh}. This is very comparable to many of the SFO sources
discussed above, making CB~224 a good isolated source to compare with the
potentially triggered SFO sources. CB~224, however, has very weak molecular
emission lines. 

The CO~(\jequals{1}{0}) central line profile shows evidence of a blue-shifted
wing, which is not present in the \ceio~(\jequals{1}{0}) profile. This may
indicate the presence of an outflow (Figure~\ref{cb_lbn_profiles}b). 
The \hcop~(\jequals{1}{0}) line profile is blue-shifted relative to both the
\nthp\ emission, and the \hthcop~(\jequals{1}{0}) emission profile. The
\hcop~(\jequals{1}{0}) profile shows a
knee 
right where the \hthcop\ emission peaks indicating a fairly classical
self-absorption profile typical of infall regions. The \hcop~(\jequals{3}{2})
profile shows some evidence of a slight centroid blue-shift relative to the
\hthcop~(\jequals{3}{2}) line and the \nthp~(\jequals{1}{0}) line, but it is
not as convincing an infall signature as the lower energy transition. 

\begin{figure}
\epsscale{0.5}
\plotone{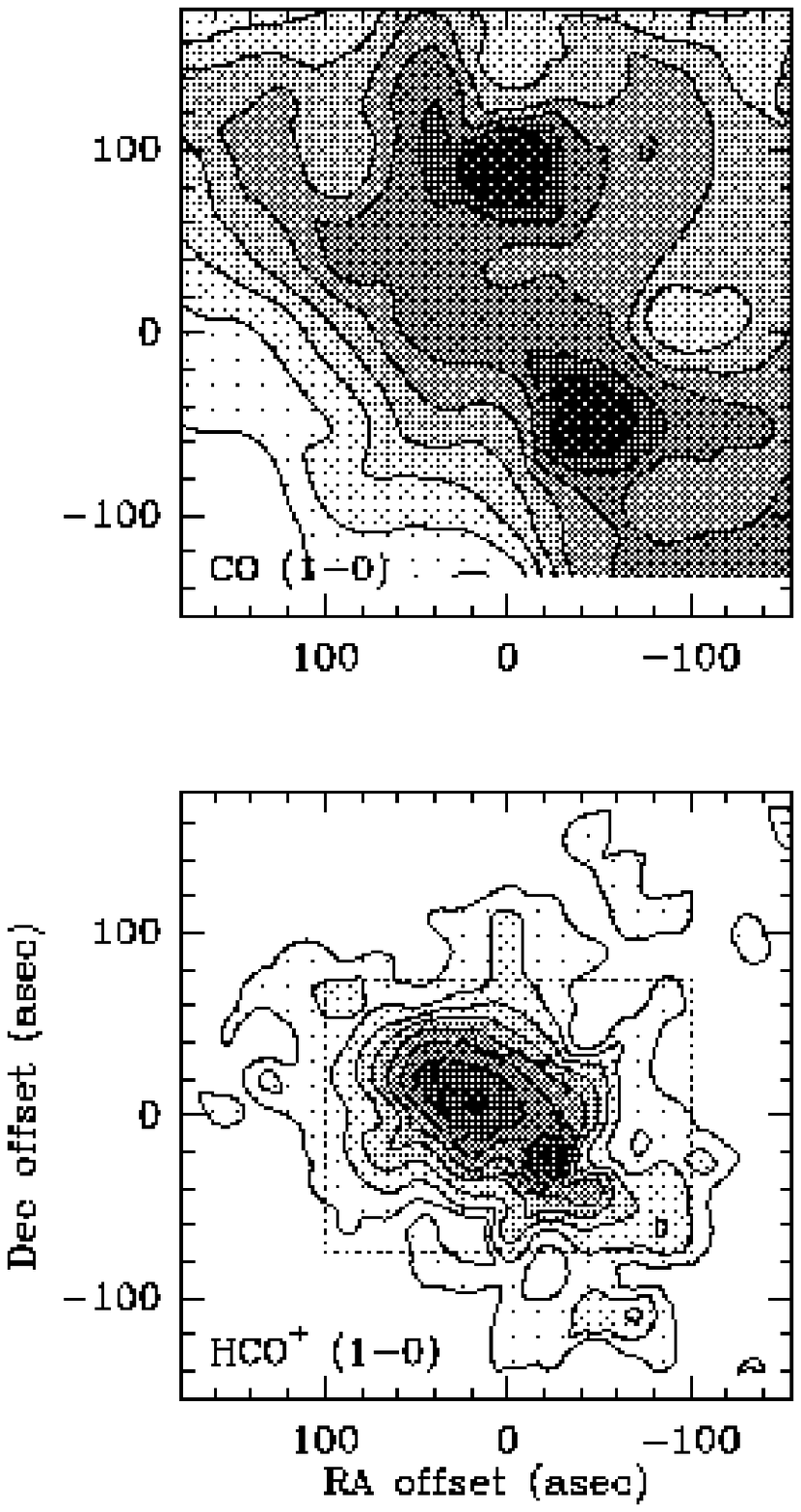}
\caption{\label{cbccxxiv_itty} Integrated intensity maps of CB~224 in various
  transitions and isotopomers or \hcop and CO. The IRAS source 20355+6343 is
  located at the center of each map. The CO~(\jequals{1}{0}) map has a lowest
  contour of 2.4 K~\mbox{km~s$^{-1}$} ($3\sigma$) and increments of 0.8
  K~\mbox{km~s$^{-1}$} ($1\sigma$). The \hcop~(\jequals{1}{0}) map has a lowest
  contour of 0.3 K~\mbox{km~s$^{-1}$} ($3\sigma$) and increments of 0.1
  K~\mbox{km~s$^{-1}$} ($1\sigma$). The dotted rectangle in the
  \hcop~(\jequals{1}{0}) map indicates the region over which the \hcop\
  centroid is shown in figure~\ref{cbccxxiv_centroid}. The dashed contour in
  the 
  \hcop~(\jequals{1}{0}) indicates the half power contour of the
  \nthp~(\jequals{1}{0}) emission.}
\end{figure}

\begin{figure}
\epsscale{1.0}
\plotone{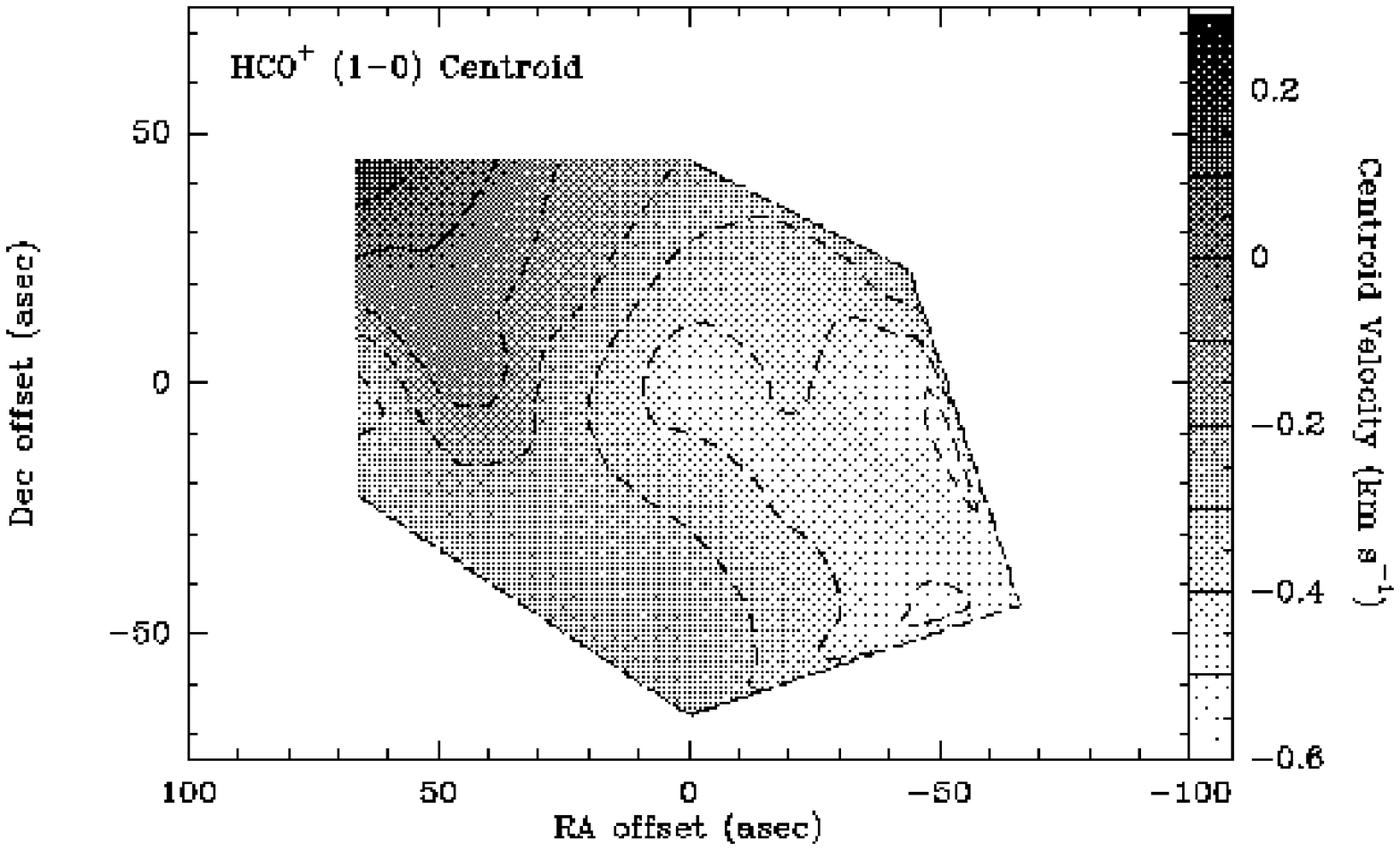}
\caption{\label{cbccxxiv_centroid} The CB~224 centroid velocity integrated over
  the line core of \hcop~(\jequals{1}{0}). The line of sight velocity has been
  subtracted out and the contours and greyscale are indicated on the wedge to
  the right of the figure.}
\end{figure}

The
CO~(\jequals{1}{0}) integrated intensity map shows molecular emission is fairly
ubiquitous around this source (Figure~\ref{cbccxxiv_itty}), however there are
two CO peaks, one north and one south of the IRAS core.
The
\hcop~(\jequals{1}{0}) integrated intensity map clearly highlights the core gas
surrounding the IRAS source at the center of the map. There is a gradient
running northeast--southwest in
the \hcop~(\jequals{1}{0}) centroid velocity which could indicate rotation in
the core (Figure~\ref{cbccxxiv_centroid}). 

\section{Discussion}
\subsection{Core Masses}
Of the seven bright rimmed clouds observed and the three Bok globules, all the
sources show evidence of a very dense core with the exception of SFO~4, which
is unusual in that it appears to be a fairly diffuse cloud. \citet{tur}
has shown that the \nthp~(\jequals{1}{0}) transition can be a good probe of
the properties of star forming cores. While CO, \hcop, and other molecules have
a tendency to freeze on to grains in the low temperature, high density
environment of cores, \nthp\ remains relatively undepleted \citep{bl}
in dark cloud cores, making it a good probe of the density
profile. \nthp~(\jequals{1}{0}) also has several fine structure lines
\citep{cmt} which can be used to derive the excitation temperature of the
\nthp~(\jequals{1}{0}) line. We use the ``hyperfine structure (hfs) method'' in
CLASS \citep{fgl} to derive these properties from the \nthp\ emission, using
the component separations observed by \citet{cmt}. The excitation
temperature and the optical depth derived from the hyperfine line ratios can be
used to derive the column density of gas within the telescope beam, which can
be related to the gas mass in the beam, assuming a relative abundance ratio
between \nthp\ and molecular hydrogen \citep{bcm}. We have performed this
analysis for the objects in our sample, and the results are shown in
Table~\ref{coretable}. The column densities of \nthp\ fall within the same
range of column densities observed in the cores of dark clouds \citep{bcm}.

The mass of the core can also be derived by comparing the emission of
\hcop~(\jequals{1}{0}) to \hthcop~(\jequals{1}{0}), under the assumption that
the \hthcop\ emission is optically thin. The optical depth can be
derived from assuming an abundance ratio between the C and $^{13}$C isotopes,
and
comparing the ratio of \hthcop\ emission to \hcop\ emission. Then the
excitation temperature is derived by assuming a filling factor of unity, and
the excitation temperature and optical depth are used to derive the column
density of gas. The results of this analysis are presented in
table~\ref{hcoptable}. It is
encouraging to see that even though \nthp\ and \hcop\ do not trace exactly the
same gas they produce fairly comparable results for the overall core
mass. These clouds tend to have cores with masses on the order of tens of solar
masses.

\subsection{Outflows}
Outflows appear to be fairly ubiquitous in our sample. We detected outflows in
nearly all the sources we surveyed. This is to be expected, since a star is
forming and the cloud must dissipate its angular momentum in order to
collapse. However, in the case of shock triggered regions, the ionization front
does
dissipate some of the angular momentum of the collapsing cloud
\citep{elm1}, though the ionization front would probably not have a noticeable
effect on the scale of the accretion disk. 

We have analyzed the energy in the detected outflows and present the results in
table~\ref{outflowtable}. We derived the mass of the lobes by averaging the
integrated intensity of the line-wing emission over the spatial extend of the
lobe. We then derived a column density from this integrated intensity assuming
that the emission is optically thin and an excitation temperature of
30K. The momentum ($P$) and kinetic energy (KE) of an
outflow can be derived by computing 
\[
P = \sum_{\rm ch} m_{\rm ch} v_{\rm ch},\ {\rm and}\ {\rm KE} = \frac{1}{2} \sum_{\rm ch}
m_{\rm ch} v_{\rm ch}^{2}
\]
along the spatially averaged spectrum of the outflow lobe. Unless the outflow
is along the line of sight, this method underestimates these quantities by a
factor of $\sin(i)$ where $i$ is the inclination of the outflow relative to the
plane of the sky. As this method tends to underestimate the momentum and
kinetic energy of the outflow, and since these estimates have fairly large
uncertainties we chose not to integrate over the averaged spectrum, but instead
to calculate the momentum to be $M_{\rm total} V_{\rm max}$ and kinetic
energy to be $\frac{1}{2} M_{\rm total} V_{\rm max}^{2}$ where $M_{\rm total}$
is the total mass of the outflow lobe, and $V_{\rm max}$ is the maximum
measured velocity of the outflow lobe relative to the line center. This assumes
that all the material in the flow is moving at the maximum observed velocity,
and that the lower velocities are the result of projection effects. This
technique results in upper limit estimates of the flow's momentum and energy
\citep{wlym}. The dynamical age ($\tau_{d}$) of the flow is found by
calculating
the distance from the core to the edge of the outflow lobe and dividing by
$V_{\rm max}$, while the kinetic luminosity ($L$) is derived by deviding the
kinetic energy of the lobe by the dynamical age of the flow.

The dynamical ages and kinetic luminosities of these outflows tend to agree
with those of previously observed outflow sources \citep{sacgm}. The outflow
ages all tend to be greater than $10^{4}$ years, which may indicate that the
embedded objects are class~I sources rather than class~0 sources, which
typically have outflow ages less than $10^{4}$ years \citep{sacgm}. 

\begin{figure}
\epsscale{1.0}
\plotone{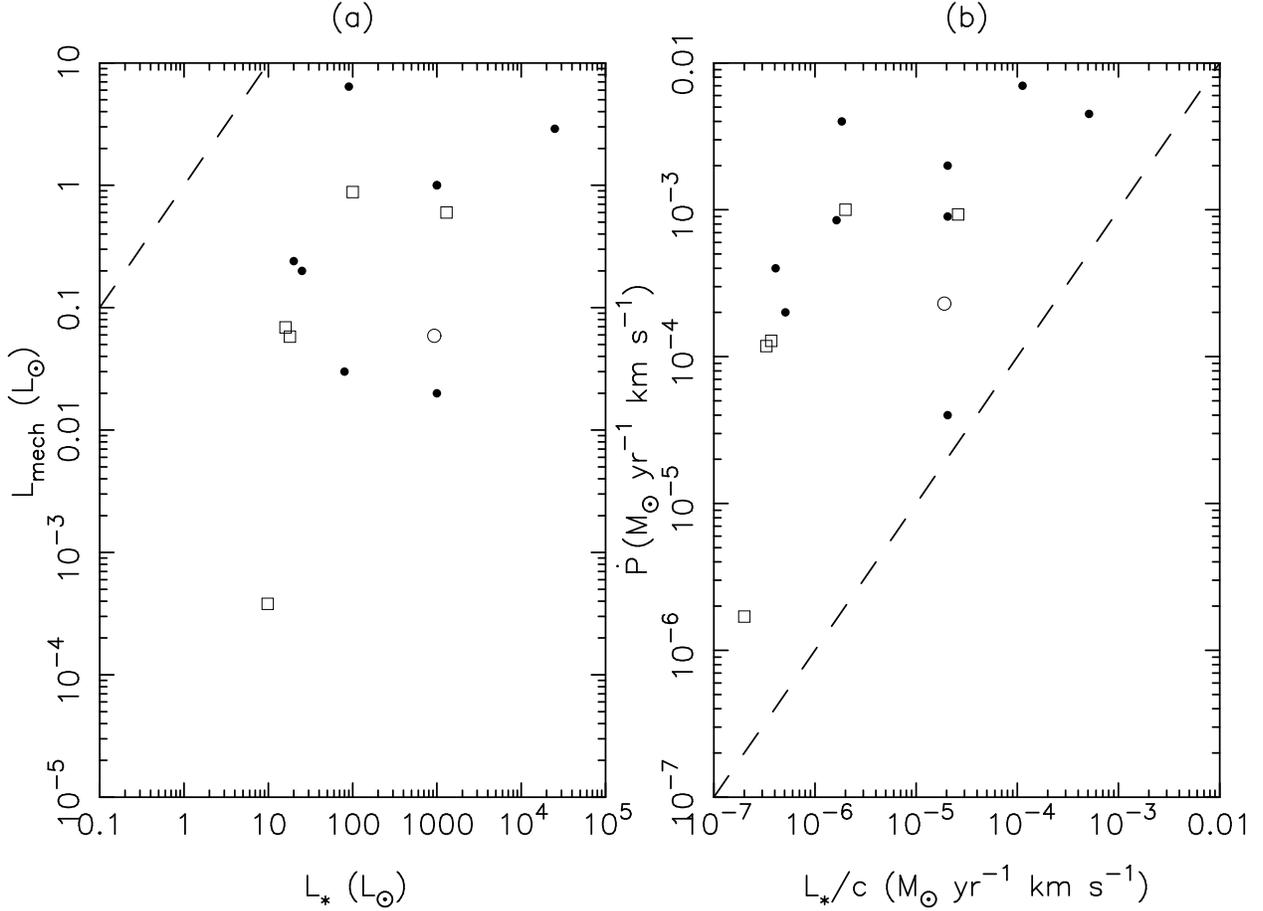}
\caption{\label{outflow} The energetics of the outflow are compared with those
  of the photons arising from the embedded IRAS sources we observed. The open
  squares represent the bright rimmed clouds we observed, the open circle is
  CB~3 and the closed circles are data from \citet{lada1}. Figure~(a) shows the
  outflow mechanical luminosity compared to the luminosity of the IRAS
  source. Figure~(b) compares the force needed to accelerate the molecular
  outflows to the total radiant pressure of the central object. The dashed
  lines indicate where forces or luminosities are equal. The bright rimmed
  clouds share similar properties with previously observed outflows, namely
  that although there exists enough energy in the central source to drive the
  outflows, radiative scattering is not the source of the outflow's
  acceleration.}
\end{figure}

We compare the mechanical luminosity and force needed to accelerate the outflow
to the luminosity and radiant pressure of the IRAS source driving the
outflow in Figure~\ref{outflow}. We find that the bright rimmed clouds occupy
similar regions in the
two plots indicating that they are probably driven by similar processes as the
other outflows around forming stars that have been observed to date. We find
that the embedded IRAS source has sufficient energy to drive the outflows we
observe, but that the radiant pressure of the IRAS source is not sufficient to
accelerate the outflows to the velocities we observe. This confirms similar
observations of \citet{lada1}.

\subsection{Infall Motion}
\label{infall}
Although several outflows were detected, and the presence of outflows implies
infalling gas, there were few line asymmetries indicative of infall in the
bright rimmed clouds. \citet{mmtwbg} have defined a parameter to quantify the
line asymmetry in terms of the line center velocity of an optically thick
line, the line center velocity of an optically thin line, and the width of the
optically thin line. The asymmetry ($\delta V$) is then defined as \[\delta V =
\left( V_{\rm thick} - V_{\rm thin} \right)/\Delta V_{\rm thin}.\]
\citet{mmtwbg} find the line center velocities of both the thick and thin lines
by fitting a gaussian to the lines, however this does not yield a very robust
estimator of the asymmetry in my opinion. As the optical depth increases, the
thick line does immediately separate into two distinguishable components. At
moderate optical depths a knee forms in the line profile as the peak becomes
blue-shifted. A gaussian fit to this type of profile is very inaccurate,
however the centroid of this profile is well determined. Therefore we use
the centroid velocity of the thick line when calculating the line asymmetry. 

We tabulate our line asymmetry values in table~\ref{deltavtable}. The
millimeter asymmetry is derived from observations of the
\hcop~(\jequals{1}{0}) optically thick line and the \hthcop~(\jequals{1}{0})
optically thin line. The submillimeter asymmetry value is derived from the
observations of the \hcop~(\jequals{3}{2}) optically thick line and the
\hthcop~(\jequals{3}{2}) optically thin line. \citet{mmtwbg} suggest that a
$\delta V$ between --0.25 and 0.25 be considered symmetric. All the bright
rimmed clouds,
with the exception of SFO~18, show no significant asymmetry in their central
line profiles according to this criterion.

\begin{figure}
\epsscale{1.0}
\plotone{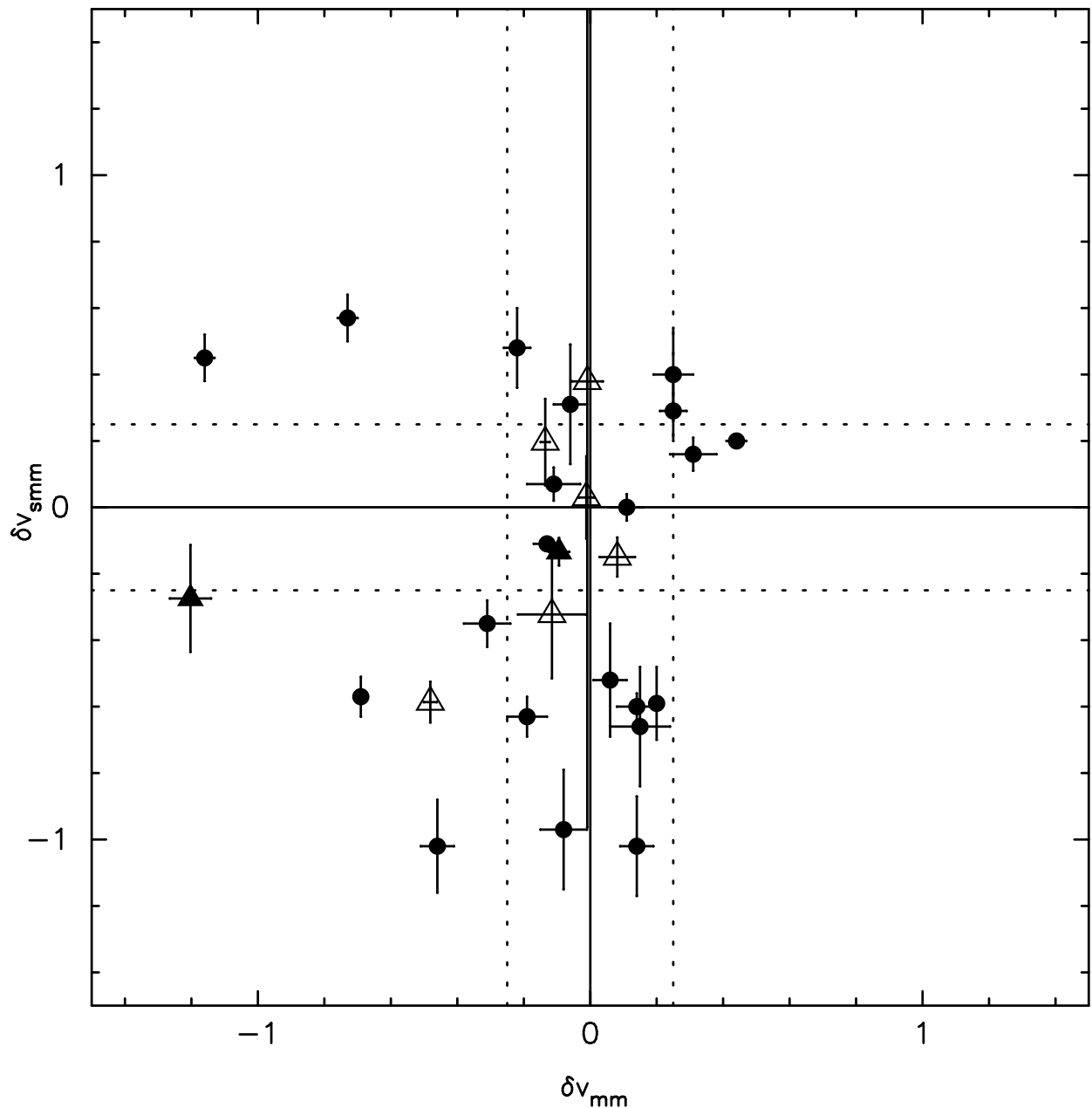}
\caption{\label{deltav} The line asymmetry parameter ($\delta V$) towards the
  star forming core measured in both millimeter (x--axis) and submillimeter
  (y--axis) molecular
  transitions. The open triangles are the bright rimmed clouds reviewed in this
  paper observed in the millimeter transition \hcop~(\jequals{1}{0}) and the
  submillimeter transition \hcop~(\jequals{3}{2}). The filled triangles are the
  Bok globules we observed in the same transitions as the bright rimmed
  clouds. The filled circles are class~I and class~0 sources observed in the
  millimeter CS~(\jequals{2}{1}) transition by \citet{mmtwbg} and in the
  submillimeter \hcop~(\jequals{3}{2}) transition by \citet{gemm}.}
\end{figure}

Previous studies of class~0 and class~I sources performed using the millimeter
CS~(\jequals{2}{1}) transition \citep{mmtwbg} and the submillimter
\hcop~(\jequals{3}{2}) transition \citep{gemm} show preferentially blue
asymmetric line profiles, thought to be the result of infall in these star
forming cores. \citet{mmtwbg} quantify the overall predilection of the observed
sources to have blue asymmetric line profiles in terms of a parameter they call
the ``blue excess'' which is defines as\[\mbox{blue excess} = \frac{N_{\rm blue}
  - N_{\rm red}}{N_{\rm total}}.\] where $N_{\rm blue}$ is the number of
sources with $\delta V < -0.25$, $N_{\rm red}$ is the number of sources with
$\delta V > 0.25$, and $N_{\rm total}$ is the total number of
sources. \citet{gemm} find a blue excess of 0.28 for the class~0 and class~I
sources they observe, and an overall blue excess of 0.31 for all the sources in
the literature. The bright rimmed clouds have a blue excess of 0.2 measured
in both the millimeter and submillimeter transitions, however that is measured
with a sample of only 6 sources. A scatter plot, comparing the asymmetry of the
bright rimmed clouds and Bok globules we observed with those observed by
\citet{mmtwbg} and \citet{gemm} is shown in Figure~\ref{deltav}. We see that
the bright rimmed clouds show asymmetry values comparable to those of other
star forming regions. The bright rimmed clouds, however, do not show as wide a
deviation of asymmetry values, and tend to have $\delta V$s which are closer to
0 than typical star formation regions, whose $\delta V$s tend to be negative. A
wider survey of bright
rimmed clouds is required to determine if they indeed have significantly less
blue excess than other YSOs.

\begin{figure}
\epsscale{1.0}
\plotone{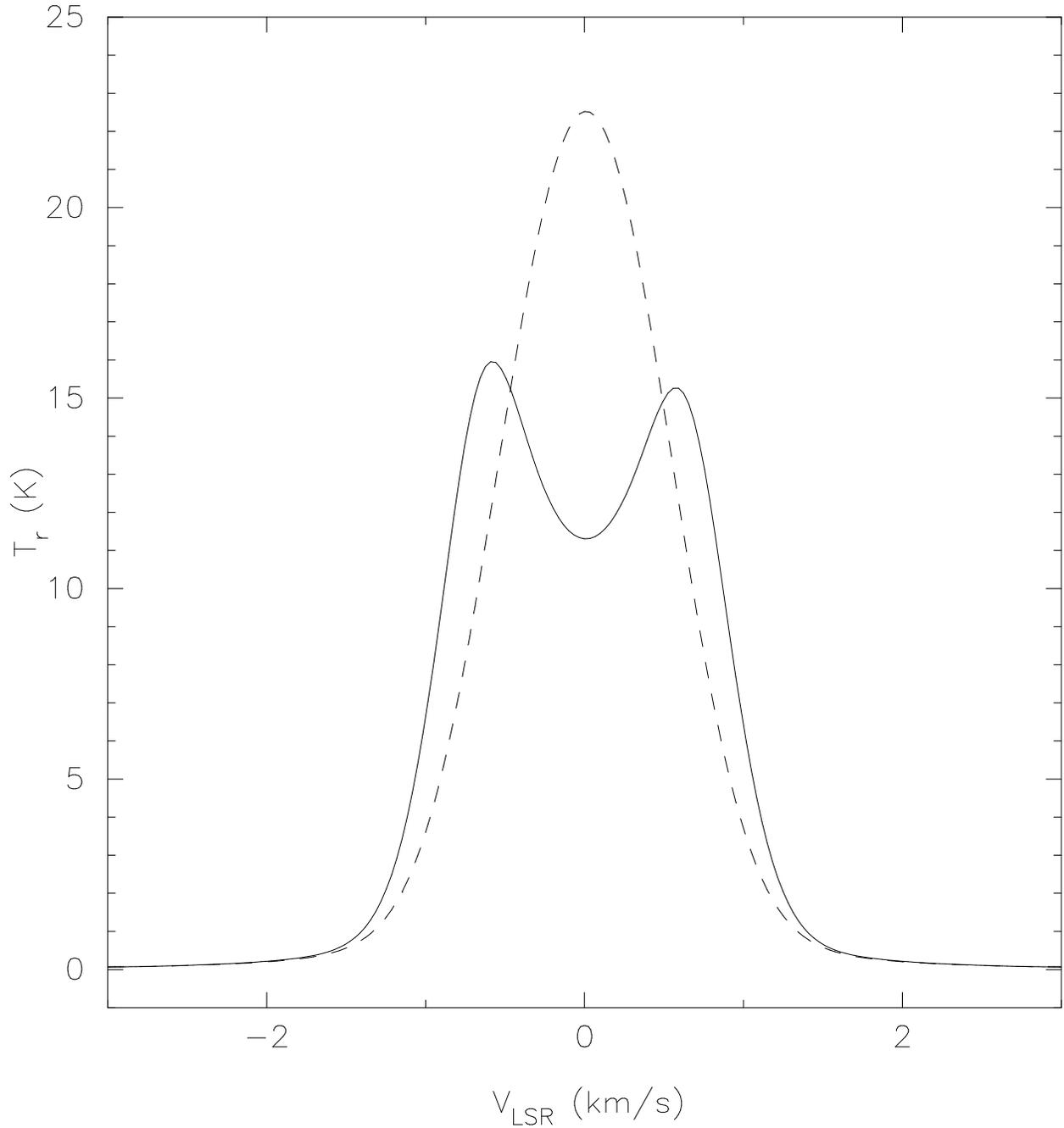}
\caption{\label{lteinfall} The blue-asymmetric line profile is very dependent
  on temperature. The solid line is an LTE radiative transfer model of a
  spherical cloud in free fall collapse with a linear temperature gradient of
  40~K at the
  center and 10~K at the edge. The dashed line is the same collapsing sphere
  with a temperature gradient of 40~K at the center and 300~K at the edge. The
  characteristic blue-asymmetric two-humped line profile disappears when the
  temperature gradient is reversed, and in fact is replaced with a slight
  red-asymmetry.}
\end{figure}

Why do we not detect infall in more bright rimmed clouds? Does the incoming
shock front produce conditions that alter the shape of the emergent line
profile? An isolated Bok
globule usually has an excitation temperature profile which
drops from a value on the order of 20--50K at the core center to 10K at the
edge of the core. Bright rimmed clouds, however, are stripped of their
molecular envelopes and heated by UV flux from O stars. This may be expected to
flatten or in
some cases invert the temperature profile, so that the cloud core center may
still have a physical temperature around 20K, but the edge of the cloud core
may have a physical temperature of several hundred K. Since the gas
throughout the core is fairly excited, it would eliminate the self-absorption
which characterizes the blue line asymmetry typically associated with
infall. This inversion of the excitation temperature gradient could even lead
to absorption of the blue-shifted gas by denser, yet less excited gas closer to
the core. Figure~\ref{lteinfall} illustrates the effect of inverting the
temperature gradient on a free-falling infall region. The model presented in
that figure is a collapsing region, with a peak density of
$10^{7}$~\mbox{cm$^{-3}$} which drops down to a density of
$10^4$~\mbox{cm$^{-3}$} near the edge of the beam. The infall
region is confined to an area which is roughly
50\% of the beam size. We assume a free-fall collapse models, with infall
velocities increasing towards the center of the collapsing sphere. The only
difference between the model which generates
the double peaked solid line and the single peaked dashed line is the
temperature gradient across the infalling region. The solid line is the result
of a linear temperature gradient which peaks in the center at 40~K and drops
off towards the edge of our beam to 10~K. The dashed line also is a linear
temperature gradient, but the temperature increases from 40~K at the center of
the sphere to 300~K at the outer edge. The approximates more closely the
behavior of infalling gas clouds under the influence of wind-triggered collapse
as modeled by VC. Our model shows that heating of the envelope of a collapsing
cloud can lessen or remove entirely the blue-asymmetric line signature of a
collapsing molecular cloud.

\begin{figure}
\epsscale{0.8}
\plotone{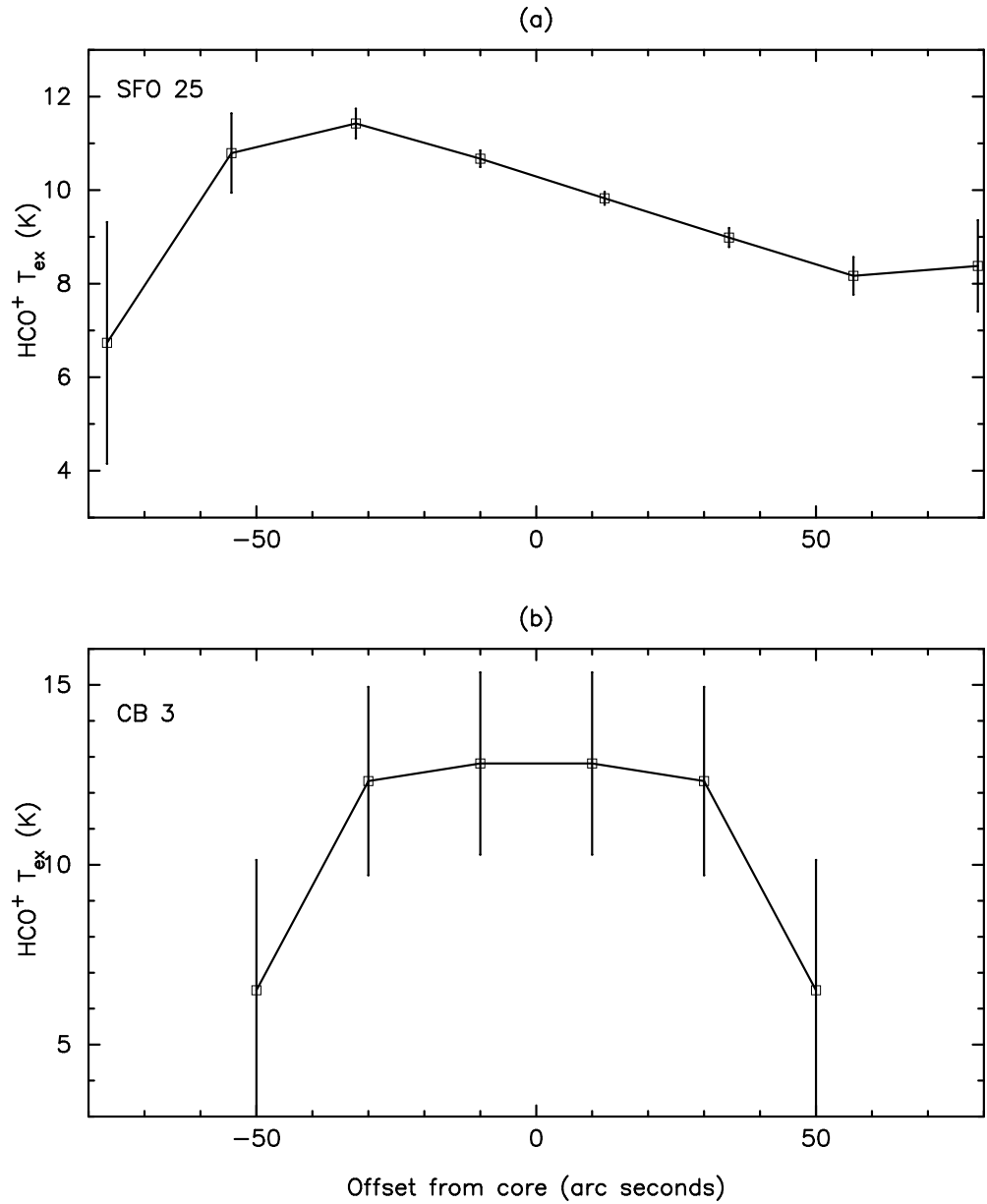}
\caption{\label{texprofiles} Under the assumption of thermodynamic equilibrium
  and low optical depth we derive the excitation temperature profiles across
  both a bright rimmed cloud (SFO~25, a) and a bok globule (CB~3, b). For the
  bright rimmed cloud our cut ran along the line from the ionizing source to
  the core, with the shock front on the negative side of the core. For the Bok
  globule we averaged annuli around to core and calculated the excitation
  temperature in each annulus.}
\end{figure}

In order to test the assumption that the ionization front may be raising the
excitation temperature near the edge of the cloud, we have smoothed the
\hcop~(\jequals{3}{2}) map to the same resolution as the \hcop~(\jequals{1}{0})
in both a bright rimmed cloud (SFO~25) and a Bok globule (CB~3) which we
observed. We assume thermodynamic equilibrium between these transitions, as
well as a low optical depth in order to derive the excitation temperature
across the core. SFO~25 has fairly gaussian lines, centered at the same
velocity as their isotopic counterparts, indicating that the \hcop\ emission
for this source may be optically thin. In CB~3, as in all the Bok globules we
observed, we know the \hcop\ emission is not optically thin, however this would
tend to wash out the excitation temperature to the cloud core, so in regions
where the emission is optically thick we can consider the derived excitation
temperature to be a lower limit. Towards the edge of the CB~3 core the \hcop\
emission does become optically thin making this a good assumption. In the case
of SFO~25 we derive the excitation temperature along a line which cuts through
both the ionization front and the star forming core, and plot that excitation
temperature profile in Figure~\ref{texprofiles}a. The center (0\sec) offset
represents the star forming core position, and the ionization front is in the
negative direction. Figure~\ref{texprofiles}b shows the excitation temperature
profile around the core of CB~3, however we averaged annuli around the central
core in order to derive the plotted values, rather than take a cut straight
across in order to maximize our signal to noise ratio. In the case of SFO~25,
the excitation temperature peaks near the edge of the cloud core, while in CB~3
the excitation temperature peaks near the center.
Although it is likely that the cores of these bright rimmed clouds
are collapsing in a similar manner to other class~0 or class~I sources, the
heating due to the nearby HII region may dampen their spectral line infall
signature.

\section{Conclusions}
Our observations constitute the first detailed millimeter and submillimeter
multitransition study of bright rimmed clouds. Among the 7 bright rimmed
clouds we observe, 6 seem to share traits similar
with other low to intermediate mass star forming regions. Our analysis of these
bright rimmed clouds has yielded the principal results that follow.
\begin{enumerate}
\item New FCRAO CO~(\jequals{1}{0}), \ceio~(\jequals{1}{0}),
  \hcop~(\jequals{1}{0}), \hthcop~(\jequals{1}{0}), and \nthp~(\jequals{1}{0})
  observations along with new HHT CO~(\jequals{2}{1}), \hcop~(\jequals{3}{2}),
  \hcop~(\jequals{4}{3}), \hthcop~(\jequals{3}{2}), and
  \hthcop~(\jequals{4}{3}) observations of 7 bright rimmed clouds and 3 Bok
  globules were presented. These observations constitute the most detailed
  millimeter and submillimeter study of bright rimmed clouds to date.
\item The millimeter CO and \hcop\ emission tends to terminate abruptly at the
  ionization front. As a result, the overall morphology of the CO and \hcop\
  millimeter integrated intensity maps are similar with the optical
  morphologies identified by SFO.
\item The millimeter \hcop\ tends to show the dense swept up ridge behind the
  ionization front, as well as the star forming core around the embedded IRAS
  source. In some of the bright rimmed clouds the \hcop~(\jequals{1}{0})
  emission also traces other overdense clumps which may later be triggered to
  collapse  by the ionization front, resulting in sequential star formation.
\item The millimeter and submillimeter \hcop\ lines from many of the bright
  rimmed clouds appear nearly gaussian, with little evidence of infall
  asymmetry. The only exceptions to this are SFO~18, which shows significant
  blue asymmetry, and SFO~16 which shows a slight red asymmetry relative to
  optically thin tracers.
\item The core masses derived for the bright rimmed clouds using both \nthp\
  and \hcop\ are typical for low and intermediate mass star formation
  regions. The \nthp\ and \hcop\ results also tend to agree to within an order
  of magnitude.
\item The overall blue excess of the sample of bright rimmed clouds is slightly
  less than that of the class~0 and class~I sources observed by \citet{mmtwbg}
  and \citet{gemm}, though the small number of bright rimmed clouds we observed
  does not make this difference statistically significant. A larger survey of
  bright rimmed clouds is required to determine if this is a significant
  finding. We do however make a case for the fact that the heating of the
  collapsing cloud by the adjacent HII region could dampen the infall
  signature, lowering the blue excess of bright rimmed clouds.
\item We observed outflows around 5 of the 7 bright rimmed clouds, including
  new detections of outflows around SFO~13 and SFO~25. These outflows appear to
  have similar properties to other outflows detected in millimeter and
  submillimeter emission.
\end{enumerate}

We do not see direct evidence of triggering in these sources. We can not
determine if star formation was induced in these clouds or if we are seeing the
collapse of pre-existing clumps. We do know that the environment has a profound
effect on these regions.
Although we have found similarities and differences between Bok globules and
bright rimmed clouds, 
a detailed understanding of the effects of an ionization front on star
formation can only be achieved by theoretically modeling this process, and then
comparing that model to observations. In a forthcoming paper we will compare
these observations with models of shock driven collapse derived by VC. In
addition to bright rimmed clouds, these models could explain the effect of
outflows and 
other environmental effects on star formation. Development of the
techniques and models which will help us understand star formation in complex
environments is in progress.

We gratefully acknowledge the staff of the HHT for their excellent support. In
particular we wish to thank Harold Butner for his assistance with setting up
our OTF observations at the HHT and Mark Heyer for his many helpful
comments. C. H. De Vries and G. Narayanan are supported
by the FCRAO under the National Science Foundation grant AST 01-00793.

\clearpage

\clearpage

\clearpage
\begin{deluxetable}{lrrccllc}
\rotate
\tablecaption{\label{sources}Source List}
\tablecolumns{8}
\tablewidth{0pt}
\tablehead{\colhead{} & \colhead{} & \colhead{} & \colhead{Distance} &
  \colhead{} & \colhead{Associated} & \colhead{} & \colhead{IRAS Luminocity} \\
\colhead{Source} & \colhead{$\alpha$(2000)} & \colhead{$\delta$(2000)} &
\colhead{(pc)} & \colhead{Type} & \colhead{HII Region} & \colhead{IRAS Source}
& \colhead{($L_{\odot}$)}}
\startdata
SFO 4 & 00h 59m 04.1s & 60\deg53\min32\sec & 190\tablenotemark{a} & B & S185 &
00560+6037 & 5.2 \\
SFO 13 & 03h 01m 00.7s & 60\deg40\min20\sec & 1900\tablenotemark{a} & B &
IC1848 & 02570+6028 & 1300 \\
SFO 16 & 05h 19m 48.9s & --05\deg52\min05\sec & 400\tablenotemark{a} & A &
Barnard's Loop & 05173--0555 & 16 \\
SFO 18 & 05h 44m 29.8s & 09\deg08\min54\sec & 400\tablenotemark{a} & A &
$\lambda$ Ori & 05417+0907 & 18 \\
SFO 20 & 05h 38m 04.9s & --01\deg45\min09\sec & 400\tablenotemark{a} & C &
IC434 & 05355--0146 & 9.8 \\
SFO 25 & 06h 41m 03.3s & 10\deg15\min01\sec & 780\tablenotemark{a} & B &
NGC2264 & 06382+1017 & 100 \\
SFO 37 & 21h 40m 29.0s & 56\deg35\min58\sec & 750\tablenotemark{a} & C & IC1396
& 21388+5622 & 110 \\ \hline
B335 & 19h 37m 01.4s & 07\deg34\min02\sec & 250\tablenotemark{b} & Bok & none &
19345+0727 & 2.6 \\
CB 3 & 00h 28m 46.0s & 56\deg42\min06\sec & 2500\tablenotemark{c} & Bok & none
& 00259+5625 & 930 \\
CB 224 & 20h 36m 17.2s & 63\deg53\min15\sec & 450\tablenotemark{c} & Bok & none
& 20355+6343 & 3.9 \\
\enddata
\tablenotetext{a}{\citet{sfo}}
\tablenotetext{b}{\citet{tso}}
\tablenotetext{c}{\citet{lh}}
\end{deluxetable}

\begin{deluxetable}{llccc}
\tablecaption{\label{transitions}Observed Transitions}
\tablecolumns{5}
\tablewidth{0pt}
\tablehead{\colhead{} & \colhead {} & \colhead{Frequency} &
  \colhead{Velocity Res.} & \colhead{Beam Size} \\
\colhead{Transition} & \colhead{Telescope} & \colhead{(GHz)} & \colhead{(km
  s$^{-1}$)} & \colhead{(\arcsec)}}
\startdata
\hthcop (\jequals{1}{0}) & FCRAO & 86.754330 & 0.068 & 56 \\
\hcop (\jequals{1}{0}) & FCRAO & 89.188523 & 0.066 & 54 \\
\nthp (\jequals{1}{0}) & FCRAO & 93.173777 & 0.063 & 52 \\
CO (\jequals{1}{0}) & FCRAO & 115.271203 & 0.203 & 46 \\ \hline
CO (\jequals{2}{1}) & HHT & 230.538471 & 0.325 & 32 \\
\hthcop (\jequals{3}{2}) & HHT & 260.557619 & 0.288 & 29 \\
\hcop (\jequals{3}{2}) & HHT & 267.557625 & 0.280 & 28 \\
\hthcop (\jequals{4}{3}) & HHT & 346.998531 & 0.216 & 22 \\
\hcop (\jequals{4}{3}) & HHT & 356.734281 & 0.210 & 21 \\ \hline
\hthcop(\jequals{4}{3}) & CSO & 346.998531 & 0.042 & 22 \\
\hcop (\jequals{4}{3}) & CSO & 356.734281 & 0.040 & 21 \\
\enddata
\end{deluxetable}

\begin{deluxetable}{lcccccc}
\tablecaption{\label{coretable}Observations of N$_{2}$H$^{+}$}
\tablecolumns{7}
\tablewidth{0pt}
\tablehead{\colhead{} & \colhead{$T_{\rm ex}$} & \colhead{$V_{\rm LSR}$} &
  \colhead{$\Delta V$} & \colhead{} & \colhead{$N$} & 
  \colhead{$M_{\rm H_{2}}$} \\
\colhead{Source} & \colhead{(K)} & \colhead{(km s$^{-1}$)} & 
  \colhead{(km s$^{-1}$)} & \colhead{$\tau$} & \colhead{(cm$^{-2}$)} &
  \colhead{(M$_{\odot}$)}}
\startdata
SFO 13\tablenotemark{a} & \nodata & \nodata & \nodata & \nodata & \nodata & \nodata \\
SFO 16 & 5.4 (1.0) & 8.190 (0.002) & 0.589 (0.017) & 2.3 (0.9) & $5\times
10^{12}$ & 12 \\
SFO 18 & 4.2 (0.2) & 11.912 (0.015) & 0.964 (0.038) & 4.1 (0.6) & $10\times
10^{12}$ & 25 \\
SFO 20\tablenotemark{b} & \nodata & 13.007 (0.036) & 0.589 (0.093) & 0.6 (0.2) & $1\times
10^{12}$ & 3 \\
SFO 25 & 5.0 (1.3) & 7.287 (0.025) & 1.832 (0.082) & 0.8 (0.5) & $5\times
10^{12}$ & 50 \\
SFO 37\tablenotemark{b} & \nodata & 0.829 (0.103) & 1.499 (0.503) & 0.3 (0.2) &
$1\times 10^{12}$  & 13 \\
B335 & 3.6 (0.2) & 8.362 (0.005) & 0.376 (0.017) & 11 (2) & $8\times 10^{12}$
& 8 \\
CB~3 & 5.2 (8.6) & -38.843 (0.104) & 1.789 (0.367) & 0.459 (1.603) &
$3\times 10^{12}$ & 270 \\
CB 224 & 8.0 (6.4) & -2.729 (0.007) & 0.450 (0.020) & 0.785 (0.946) & $2\times
10^{12}$ & 8 \\
\enddata
\tablenotetext{a}{\nthp~(\jequals{1}{0}) was not detected in this source down
  to an RMS of 0.067~K.}
\tablenotetext{b}{The excitation temperature and optical depth were not well
  contrained due to the noise in this observation, however their combination
  was, so assuming an excitation temperature of 5 K yields the above mass
  estimates} 
\tablecomments{These are the star formation core properties. $\eta_{\rm MB}$ is
  0.5, $X$(N$_{2}$H$^{+}$) is $7\times 10^{-10}$, beamwidth is 52 arc seconds.}
\end{deluxetable}

\begin{deluxetable}{lcccccc}
\tablecaption{\label{hcoptable}Observations of HCO$^{+}$}
\tablecolumns{7}
\tablewidth{0pt}
\tablehead{\colhead{} & \colhead{$T_{\rm main}$} & \colhead{$T_{\rm iso}$} &
  \colhead{$\Delta V$} & \colhead{$T_{\rm ex}$} & \colhead{$N_{\rm main}$} &
  \colhead{$M_{\rm H_{2}}$} \\
  \colhead{Source} & \colhead{(K)} & \colhead{(K)} & \colhead{(km s$^{-1}$)} &
  \colhead{(K)} & \colhead{(cm$^{-2}$)} &
  \colhead{(M$_{\odot}$)}}
\startdata
SFO 4 & 0.221 (0.023) & $< 0.037$ & 3.0 & \nodata & $< 2\times 10^{13}$ & $< 2$ \\
SFO 13 & 0.464 (0.020) & 0.012 (0.014) & 4.0 & 4.2 & $1\times 10^{13}$ & 54 \\
SFO 16 & 0.717 (0.007) & 0.234 (0.007) & 2.0 & 4.3 & $1\times 10^{14}$ & 28 \\
SFO 18 & 0.980 (0.007) & 0.272 (0.006) & 2.0 & 4.9 & $1\times 10^{14}$ & 27 \\
SFO 20 & 0.841 (0.013) & 0.082 (0.008) & 1.5 & 4.6 & $3\times 10^{13}$ & 6 \\
SFO 25 & 1.241 (0.015) & 0.071 (0.012) & 5.0 & 5.5 & $7\times 10^{13}$ & 55 \\
SFO 37 & 0.436 (0.014) & 0.047 (0.006) & 3.0 & 3.7 & $5\times 10^{13}$ & 34 \\
B335   & 0.786 (0.012) & 0.280 (0.017) & 1.0 & 4.4 & $8\times 10^{13}$ & 6 \\
CB 3 & 0.222 (0.043) & 0.017 (0.019) & 2.5 & 3.2 & $2\times 10^{13}$ & 185 \\
CB 224 & 0.664 (0.018) & 0.302 (0.011) & 0.8 & 4.2 & $8\times 10^{13}$ & 21 \\
\enddata
\tablecomments{These are the star forming core properties. $\eta_{\rm MB}$ is
  0.46, $X$(HCO$^{+}$) is $1\times 10^{-8}$, beamwidth is 54.0 arc seconds,
  C/$^{13}$C ratio is 64.0 (TMC-1 paper).}
\end{deluxetable}

\begin{deluxetable}{lcccccccc}
\rotate
\tablecaption{\label{outflowtable}Outflow Characteristics}
\tablecolumns{9}
\tablewidth{0pt}
\tablehead{\colhead{} & \colhead{$M_{\rm tot}$} & \colhead{$P$} & \colhead{KE}
  & \colhead{$V_{\rm max}$} & \colhead{$\tau_{d}$} & \colhead{$\dot{M}$} &
  \colhead{$\dot{P}$} & \colhead{$L$} \\
  \colhead{Source} & \colhead{($M_{\odot}$)} & \colhead{($M_{\odot}$ km
    s$^{-1}$)} & \colhead{($M_{\odot}$ km$^{2}$ s$^{-2}$)} & \colhead{(km
    s$^{-1}$)} & \colhead{(yr)} & \colhead{($M_{\odot}$ yr$^{-1}$)} &
  \colhead{($M_{\odot}$ yr$^{-1}$ km s$^{-1}$)} & \colhead{($L_{\odot}$)}}
\startdata
SFO 13 RL & 8.9 & 75 & 310 & 8.4 & 80000 & $1.1\times 10^{-4}$ & $9.3\times
10^{-4}$ & 0.6 \\ \hline
SFO 16 BL & 0.3 & 2.4 & 10 & 8.2 & 35000 & $8.6\times 10^{-6}$ & $7.1\times
10^{-5}$ & $4.7\times 10^{-2}$ \\
SFO 16 RL & 0.4 & 2.3 & 6.7 & 5.8 & 49000 & $8.1\times 10^{-6}$ & $4.7\times
10^{-5}$ & $2.2\times 10^{-2}$ \\ \hline
SFO 18 BL\tablenotemark{a} & 0.5 & 2.4 & 6.0 & 4.9 & 39000 & $1.3\times 10^{-5}$ & $6.3\times
10^{-5}$ & $2.5\times 10^{-2}$ \\
SFO 18 RL\tablenotemark{a} & 0.4 & 2.4 & 7.4 & 6.1 & 37000 & $1.1\times 10^{-5}$ & $6.5\times
10^{-5}$ & $3.3\times 10^{-2}$ \\ \hline
SFO 20 BL & 0.01 & 0.03 & 0.04 & 3.0 & 25000 & $4.0\times 10^{-7}$ & $1.2\times
10^{-6}$ & $2.9\times 10^{-4}$ \\
SFO 20 RL & 0.01 & 0.02 & 0.02 & 2.0 & 38000 & $2.6\times 10^{-7}$ & $5.3\times
10^{-7}$ & $8.6\times 10^{-5}$ \\ \hline
SFO 25 RL & 2.3 & 24 & 130 & 10.7 & 24000 & $9.4\times 10^{-5}$ & $1.0\times
10^{-3}$ & 0.88 \\ \hline
CB 3 BL & 7.0 & 22 & 36 & 3.2 & 130000 & $5.4\times 10^{-5}$ & $1.7\times
10^{-4}$ & $4.5\times 10^{-2}$ \\
CB 3 RL & 3.2 & 9.0 & 13 & 2.8 & 150000 & $2.2\times 10^{-5}$ & $6.1\times
10^{-5}$ & $1.4\times 10^{-2}$ \\
\enddata
\tablenotetext{a}{SFO 18 values are based on CO~(\jequals{1}{0}) observations,
  $\eta_{\rm MB}$ is 0.45 in this case.}
\tablecomments{Assumes $\eta_{\rm MB}$ is 0.8, $X$(CO) is $1\times 10^{-4}$,
  $T_{\rm ex}$ is 30 K.}
\end{deluxetable}

\begin{deluxetable}{lcc}
\tablecaption{\label{deltavtable}Line Asymmetries}
\tablecolumns{3}
\tablewidth{0pt}
\tablehead{\colhead{Source} & \colhead{Millimeter Asymmetry} &
  \colhead{Submillimeter Asymmetry}}
\startdata
SFO 13 & $-0.11 \pm 0.10$ & $-0.32 \pm \phn0.19$ \\
SFO 16 & $-0.14 \pm 0.01$ & $\phantom{-}0.20 \pm \phn0.13$ \\
SFO 18 & $-0.48 \pm 0.02$ & $-0.59 \pm \phn0.06$ \\
SFO 20 & $-0.01 \pm 0.05$ & $\phantom{-}0.38 \pm \phn1.34$ \\
SFO 25 & $-0.01 \pm 0.02$ & $\phantom{-}0.03 \pm \phn0.12$ \\
SFO 37 & $\phantom{-}0.08 \pm 0.05$ & $-0.15 \pm \phn0.06$ \\ \hline
B 335 & $-0.09 \pm 0.03$ & $-0.13 \pm \phn0.04$ \\
CB 3 & $-0.84 \pm 0.14$ & $-3.79 \pm 11.83$ \\
CB 224 & $-1.20 \pm 0.06$ & $-0.27 \pm \phn0.16$ \\
\enddata
\end{deluxetable}
\end{document}